\documentclass[article,longauth,traditabstract]{aa} 
\usepackage{graphicx}
\usepackage{txfonts}
\usepackage{natbib}
\usepackage{epstopdf}
\usepackage{float}

\usepackage{longtable}
\usepackage{rotating}
\usepackage{pdflscape}
\usepackage{multirow}
\usepackage{color}

\begin{document}
\title{The dense cores and filamentary structure of the molecular cloud in Corona Australis: \emph{Herschel}\thanks{$Herschel$ is an ESA space observatory with science instruments provided by European-led Principal Investigator consortia and with important participation from NASA.} SPIRE and PACS observations from the \textit{Herschel} Gould Belt Survey}

\subtitle{}

\author{
	D. Bresnahan\inst{1}
		\and
		D. Ward-Thompson\inst{1}
		\and
		J. M. Kirk\inst{1}
		\and
		K. Pattle\inst{1}
		\and
		S. Eyres\inst{1}
		\and
		G. J. White\inst{2,5}
		\and
		V. K\"{o}nyves\inst{3}
		\and
	    A. Men'shchikov\inst{3}
		\and
		Ph. Andr\'{e}\inst{3}
		\and
		N. Schneider\inst{7,8}
		\and
		J. Di Francesco\inst{9,10}
		\and
		D. Arzoumanian\inst{3,11}
		\and
		M. Benedettini\inst{6}
		\and
		B. Ladjelate\inst{3}
		\and
		P. Palmeirim\inst{4}
		\and
		A. Bracco\inst{3}
		\and
		S. Molinari\inst{6}
		\and
		S. Pezzuto\inst{6}
		\and
		L. Spinoglio\inst{6}
		}

\institute{University of Central Lancashire, Preston, Lancashire, PR1 2HE, United Kingdom\\
		\email{dwbresnahan@uclan.ac.uk}
		\and
		Department of Physical Sciences, The Open University, Milton Keynes MK7 6AA, United Kingdom
		\and
		Laboratoire AIM, CEA/DSM-CNRS-Université Paris Diderot,IRFU/Service d’Astrophysique, CEA Saclay, 91191 Gif-sur-Yvette, France
		\and
		Institut d’Astrophysique Spatiale, UMR8617, CNRS/Université Paris-Sud 11, 91405 Orsay, France
		\and
		RAL Space, STFC Rutherford Appleton Laboratory, Chilton, Didcot, Oxfordshire, OX11 0QX, England
		\and
		INAF - Istituto Fisica Spazio Interplanetario, via Fosso del Cavaliere 100, 00133 Roma, Italy
		\and
		Universit\'{e} Bordeaux, LAB, UMR 5804, F-33270 Floirac, France
		\and
		CNRS, LAB, UMR 5804, F-33270 Floirac, France
		\and
		Department of Physics and Astronomy, University of Victoria, P.O. Box 355, STN CSC, Victoria, BC, V8W 3P6, Canada
		\and
		National Research Council Canada, 5071 West Saanich Road, Victoria, BC, V9E 2E7, Canada
		\and
		Department of Physics, Graduate School of Science, Nagoya University, Furo-cho, Chikusa-ku, Nagoya 464-8602, Japan
		}
		
\date{Accepted 2017 February ??. Received 2017 January 27 }
\titlerunning{Dense cores and filaments in CrA}

\label{firstpage}

\abstract{
We present a catalogue of prestellar and starless cores within the Corona Australis molecular cloud using photometric data from the \textit{Herschel} Space Observatory. At a distance of $d \sim 130$ pc, Corona Australis is one of the closest star-forming regions. \textit{Herschel} has taken multi-wavelength data of Corona Australis with both the SPIRE and PACS photometric cameras in a parallel mode with wavelengths in the range 70~$\mu$m to 500~$\mu$m. A complete sample of starless and prestellar cores and embedded protostars is identified. Other results from the \textit{Herschel} Gould Belt Survey have shown spatial correlation between the distribution of dense cores and the filamentary structure within the molecular clouds. We go further and show correlations between the properties of these cores and their spatial distribution within the clouds, with a particular focus on the mass distribution of the dense cores with respect to their filamentary proximity. We find that only lower-mass starless cores form away from filaments, while all of the higher-mass prestellar cores form in close proximity to, or directly on the filamentary structure. This result supports the paradigm that prestellar cores mostly form on filaments. We analyse the mass distribution across the molecular cloud, finding evidence that the region around the Coronet appears to be at a more dynamically advanced evolutionary stage to the rest of the clumps within the cloud.}

\maketitle
\section{Introduction}
Deeper insights into the physical processes behind star formation come to light when we observe the large scale structure of molecular clouds. The \textit{Herschel} space telescope \citep{pilbratt2010} had sensitivity into the realms of large scale structure, and this structure has been until recently a much unseen and unanalysed prologue in the story of star formation at these wavelengths. With \textit{Herschel} data of a typical example of a molecular cloud, we assess how filamentary structure plays its part across the complex; and unveils how the cloud might have been influenced by the surrounding environment. Key to this paper are the dense cores of which we present a survey, and the link between them and the filaments. Filamentary structure has been previously found within molecular clouds \citep{schneiderelmegreen1979,hartmann2002,myers2009,flagey2009}. However \textit{Herschel} results have evidenced a close link between the filamentary structure, and the prestellar dense cores, the latter having been found to reside within or very close to the filaments \citep{andre2010HGBS,andre2014ppvi,aquila2015HGBS,marsh2016}.\par
We present the "first-generation" catalogue of dense cores obtained using \textit{Herschel} data of the Corona Australis molecular cloud. This paper is laid out as follows: in Section 2, we discuss some of the previous observations and properties of this star forming region. In Section 3 we discuss the observations and data reduction. In Section 4 we derive a dust temperature map, column density maps, and identify filamentary structure within the data products. We also provide mass estimates with comparisons to previous surveys. In Section 5 we discuss the extraction method utilised within this work, as well as the classification procedures for starless cores and prestellar cores. We derive core properties using the \textit{Herschel} data and identify populations of bound and unbound cores.  In Section 6, we discuss the spatial distribution of sources, as well as the subregional properties across the star forming region. In Section 7 we discuss the global properties of these cores, and the subregional differences of their properties. In Section 8 we discuss the global filamentary links with the core sample, and profile one of the filaments to the west of the nucleus. In Section 9 we derive the mass distribution of column density across the cloud, and discuss the properties of the empirical probability distribution function of the column density map. Section 10 summarises the conclusions of this paper.\par
The Corona Australis molecular cloud (CrA) is a star-forming region located $129\pm11$ pc away (\citealp{casey98}; within this paper, we take $d = 130$ pc), and is a formation site for low-mass stars \citep{wilking1985, wilking1986,nutter2005}. This region has been imaged with PACS \citep{Poglitsch2010} and SPIRE \citep{Griffin2010} as part of the \textit{Herschel} Gould Belt Survey (HGBS -- \citealt{andre2010HGBS}). The Coronet is a small open cluster in CrA containing the variable star R CrA (cf. \citealt{siciliaaguilar2013}). R CrA is used in other texts to refer to the nucleus of the star-forming region.
\section{The Corona Australis molecular cloud}
\begin{figure*}
	\centering
	\resizebox{\hsize}{!}{
	\includegraphics{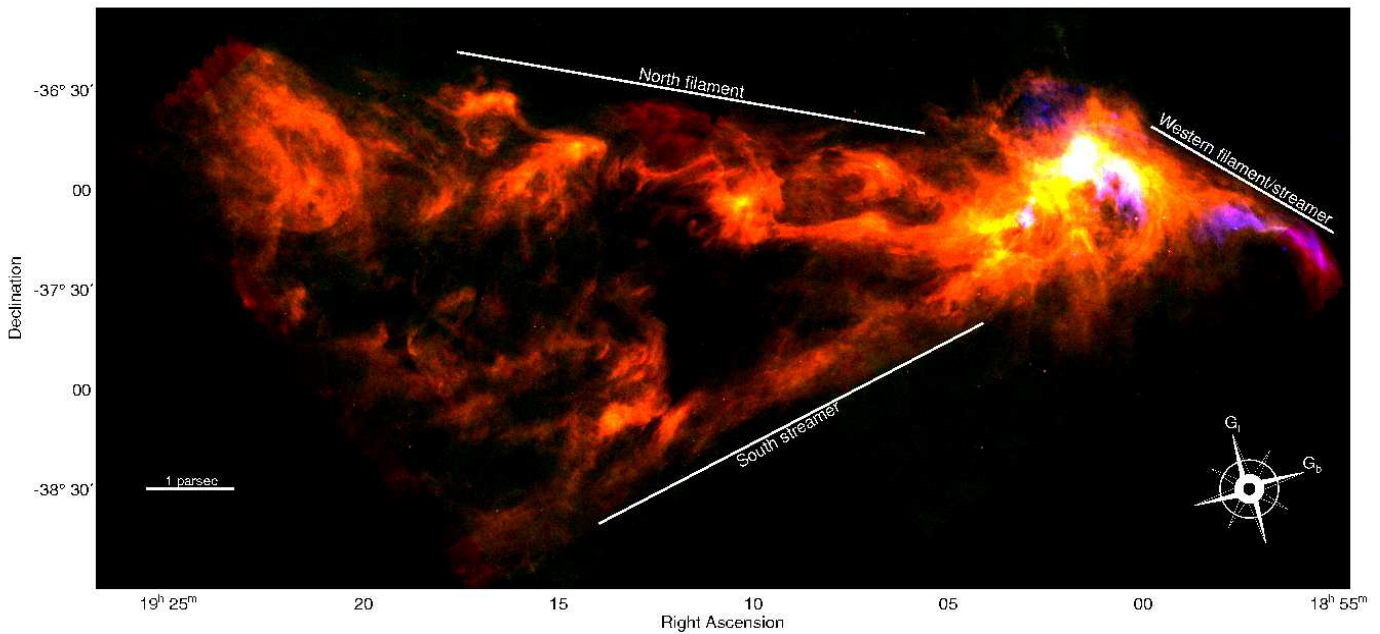}
	}
\caption{Three colour image of the Corona Australis molecular cloud. In purple/blue is \textit{Spitzer} 24-$\mu$m emission. Orange/yellow emission denotes SPIRE 160-$\mu$m emission. Red indicates SPIRE 250-$\mu$m emission. The northern tail, southern streamer, and westernmost filament/streamer are labelled. A bar indicating the length of 1 pc has been included to the lower left, and the relative orientation of the galactic coordinate system is shown to the lower right.}
\label{fig:Fig.1.}
\end{figure*}
Figure~\ref{fig:Fig.1.} shows a \textit{Spitzer-Herschel} three colour image of Corona Australis and its many interesting morphological features, the most striking of which is the Coronet region; the brightest area on Fig.~\ref{fig:Fig.1.}. The Coronet houses multiple variable stars including R, S and T CrA, and the variable nebula NGC 6729, associated with R CrA \citep{reynolds1916,graham1987}. \citet{peterson2011} made a comprehensive list of young stellar objects (YSOs) for the Coronet and its locale. Figure~\ref{fig:Fig.1.} shows \textit{Spitzer} \citep{werner2004} 24-$\mu$m emission in blue. The obvious blue emission in the west shows the warmer dust component that forms a streamer also highlighted by \citet{peterson2011}.  \citet{juvela2008,juvela2009,juvela2012} assessed part of this long streamer (referred to as a filament in those papers); and discussed the asymmetric column density profile of this filament. \citet{juvela2012} discussed the possibility of a stronger interstellar radiation field to the south of the streamer as a possible cause for the \textit{Spitzer} emission to be offset from the \textit{Herschel} emission. Along a small `stripe' perpendicular to a section of the filament, an asymmetry was seen in the column density calculated from \textit{Herschel} maps \citep{juvela2012}.\par
\citet{mamjek2001} calculated the 3D space motion of the CrA complex, concluding that the cloud is radially moving away from the Upper Centaurus Lupus (UCL) association at \textasciitilde7 km~s$^{-1}$, with the influence of UCL reaching beyond that of the CrA complex \citep{harju1993}. \citet{harju1993} proposed that the recession of CrA from Upper Centaurus Lupus could be an indication that CrA formed close to this OB association. Another possibility of CrA's origin, presented by \citet{neuhauser2000}, is a high velocity cloud collision. Their calculated velocity dispersions for the young stars around CrA, for which there are radial velocity measurements available, were low. These low dispersions give support to the scenario that CrA was formed by high-velocity cloud collisions, as presented by \citet{lepine1994}. The long filaments in CrA appear somewhat more wispy than some of the filaments in Taurus \citep{palmeirim2013}, which is expected due to the lower mass of the cloud overall. \citet{yonekura1999} mapped the complex with the NANTEN telescope in C$^{18}$O, locating several clumps along the northern filament and eight dense cores in total. \citet{cambresy1999}, \citet{andreazza1996} and \citet{vilas-boaes2000} also mapped the large-scale structure of CrA using extinction mapping, but there is a lack of the comprehensive, multi-wavelength data given by \textit{Herschel}, especially beyond the extents of the Coronet.\par
We here define the northern tail as being a filament, and the southern tail as being a streamer. We identify a filament as having a well defined crest, likely with fragmenting dense cores along it. Meanwhile a streamer is a broad, fainter structure which does not contain fragments (yet), and does not belong to a parent filament like the striations present in Taurus \citep{palmeirim2013} or Chamaeleon \citep{oliveirachameleon2014}. This approach seems reasonable as the filament clumps are often found in various surveys using different mapping techniques, but streamers may often be filtered out as part of the large-scale structure in ground-based telescope observations. The north filament appears to show sequential weakening in prominence between the successive dense cores to the east of the Coronet, with the clumps also becoming smaller. The southern streamer shows a faint assembly of structure. There is a galaxy, 2MASX J19065170-3657305, located directly behind the prominent crescent-shaped feature that lies to the north of the first clump east of the Coronet. We assume a constant distance to both CrA tails since they join up to the Coronet region. The northern filament is \textasciitilde$2.5^\circ$ across,which at a distance of $130$ pc, corresponds to \textasciitilde$5.7$~pc. The streamer, using the same argument, is  \textasciitilde$4.5$~pc across. The angle between the two tails is $~30-40^\circ$ in the plane of the sky.
\section{Observations and data reduction}
Observations of Corona Australis (OBSIDs: 1342206677--80) were conducted on 17 October 2010 with the common area mapped by the Spectral and Photometric Imaging REceiver (SPIRE; \citealt{Griffin2010}) and the Photodetector Array Camera and Spectrometer (PACS; \citealt{Poglitsch2010}) being \textasciitilde29 deg\textsuperscript{2}. Two sets of observations were taken simultaneously using the parallel mode. For each observation, \textit{Herschel} conducted two orthogonal scans of the field with a scanning speed of 60\arcsec s\textsuperscript{-1}.\
\subsection{PACS Data reduction}
The individual scan directions of the parallel-mode PACS data at 70~$\mu$m and 160~$\mu$m were reduced with \textsl{HIPE} \citep{ott2011} version 10.0, provided by the \textit{Herschel} Science Center.\par
Starting from the raw data (level-0) and up to the level-1 stage, standard steps of the default pipeline were applied. The PACS photometer flux calibration scheme was applied using the up-to-date responsivity and correction factors (PACS ICC report, \citeauthor{balog2014}) of the executed \textsl{HIPE} version with the calibration file set {PACS\_CAL\_45\_0. Processing requires several steps including trend correction, deglitching, and cleaning. The maps were converted from volts to Janskys. The PACS bolometers undergo a non-linear regime around and above the $100$ Jy level in all channels, leading to underestimation of flux densities for brighter targets. Columns 0 and 15 of the red bolometer array, and columns 0, 16, 32 and 48 of the blue bolometer array were masked to exclude pixels that had failed. Cosmic rays hitting the detectors were removed using the second-level deglitching method in \textsl{HIPE}.\par
The final step was completed using \textsl{Scanamorphos}, version 20 \citep{roussel2013scanamorphos}. Removal of long artefact glitches was completed following \citet{aquila2015HGBS}. The PACS data flux calibration error is given as $3\%$ and $\lesssim 5\%$ for 70~$\mu$m and 160~$\mu$m respectively. After individually reducing these observations and stacking them, the two observations were combined into a single mosaic. The final map products have a pixel size of 3\arcsec. The PACS half-power beam width (HPBW) sizes are shown in Table \ref{table:her_beams_pk_med}. We adopt conservative calibration errors of $12\%$ following \citet{paladini2012} and \citet{kirk2013}.
\subsection{SPIRE data reduction}
The data for all SPIRE channels were reduced using \textsl{HIPE} version 10.0. Similar to \citet{aquila2015HGBS}, the nominal and orthogonal scan directions were processed individually and then combined. By combining these observations individually, the slowly varying $1/f$ noise component is eliminated.\par
The raw level-0 data were processed to level-0.5 using the calibration tree SPIRE\_CAL\_10\_1 built into \textsl{HIPE}. Electrical cross-talk must be eliminated, along with correcting temperature drifts, deglitching and eliminating the effects of the cooler-outgassing caused by recirculating coolant. We do not utilise a colour correction at this level, but do calibrate for this effect in later after the source extraction process. The estimated difference on the column density is negligible at \textasciitilde$2\%$.\par
The destriper module in \textsl{HIPE} was used to correct the maps using baseline subtraction. The destriper has larger effects on the structured non-uniform backgrounds. Without correction, effects of the striping due to the calibration of the individual detectors are significant. Cross calibrating the detectors and referencing against a common median removes much of this effect. The final map products have pixel sizes of $6'', 10''$ and $14''$ for SPIRE 250~$\mu$m, 350~$\mu$m, and 500~$\mu$m, respectively. The SPIRE half-power beam width are shown in Table \ref{table:her_beams_pk_med}. After reducing the two observations, these observations were also combined in a mosaic. The calibration accuracy is given as \textasciitilde$5\%$ for point sources \citep{bendospirecal2013} and better than \textasciitilde$10\%$ for extended sources across all channels (cf. Konyves, 2015, for further discussion on map making and comparisons of PACS and SPIRE). The final reduced data are shown in Appendix~\ref{sec:app.data}. 

\section{Results and analysis}
\subsection{Dust temperature and column density maps}\label{sec:derivmaps}
We calculated column densities in the same manner as previous HGBS papers \citep{arzoumanian2011fils, kirk2013,palmeirim2013,aquila2015HGBS}.
Possessing multi-wavelength data allows calculation of cold dust properties of the complex.
\textit{Herschel} beams may be approximated by Gaussians (see the SPIRE and PACS observers manuals regarding details about the photometric calibrations). However \citet{aniano2011} highlighted the case for using optimised kernels to preserve the colour of extended sources when looking at multi-wavelength data. We therefore used the kernels of \citet{aniano2011} that have been made specifically for each instrument waveband on both \textit{Herschel} instruments to take their resolution to that of the lowest resolution. The procedure was limited by the resolution of the SPIRE 500-$\mu$m data (\textasciitilde$36''$) so all other wavebands were convolved to that resolution. We then converted the maps to units of MJy/sr, making the rescaling of flux not a function of the pixel size. The projections used in each mosaic for CrA were not centred on the pixel grid, instead being centred on the western-most observed field. We regridded the 500-$\mu$m data to a projection that is grid centred and then re-projected the other maps to this new regridded 500-$\mu$m map, such that the pixels are the same size and occupy the same sky position. We used the  \textsl{AstrOmatic TERAPIX} software \textsl{SWARP} \citep{SWARP2002} to do the regridding. This essentially creates a cube of data with two spatial axes and one wavelength axis.\par
\begin{table}
	\caption{\label{tab}Properties of \textit{Herschel} maps over all wavelengths. The first line lists the instrumental half-power beam width (HPBW). The second line lists the offset \textit{Planck/IRAS} monochromatic intensities about the CrA field (see text for details).}
\centering
\begin{tabular}{ l ccccc }
\hline\hline
Property & \multicolumn{5}{c}{Wavelength ($\mu$m)} \\
&	500	&	350	&	250	& 	160	&	70\\
\hline
HPBW ($\theta_{\textrm{beam}}$) (\arcsec)	&	36.3	&	24.9	&	18.2	&	13.5 &	8.4\\
Offset (MJy sr\textsuperscript{-1})	&	4.0	&	9.0	&	16.1	&	19.1	&	2.7\\	\hline
\end{tabular}
\label{table:her_beams_pk_med}
\end{table}
The \textit{Planck} observatory \citep{ade2011} High Frequency Instrument (HFI; \citealt{lamarre2010}) shared wavebands with the SPIRE instrument at 350~$\mu$m and 500~$\mu$m. \textit{Planck} has the advantage of being an all-sky survey telescope, and can be calibrated to absolute surface brightness relative to the cold microwave background (CMB). \textit{Herschel} images are created such that they have an arbitrary median, and we are able to bring the backgrounds to the same level as \textit{Planck} using \textit{Planck} data, following the method given by \citet{bernard2010}. Table~\ref{table:her_beams_pk_med} shows the offset values that were added to each of the maps within the \textit{Herschel} data.\par
With all wavelengths regridded onto the 500-$\mu$m data, we fitted spectral energy distributions (SEDs) to each pixel stack. The 70-$\mu$m data were not included in this fitting process. We adopted a greybody function to fit data at each pixel of the form
\begin{equation}
F_{\nu} = \frac{MB_{v}(T)\kappa_{\nu}}{D^{2}}
\end{equation}
where $F_{\nu}$ is the flux density at frequency $\nu$, $M$ is the mass per pixel, $B_{v}(T)$ is the Planck function with a temperature $T$; $\kappa \propto \nu^{\beta}$  is the dust mass opacity, and $D$ is the distance to the source (cf. \citealt{kirk2013}). We adopted a power-law approximation to the dust opacity law per unit mass (dust+gas) at submillimetre wavelengths, given by the relation $\kappa_{\lambda}=0.1 \times (\lambda /$ 300 $\mu$m$)^{-\beta}$ cm$^{2}/$g \citep{aquila2015HGBS}. \citet{roy2014} estimated that this value for the opacity law is appropriate to better than $50\%$. The dust emissivity index $\beta$ remained fixed at 2 \citep{hildebrand1983}. Our choice of a fixed value of $\beta$ is to avoid the degeneracy between temperature and $\beta$ during the SED fitting process. \citet{sadavoy2013} found that using \textit{Herschel} only wavelengths provides a good estimate of core mass and temperature, and that for values of $1.5\leq \beta \leq 2.5$, the temperatures vary by less than 2 K. We used the \textsl{IDL} Levenberg–Marquardt least squares fitting procedure \textsl{MPFIT} \citep{markwardt2009} to fit the greybody to the regridded fluxes.\par
Each data point has an associated error calculated such that $\sigma^{2} = \sigma_{\text{rms}}^{2} + (F_{\text{H}}C_{\text{H}})^{2} + (F_{\text{B}}C_{\text{B}})^{2}$ where $\sigma_{\text{rms}}$ is the rms pixel intensity variation within the regridded maps, following \citet{kirk2013}. We calculated the rms by subtracting a version of the map that is smoothed by a Gaussian of $2\theta_{\text{beam}}$, where $\theta_{\text{beam}}$ is the half-power beam width at the given wavelength. $F_{\textrm{H}}C_{\textrm{H}}$ is the fractional uncertainty within the \textit{Herschel} calibration multiplied by the flux density at that pixel; this is \textasciitilde$10$\% for SPIRE bands \citep{spireom0798} and \textasciitilde$12$\% for PACS bands \citep{paladini2012}. $F_{\text{B}}C_{\text{B}}$ is the calibration error in the \textit{Planck} instruments multiplied by the \textit{Planck} offset value. The calibration error in the \textit{Planck} offset values is $C_{\text{B}} = 5\%$ \citep{bernard2010}. \par
The mass-per-pixel was then converted to a hydrogen column density using $N(\textrm{H}_{2}) = M / m_{\text{H}} \mu A$; where $N(\textrm{H}_{2})$ is the column density of H$_{2}$, $M$ is the mass, $m_{\textrm{H}}$ is the mass of a hydrogen atom, $\mu$ is the mean particle mass ($\mu = 2.86$) assuming \textasciitilde$70\%$ H$_{2}$ by mass, and $A$ is the area of each pixel. \par
The result is a map at the $36''$ resolution of the SPIRE 500-$\mu$m data. A `high resolution' column density map was then derived using the unsharp masking method described by \citet{palmeirim2013}, to retrieve a map that has the same resolution as the \textit{Herschel} 250-$\mu$m data. An important note to make is that while the initial column density maps were all created with the use of the kernels resulting from the work of \citet{aniano2011}, the fundamental assumption of an unsharp mask is that Gaussian kernels are used. We hence used Gaussian kernels, which were created using the effective beams, in the procedure that produces the high-resolution column density map. Furthermore, it is also important to test the correspondence between the two column density maps. To this end, we employed the same technique as \citet{palmeirim2013} by smoothing the high-resolution column density maps back to the lower resolution column density map \citep[Appendix A]{palmeirim2013}. The mean ratio between the low- and high-resolution column density maps was approximately unity in areas above the $1\sigma$ rms level of the high-resolution map, indicating excellent agreement between the two maps. To check the robustness of the column density maps, we compared the low-resolution column density map to optical extinction maps created by \citet{dobashi2005}. The procedure and results of this test are shown in Appendix~\ref{sec:app.dobashi}. The high-resolution column density map is shown in Fig.~\ref{fig:hi_res_coldens}. The $36''$ resolution dust temperature map is shown in Fig.~\ref{fig:lo_res_temp}.\par

\begin{figure*}[!]
	\centering
		\includegraphics[width=\hsize, trim={1cm 2cm 1cm 0.5cm}]{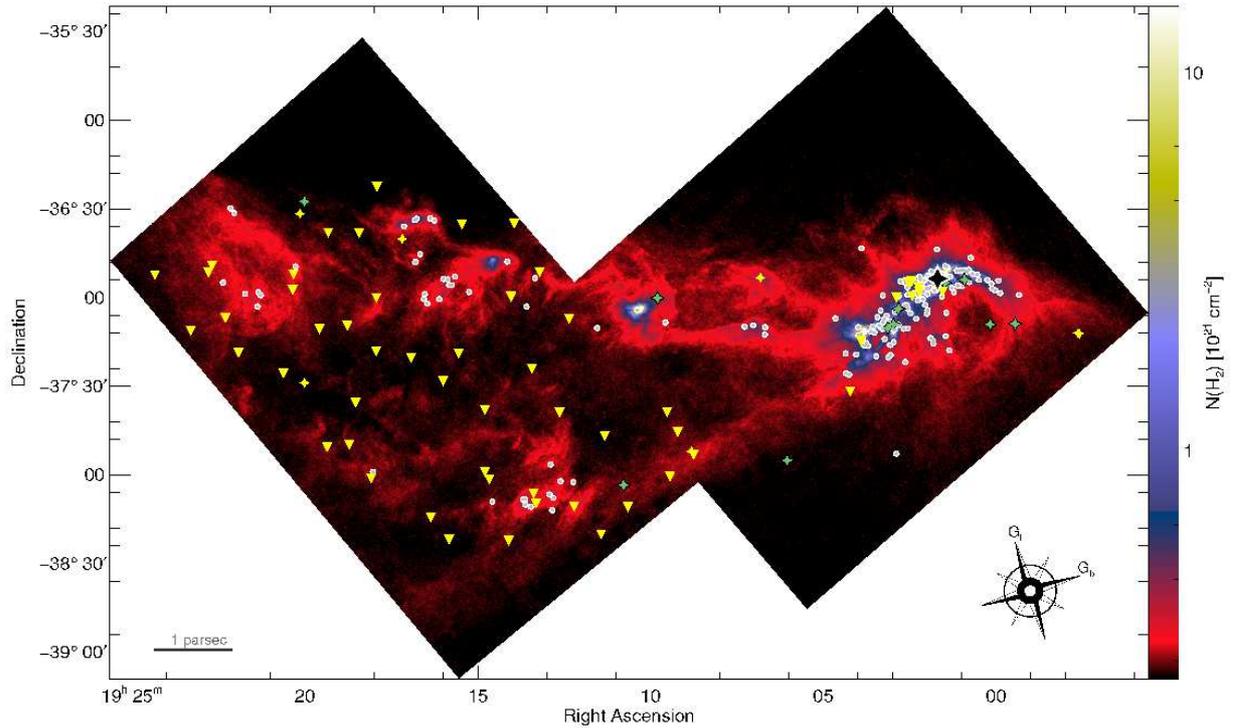}
\caption{H$_2$ column density map of the Corona Australis molecular cloud at $18.2''$ angular resolution, derived from the \textit{Herschel} data using the method discussed in Sect. \ref{sec:derivmaps}. The grey filled circles with white outlines, and filled green stars with back outlines, are the locations of the 163 starless cores and 14 protostellar cores identified using \textit{Herschel} images with \textsl{getsources} (see Sect. \ref{sec:extraction}). The black star towards the west is the location of the Coronet cluster, containing the well studied variable star R CrA. There are 62 faint sources, shown by yellow downward triangles and stars, which are contained within the catalogue, but excluded from the scientific discussion on the ground that they are likely to be extragalactic (see Appendix~\ref{sec:completeness} for details). The relative orientation of the galactic coordinate axes is indicated at the lower right.}
\label{fig:hi_res_coldens}
\end{figure*}
\begin{figure*}[!]
	\centering
		\includegraphics[width=\hsize,  trim={1cm 2cm 1cm 0.5cm}]{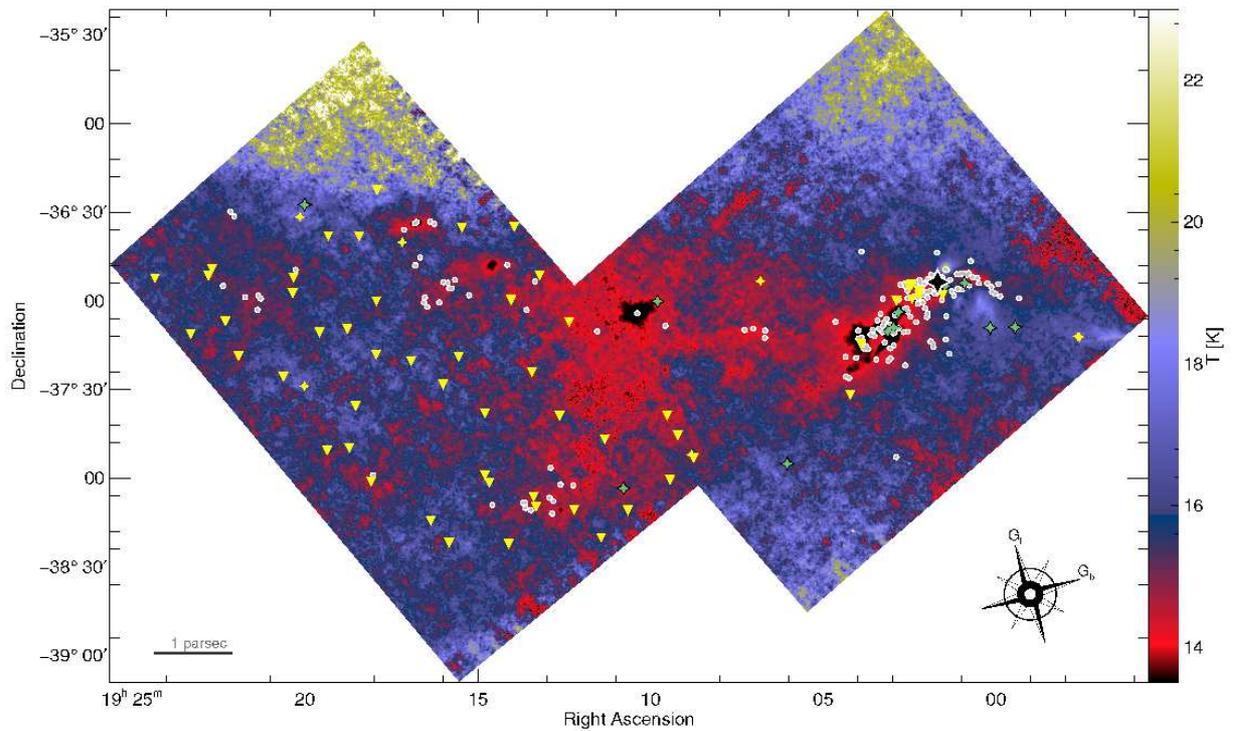}
\caption{Dust temperature map of the Corona Australis molecular cloud at $36''$ resolution, as derived from HGBS data (see Sect \ref{sec:derivmaps}). The black star towards the west is the location of the Coronet cluster. The relative orientation of the galactic coordinate axes is indicated at the lower right.}
\label{fig:lo_res_temp}
\end{figure*}
\begin{figure*}
\centering
		\includegraphics[width=\hsize, trim={1cm 4cm 1cm 4cm}]{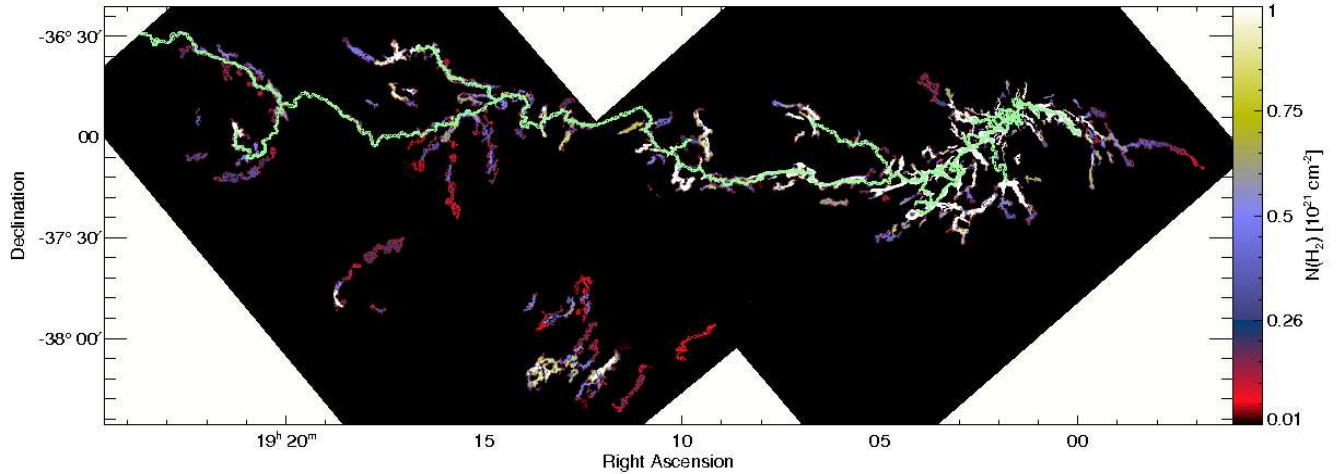}
\caption{The network of filaments as traced by \textsl{getfilaments} \citep{getfilaments2012} in the \textit{Herschel} high-resolution column density map of the Corona Australis cloud. Angular scales up to \textasciitilde$145''$ (\textasciitilde$0.37$~pc at $d=130$~pc are shown in the above image for visualisation purposes). Overlaid in green are filaments found by \textsl{DisPerSE}, using the parameters mentioned in Section \ref{sec:filamentstruc}, and run on the high resolution column density map.}
\label{fig:hi_res_fils}
\end{figure*}

\subsection{Filamentary structure of Corona Australis}\label{sec:filamentstruc}
\textit{Herschel} has so far provided much evidence for the ubiquity of filamentary structure within molecular clouds \citep{andre2010HGBS,konyves2010, dwt2010polaris,hill2011,schneider2012rosette,kirk2013,palmeirim2013,andre2014ppvi,aquila2015HGBS,benedettini2015}, and Corona Australis is no different. Filamentary structure identification can be difficult due to the dynamical range of the \textit{Herschel} data, so it is necessary to employ algorithmic approaches to extract this structure. What constitutes a filament in the \textit{Herschel} data is not only a semantic argument, but a mathematical one, as shown for example in the different approaches of \citet{getfilaments2012} and \citet{sousbie2011}. \citet{aquila2015HGBS} define a filament for the purposes of the HGBS in Aquila as having a minimum aspect ratio of \textasciitilde3 and a minimum average column density excess of \textasciitilde$10\%$ with respect to the local background along the length of the filament.\par
We first employed \textsl{getfilaments} \citep{getfilaments2012} to trace the filamentary structure. \textsl{getfilaments} is able to analyse \textit{Herschel} data over multiple wavelengths and at multiple scales. By assessing the highly filtered single-scale images, \textsl{getfilaments} is able to select filamentary structure that is represented in footprints at variable transverse angular scales as well as produce smooth, single-pixel-wide skeletons.\par
Next we employed the \textsl{DisPerSE} algorithm \citep{sousbie2011} to trace filaments. \textsl{DisPerSE} operates by locating persistent topological structures such as voids, filaments, and peaks within the filamentary structure. \textsl{DisPerSE} has already been used in many papers for the HGBS (e.g.,  \citealt{arzoumanian2011fils,peretto2012,palmeirim2013,schneider2012rosette}). Some visual analysis is required with all extraction algorithms, so to find the most `visually optimal' set of filaments we ran \textsl{DisPerSE} over two parameter spaces; smoothness and assembly angle, over the high resolution map.\par 
The smoothness forces the single-pixel-wide skeletons to be smoothed over $n$ pixels, while the assembly angle collapses multiple skeletons into single skeletons if they have an angle between them of less than the assembly angle. Running \textsl{DisPerSE} for skeleton smoothnesses of $4\leq n \leq 10$ and assembly angles of $40^{\circ} \leq \theta \leq 70^{\circ}$ with the order being to \textit{smooth}, \textit{breakdown} and then \textit{assemble}. We use the parameters here where $n = 6$ and $\theta = 45^{\circ}$ based on visual inspection. Following \citet{aquila2015HGBS}, we use a 5$\sigma_{\textrm{rms}}$ as a persistence value which in CrA is calculated by the method in Sect. 4.1 with a value of $1.6\times10^{21}$~N(H$_{2}$)~cm$^{-2}$. Reconstructing filaments can be approached using the `toFITS' option, constructing single-pixel-wide skeletons on the same grid as the original column density map, to assess the structure. After reconstructing the filaments on these grids, we then began `cleaning' the filaments. As discussed by \citet{palmeirim2013}, the creation of a high-resolution map heightens the noise level. In Corona Australis, for example, \textsl{DisPerSE} does not highlight the southern tail as a filament. We gain advantages, however, in that some areas become more well-defined for the purposes of filament finding. The filaments found by both \textsl{getfilaments} and \textsl{DisPerSE} are shown in Fig.~\ref{fig:hi_res_fils}. For \textsl{getfilaments} identified filaments, the angular scales up to 145$\arcsec$ are shown. The westernmost filament found by \textsl{DisPerSE} has been truncated by hand because the remainder appeared to have been affected by cirrus in this region. For clarity, a zoom-in of filaments in the region around the Coronet is in Appendix~\ref{sec:app.coronetfils}.
\subsection{Mass estimation}
The total cloud mass is given directly from the SED fitting process, as we define $M$ to be in solar masses with the column density calculated later. With an adopted distance of $D=130$ pc, the total cloud mass is \textasciitilde$820$ M$_{\odot}$. Our estimate of mass supports that of \citet{yonekura1999} who used NANTEN C$^{18}$O emission over a much wider area, almost matching the size scale of the \textit{Herschel} maps, to produce a mass estimate of \textasciitilde$900$ M$_{\odot}$. We can also compare with the results of \citet{harju1993} who used C$^{18}$O measurements from the Swedish-ESO Submillimetre Telescope (SEST) telescope to measure the mass of the large scale structure about the Coronet \citep[Fig.~1]{harju1993}. Our result is \textasciitilde$160$ M$_{\odot}$ which compares well with their mass estimate of \textasciitilde$110$ M$_{\odot}$, bearing in mind \textit{Herschel}'s greater sensitivity to the large-scale structure, particularly the cirrus and striations around the filaments. 
\begin{figure*}[!]
	\centering
	\resizebox*{0.90\hsize}{!}{
		\includegraphics[angle=0, trim={0cm 1.2cm 0cm 0cm}]{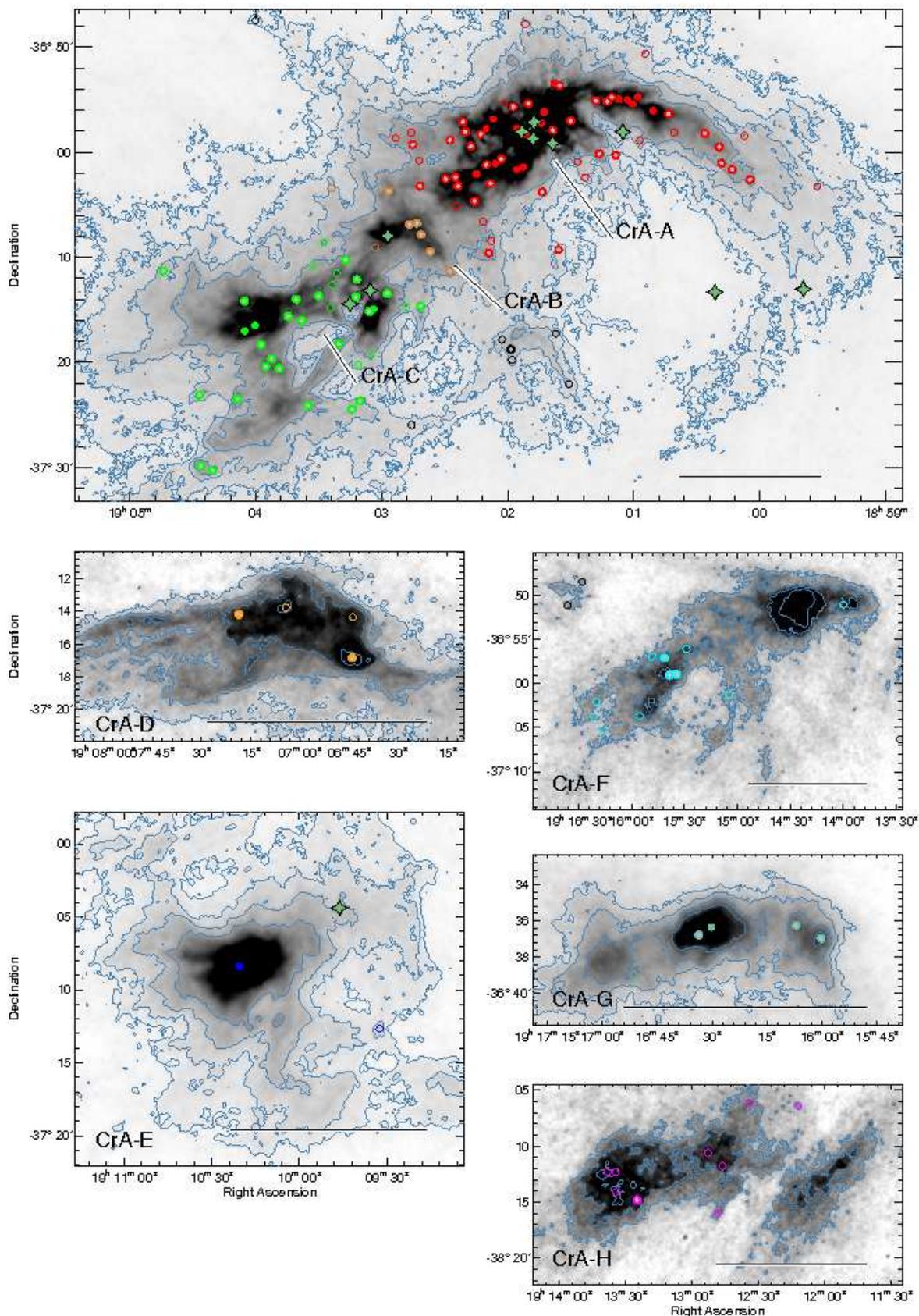}
	}
\caption{Division of the Corona Australis molecular cloud into significant 'clumps' of emission that are prominent in all emission maps and the column density maps. The hollow, grey-filled, and colour-filled circles are starless cores, candidate prestellar cores, and robust prestellar cores, respectively. The following colours are used for the symbols; red for CrA-A; brown for CrA-B; green for CrA-C, orange for CrA-D; dark blue for CrA-E; cyan for CrA-F; sea green for CrA-G and magenta for CrA-H. The green stars represent YSO/protostellar candidates. Cores with no defined subregion are black. The contour levels start from 3$\sigma\sim10^{21}$~N(H$_{2}$)~cm$^{-2}$, and increase in levels of 1.5 times the previous level. The black bars towards the lower-right of each cut-out show a 0.5 pc transverse distance at the adopted distance of CrA.}
\label{region_map}
\end{figure*}
\section{Multiwavelength core selection with getsources} \label{sec:extraction}
Like the rest of the HGBS, we used \textsl{getsources} to extract the starless core and YSO sources from the \textit{Herschel} data. \textsl{getsources} is a multi-wavelength, multi-scale source extraction algorithm \citep{getsources2012}. The HGBS first-generation catalogue of cores is given in Appendix~\ref{sec:app.catalogue} and was produced with the "November 2013" major release of \textsl{getsources} (version~1.140127). We give a brief overview of the methodology employed by \textsl{getsources} to extract sources. The algorithm comprises two stages: detection and measurement.\par
At the detection stage, \textsl{getsources} analyses `single-scale' images, which are fine spatial decompositions of the original `observed' images. This occurs for all of the input wavebands, across a range of spatial scales. The decomposition filters out the spatial scales that are irrelevant at the scale which is being analysed. This works particularly well in crowded regions, and for extended sources. For each single-scale image, \textsl{getsources} finds the 3$\sigma$ and 6$\sigma$ intensity levels in order to separate significant sources from the background fluctuations and the noise. An advantage to this is that filamentary structure, which is ubiquitously associated with sources, can also be excluded from the source itself. \textsl{getsources} is able to combine multi-wavelength data in order to construct fully wavelength-independent detection catalogues. By using these methods, \textsl{getsources} is able to extract sources which are weak, and may sometimes be excluded using only a single-wavelength.\par
Sources are detected by \textsl{getsources} in the combined single-scale detection images. This involves tracking the evolution of the source intensity as well as its segmentation mask (see Sect. 2.5 of \citealt{getsources2012}). The spatial scale on which the source is brightest determines its characteristic size, and the footprint size. The footprint is defined as the group of pixels which give a non-negligible contribution to the integrated source flux. The first moments of intensity are used to determine the source position. \textsl{getsources} analyses the wavelength-combined single-scale detection images, between the scale at which the source first appears, and that at which it is brightest. The source positions are therefore weighted towards the wavebands with higher angular resolution, and the larger scale emission does not contribute to these derived positions.\par
The measurement stage involves the use of the observation image at each wavelength. A more sophisticated scheme of background estimation as well as source deblending is employed by \textsl{getsources}. The background is subtracted by using the border pixels surrounding the source footprint. A linear interpolation is made under the source footprint using these pixels, and is constrained by the angular resolution of each waveband. The footprint of the source must be at least as large as the observational beam size, and the extent must correspond to the source intensity distribution at each wavelength.\par
Overlapping sources are deblended iteratively. The pixel intensity is split between sources by assuming a simple shape for their intensity distributions. The deblending shape is defined like the Moffat distribution (see equation 14 of \citealt{getfilaments2012}, and \citealt{moffat1969}). This particular distribution has a Gaussian-like circular profile, with stronger power-law wings, which should approximate the intensity distributions of observed sources. The peak and integrated intensity uncertainties are given by standard deviations, estimated within elliptical annuli, just outside of the source footprints. In crowded regions, the annuli are taken outside of any of the overlapping sources up to a usable area of at least 20 beam sizes. Aperture corrections are applied by \textsl{getsources}, using tables of encircled energy fraction values for the PSFs provided by the PACS and SPIRE ICCs (\citealt{balog2014,bendo2011}).\par
This `first generation' catalogue of starless, prestellar, and protostellar cores employs two methods to extract sources reliably. The peak continuum emission at 160~$\mu$m can be offset from peaks of column density by an illuminating, anisotropic radiation field (see, e.g. \citealt{nutter2009,juvela2012}). In order to locate sources we followed the approach used by \citet{aquila2015HGBS}, who derived a temperature-corrected 160-$\mu$m map. An approximate map of column density is calculated using the colour-temperature map derived by using the intensity ratio of the 160-$\mu$m and 250-$\mu$m data (cf. \citealt[Appendix~A]{palmeirim2013}). The high-resolution column density map is used at the detection stage along with the temperature-corrected 160-$\mu$m map, and the 250-$\mu$m, 350-$\mu$m, and 500-$\mu$m data in \textsl{getsources}, improving the detectability of dense cores. Corona Australis has a very diffuse tail, and several artefacts were blocked out of the \textsl{getsources} extraction process by masking areas with reasonably low emission. Appendix D shows the two fields that were used for the extractions. Each field was extracted separately. \par
The next stage is to perform an independent extraction to detect YSOs and protostars. This essentially creates two complementary extractions; one for the dense cores, and one for the protostars and YSOs. The 70-$\mu$m emission traces these point-like objects (e.g., \citealt{lindberg2014,peterson2011}). Hence, we are able to compare the results of extracting sources from 70-$\mu$m data with those of the previous dense core extraction. The ability to separate YSOs from starless cores well in this manner is highlighted by \citet{aquila2015HGBS}.
\subsection{Selection and classification of reliable core detections}
Across the HGBS, several criteria are used in order to filter and categorise the populations of sources from the extractions. These selections are based on the raw catalogue products of \textsl{getsources} (see \citealt{aquila2015HGBS} for more details). \par
\subsubsection{Selection of candidate dense cores (either starless or protostellar)
from the “core” set of extractions}
Several criteria were used to clean the catalogue, taking advantage of the multi-wavelength data available from \textit{Herschel}.
Firstly, a column density detection significance greater than 5, where the detection significance refers to a single-scale analogue of a classical signal-to-noise ratio (S/N) [see Eq.~(17) of \citet{getsources2012}] in the high-resolution column density map. 
There was global detection significance over all wavelengths [see Eq.~(18) of \citet{getsources2012}] greater than 10. The sources' global “goodness” were $\geq$~1, where goodness is an output quality parameter of getsources, combining global signal-to noise ratio and source reliability, and defined in Eq.~(19) of \citet{getsources2012}. The column density measurement signal-to-noise ratio (S/N) was greater than 1 in the high-resolution column density map. The monochromatic detection significance was greater than 5 in at least two bands between 160~$\mu$m and 500~$\mu$m. The flux density measurement had a S/N~$>$~1 in at least one band between 160~$\mu$m and 500~$\mu$m, for which the monochromatic detection significance is simultaneously greater than 5. All of the above criteria had to be met \citep{aquila2015HGBS}.
\subsubsection{Selection of candidate YSOs from the “protostellar” set of
extractions}
Once again, a set of criteria was established. Firstly, the monochromatic detection significance was greater than 5 in the
70-$\mu$m band. There was a positive peak and integrated flux density at 70~$\mu$m.
The global “goodness” was greater than or equal to 1. The flux density measurements had a S/N~$>$~1.5 in the 70-$\mu$m band. The FWHM source size at 70~$\mu$m is smaller than 1.5 times the 70-$\mu$m beam size.
Finally, the estimated source elongation is less than 1.30 at 70~$\mu$m, where the source elongation is defined as the ratio of the major and minor FWHM sizes. Again, all of these criteria had to be met \citep{aquila2015HGBS}. 

\subsubsection{Selection of candidate starless cores and protostellar cores}

After cross-matching the selected dense cores with the candidate
YSOs/protostars, a selected dense core is classified as
`starless’ if there is no candidate 70-$\mu$m YSO within its half-power
(high-resolution) column density contour. Conversely, a selected dense core is classified as `protostellar’ if there is a candidate 70-$\mu$m YSO within its half-power column density contour. The most reliable SED of a selected protostellar core is obtained by combining the 70-$\mu$m flux density from the “protostellar” extractions with the 160-$\mu$m, 250-$\mu$m, 350-$\mu$m, and 500-$\mu$m flux densities from the “core” extractions.
After automatic processing of the source catalogue, we conducted further checks to help categorise sources and create flags for these objects in the catalogues where required.\par
Corona Australis is \textasciitilde$17^{\circ}$ below the Galactic Plane. While this location is advantageous, in that the inclusion of more clouds along the line of sight is much less probable at this latitude, background galaxies in the catalogues must be identified. To eliminate these objects from the cores/YSO selection, we cross-matched the sources with those of the NASA Extragalactic Database (NED; \citealt{Mazzarella2007}), and identified several candidates for removal from the catalogue.\par 
An important caveat to the results in this region, particularly with regard to the number and spatial distribution of starless cores, is the uncertainty in eliminating extragalactic candidates that have no NED classification. Though we took care to crosscheck these sources with known extragalactic candidates, some compact sources may remain incorrectly classified. Optical surveys concerned with the identification of extragalactic sources are often obscured by Galactic dust emission, so the completeness of these extragalactic catalogues decreases towards Galactic star-forming molecular clouds. We found that 30 of the 540 automatically selected sources had NED extragalactic catalogue matches - approximately 6\%. We also checked within 1\arcmin~of each source for SIMBAD database matches. The SIMBAD matches are also given in the online table.\par 
The catalogue of YSO candidates given by \citet{peterson2011}, using multiple wavelengths from IRAC \citep{fazio2004} and MIPS \citep{rieke2004}, was also used. These data are especially well-suited for use around the coronet where \textit{Herschel} 70-$\mu$m sources are blended.  We cross-matched the core/protostellar catalogue, with candidates of \citet{peterson2011} which are classified as I, II, or III. In addition, the WISE catalogue provided by \citet{marton2015} of YSO candidates was also used for the wider area not covered by \textit{Spitzer}. We give the nearest \textit{Spitzer} and WISE match, should one exist in the catalogue, for each source (see Appendix~\ref{sec:app.catalogue}).\par
The reflection nebulae around the Coronet were spuriously extracted by \textsl{getsources}. We utilised the archival 24-$\mu$m data from \textit{Spitzer} MIPS to cross check peaks in the PACS emission, and identify sources that are associated with nebulosity.\par
\cite{DWT1985CRA} studied the reflection nebulae NGC 6726 and 6727, which both appear in the 70-$\mu$m emission; NGC 6726 is illuminated by TY CrA and HH 176386, NGC 6729 is illuminated by the variable stars R and T CrA, and these reflection nebulae are identified within the catalogue. \par
We also used two other source extraction algorithms, in a similar manner to \citet{aquila2015HGBS}, to generate two flags to indicate the robustness of the source detection.\par
Namely, we used \textsl{CSAR} (Cardiff Sourcefinding AlgoRithm - \citealt{kirk2013,kirk2015androm}) and the CUrvature Thresholding EXtractor (\textsl{CuTEx} - \citealt{molinari2011}). There are advantages to using both of these algorithms. For example, \textsl{CSAR} is able to preserve hierarchical information regarding the density structure of the cloud and \textsl{CuTEx} is able to deblend sources in closely packed regions such as the Coronet. \par
We cross-matched the \textsl{getsources} sources with those found by \textsl{CSAR} and \textsl{CuTEx}. \textsl{CSAR} sources have co-ordinates that are the centroids of the source masks. We used three different methods to match the sources found by \textsl{getsources} to those found by \textsl{CSAR}. The first involved identifying any \textsl{CSAR} or \textsl{CuTEx} source positions (i.e., peak coordinates) within the $6''$ of each \textsl{getsources} source. The second involved identifying any \textsl{CSAR} or \textsl{CuTEx} source positions within the $18.2''$ FWHM elliptical contour of each \textsl{getsources} source. The third, which is constrained to \textsl{CSAR}, involves identifying any \textsl{getsources} cores that are located directly on top of the \textsl{CSAR} masks for those sources. The latter method identifies sources that are regarded as single sources by \textsl{CSAR}, but are multiple sources deblended by \textsl{getsources}. Table \ref{table:crossmatchsourcefind} shows the percentages of starless and prestellar cores identified by \textsl{getsources} that are also identified by \textsl{CuTEx} and \textsl{CSAR}, using the second ellipse matching method, and the \textsl{CSAR} mask checking method.\par
The low fraction of \textit{getsources} identified unbound starless cores which are found by the \textsl{CSAR} and \textsl{CuTEx} ellipse cross-matching method can be explained by the morphology of these cores. The extended and somewhat more diffuse nature of these sources typically means they are filtered out by \textsl{CuTEx} during the process of creating the double derivative maps. Their intrinsically lower contrast and flat-topped nature means that \textsl{CSAR} does not identify these starless cores as objects with a well-defined peak. These unbound, diffuse cores also tend to appear away from the dense filaments and clumps. Inspection of Fig.~\ref{region_map} shows that many of these starless cores are located on low column density backgrounds. \par
We created `card' images, two examples of which are shown in Appendix~\ref{sec:app.catalogue}. These cards, along with the position of the source in the column density map, were used to visually inspect the sources. At this stage, we discarded ~35\% of the 540 automatically selected cores, leaving 354.\par
Contamination by unconfirmed extragalactic sources increases where the morphology of the cloud is more cirrus-like. We identified sources with 250-$\mu$m integrated flux densities below 100 mJy as being most likely extragalactic in nature, even though most of these sources do not belong to the NED extragalactic database. This was a particular problem within the eastern extraction field, where the morphology of the molecular cloud is akin to Polaris (see \citealt{dwt2010polaris}). The procedure is outlined in Appendix \ref{sec:completeness}. We also excluded sources that had spuriously low masses from the scientific discussion. Though these sources are likely true astronomical sources, the median temperature for these cores was also high, at \textasciitilde22 K. A further 115 sources were eliminated at this stage.\par
Our \textsl{getsources} selection and classification procedure resulted in a final sample of 177 dense cores (not counting 62 additional low-mass objects listed in the online catalogue which possess 250-$\mu$m integrated flux densities between 100 mJy and 150 mJy, and may well be extragalactic; or possess a mass lower than 0.001 M$_{\odot}$ after the fitting procedures described in Sect. \ref{sec:derivcoreprop}), including 163 starless cores, and 14 protostellar cores. Figure \ref{region_map} shows the sample of dense cores and protostellar cores overlaid on the high-resolution column density map. The grey and white dots show the dense cores, and the green stars show the locations of 14 protostellar cores. The yellow downward pointing symbols mark the locations of the 62 objects included within the catalogue, but excluded from further analysis. As discussed in Sect. \ref{sec:selfgrav} below (see also Fig.~\ref{mass_size}), we classified 23 of the starless cores as robust prestellar cores. A further 76 of the starless cores were classified as candidate prestellar cores. An example of the distribution of sources found by \textsl{getsources}, \textsl{CSAR} and \textsl{CuTEx} is given in Appendix~\ref{sec:app.catalogue}.
\begin{table}
	\caption{Percentages of \textsl{getsources} sources found using \textsl{CSAR} and \textsl{CuTEx}. The two rows for CSAR indicates the two methods that were used to locate corresponding sources between the two extraction algorithms.}
\centering
\begin{tabular}{ l ccc }
\hline\hline
&	Starless	& Candidate  & Robust \\
&		& prestellar & prestellar\\
\hline
\textsl{CSAR}$_{ellipse}$ &  3\% & 34\% & 47\% \\
\textsl{CSAR}$_{mask}$ & 43\% & 65\% & 91\%	\\
\textsl{CuTEx}$_{ellipse}$ & 18\% & 74 \% & 100\%	\\	\hline
\end{tabular}
\label{table:crossmatchsourcefind}
\end{table}
The observed properties of all selected cores are given in the accompanying online catalogue (Table A.1 in Appendix~\ref{sec:app.catalogue}).
\subsection{Derived core properties}\label{sec:derivcoreprop}
We fitted SEDs in a manner similar to that used to derive the column density maps in Sect. \ref{sec:derivmaps}. We made use of the integrated flux densities that \textsl{getsources} produces for each of the cores with an associated error $\sigma^{2} = \sigma_{\textrm{gs}}^{2} + (F_{\textrm{H}}C_{\textrm{H}})^{2}$; where $\sigma^{2}$ is the  quadrature sum of the errors, $\sigma_{\textrm{gs}}$ is the error in flux density due to the uncertainty in background estimation from \textsl{getsources} flux densities, and $F_{\textrm{H}}C_{\textrm{H}}$ is the error in calibration. We used MPFIT to fit the mass and temperature of the flux densities of the sources. Examples of SED fits are given in Fig.~\ref{fig_SED}. \par
Following the method discussed by \citet{aquila2015HGBS}, we tested the robustness of the SED fits by conducting two successive SED fitting runs. The first run included the 70-$\mu$m data point and the \textsl{getsources} \textit{detection errors} were used to weigh the SED data points. The detection error is defined as the flux of the source divided by the monochromatic detection significance at each wavelength, and is typically $\lesssim$15\%. The second run excluded the 70-$\mu$m data point and the measurement errors were instead used to weigh the SED points. The \textit{measurement errors} were defined as a function of the observational beam, peak flux uncertainty, and elliptical major and minor ellipse parameters (see Eq. 16 of \citealt{getsources2012}). \par
The measurement errors are a more conservative measure of the integrated flux for the extracted sources. The results of the SED fits were accepted if: i) there were three significant flux measurements for this source in \textit{Herschel} wavebands; ii) the source has larger integrated flux density at 350~$\mu$m than at 500~$\mu$m, and; iii) there was a factor of less than two difference between the core mass estimates from the two runs. For the starless cores that had an SED fit rejected, the masses were calculated using the median temperature of 13.4 K, from the cores with a valid SED fit. Cores with no valid SED fit are flagged in the catalogue. Approximately 42\% of the dense cores had valid SED fits.\par
We derived the radii of the cores as a geometric mean of the semi-major and semi-minor axes of the elliptical Gaussian fits to the cores. Both the spherical (derived as a geometric mean of the axes) and elliptical geometries are used to derive parameters of the cores.\par
Following \citet{aquila2015HGBS}, we can approximate the candidate prestellar cores using the critical Bonnor-Ebert (BE) sphere model \citep{bonnor1956,ebert1955}, where the critical radius of the sphere is given by the FWHM diameters of the sources. A peak (or central-beam) column density, an average column density, a central-beam volume density, and an average volume density were then derived for each core based on its estimated mass and radius. Central-beam column densities were estimated from the peak flux densities of the cores at the resolution of the SPIRE 500-$\mu$m observations ($\theta_{\textrm{beam,500}} = 36.3''$ or \textasciitilde$0.023$ pc at $d = 130$ pc) using an SED fitting procedure similar to that described in Sect. 4.1. The central-beam volume density $n_{0}$ (at the same resolution) was derived from the respective central-beam column densities $N_{0}$, assuming a Gaussian
spherical distribution, for which $n_{0} = N_{0}/(\sqrt{4\textrm{ln}2/\pi})(\theta_{\textrm{beam,500}\mu\textrm{m} }^{-1})$, where $\theta_{\textrm{beam,500}\mu\textrm{m}}$ is $36.3''$ at 500~$\mu$m. \par 
All of the derived properties are provided in online Table A.2 for the whole sample of selected \textit{Herschel} cores in CrA. Following \citet{kirk2013}, we colour-corrected the flux measurements using an iterative cycle method using the values for SPIRE \citep{bendo2011} and PACS \citep{muller2011}. 
\par

\subsection{Selecting self-gravitating prestellar cores}\label{sec:selfgrav}
\citet{konyves2010} and \citet{aquila2015HGBS} used the critical Bonnor-Ebert mass as a proxy for the virial mass for dense cores to define their boundedness. \citet{andre2000}, \citet{difrancesco2007ppv} and \citet{dwt2007ppv} highlight that dense cores can be categorised as prestellar if they are both starless and self-gravitating. The virial mass ratio given by $\alpha_{\textrm{vir}} = M_{\textrm{vir}}/M_{\textrm{obs}}$, where $M_{\textrm{vir}}$ is the virial mass ($M_{\textrm{vir}}=3R_{\textrm{core}}\sigma_{\textrm{tot}}^{2}/G$) for a spherical density $\rho\propto r^{-2}$ and $\sigma_{\textrm{tot}}$ is the total contribution from thermal and non-thermal velocity dispersion of the core. For self-gravitating objects, one expects $\alpha_{\textrm{vir}} \leq 2$ with objects having negligible self gravity having $\alpha_{\textrm{vir}} \gg 2$. \citet{andre2007N2Hplus} conducted measurements in N$_2$H$^+$(1--0) line emission, and found that low- to intermediate-mass cores have low non-thermal motions. Owing to this result, we adopted the thermal value of the Bonnor-Ebert mass as a proxy for the virial mass. The critical Bonnor-Ebert mass may take the form
\begin{equation}
M_{\textrm{BE,crit}} \approx 2.4 R_{\textrm{BE}}c_{\textrm{s}}(T)^{2}/G
\end{equation}
where $R_{\textrm{BE}}$ is the BE radius, $c_{\textrm{s}}$ is the isothermal sound speed, and $G$ is the gravitational constant. We equate the gas temperature $T$ with the estimated dust temperature. The radius $R_{\textrm{BE}}$ is estimated from deconvolution of the observed core radius measured from the high resolution column-density map. We used the ratio of the equivalent critical Bonner-Ebert mass of the cores to the total mass, derived using SED fits to out core flux densities.\par 
Completeness testing in CrA, Aquila \citep{aquila2015HGBS} and Taurus L1495 \citep{marsh2016} revealed that the misclassification of bound prestellar cores occurs for marginally resolved sources, and they are classified as starless. To include prestellar cores that have fallen into this category, we lowered the threshold value at which unbound starless cores are classified as bound prestellar cores. \citet{aquila2015HGBS} obtained a function dependence of $t=0.2(\theta_{\textrm{source}} / \theta_{\textrm{beam}})^{0.4}$, for $0.2\leq t\leq0.5$. $\theta_{\textrm{source}}$ and $\theta_{\textrm{beam}}$ are the FWHM and HPBW of the source, and beam, respectively. Both values are with respect to the high-resolution column density map.
Figure \ref{region_map} shows the location of the dense cores. Four populations of cores are shown. The open circles show the cores we defined as \textit{starless} cores. Prestellar cores are divided into two sub-populations. The \textit{robust prestellar} cores are shown as circles filled using their respective subregion colour. The \textit{candidate prestellar} cores are shown as coloured circles, filled in grey. Protostellar cores are shown as green filled stars. Each core is colour coded with respect to arbitrary subregions, which are labelled on the figure. Figure \ref{mass_temp} shows a plot of the masses of the cores against their estimated temperatures, both given by the SED fits. The coloured cores belong to the regions defined in Fig.~\ref{region_map}. A collection of starless cores is located in a diffuse arc of low column density material in the far east of the map. The uncertainty in misclassifying a source increases towards regions of low column density. Compact background objects become blended with foreground cirrus material, appearing as extended sources.
\begin{figure}[!h]
	\begin{center}
		\begin{minipage}{1.0\linewidth}
		\resizebox{1.0\hsize}{!}{\includegraphics[angle=0, trim={1.8cm 0cm 0cm 0cm}]{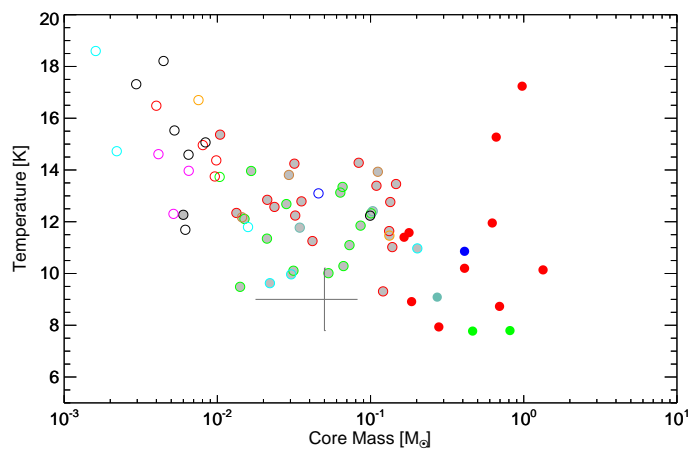}}
		\end{minipage}
	\end{center}
\caption{Plot of the estimated core mass versus the mean temperature of the core. The hollow circles indicate starless cores, while the filled circles indicate prestellar cores. The colours indicate the regions as given by Figure \ref{region_map}. The open circles represent cores that were classified as unbound starless cores. The robust prestellar cores are indicated by circles filled using the same colour as their respective subregion. Candidate prestellar cores are filled in grey, with an outer circle colour-coded to their respective subregion (see Fig.~\ref{region_map}), or black for cores with no assigned subregion}
     \label{mass_temp}
\end{figure}
\begin{figure*}[!]
		\centering
		\resizebox{1.0\hsize}{!}{\includegraphics[angle=0, trim={2.5cm 0.5cm 0cm 0cm}]{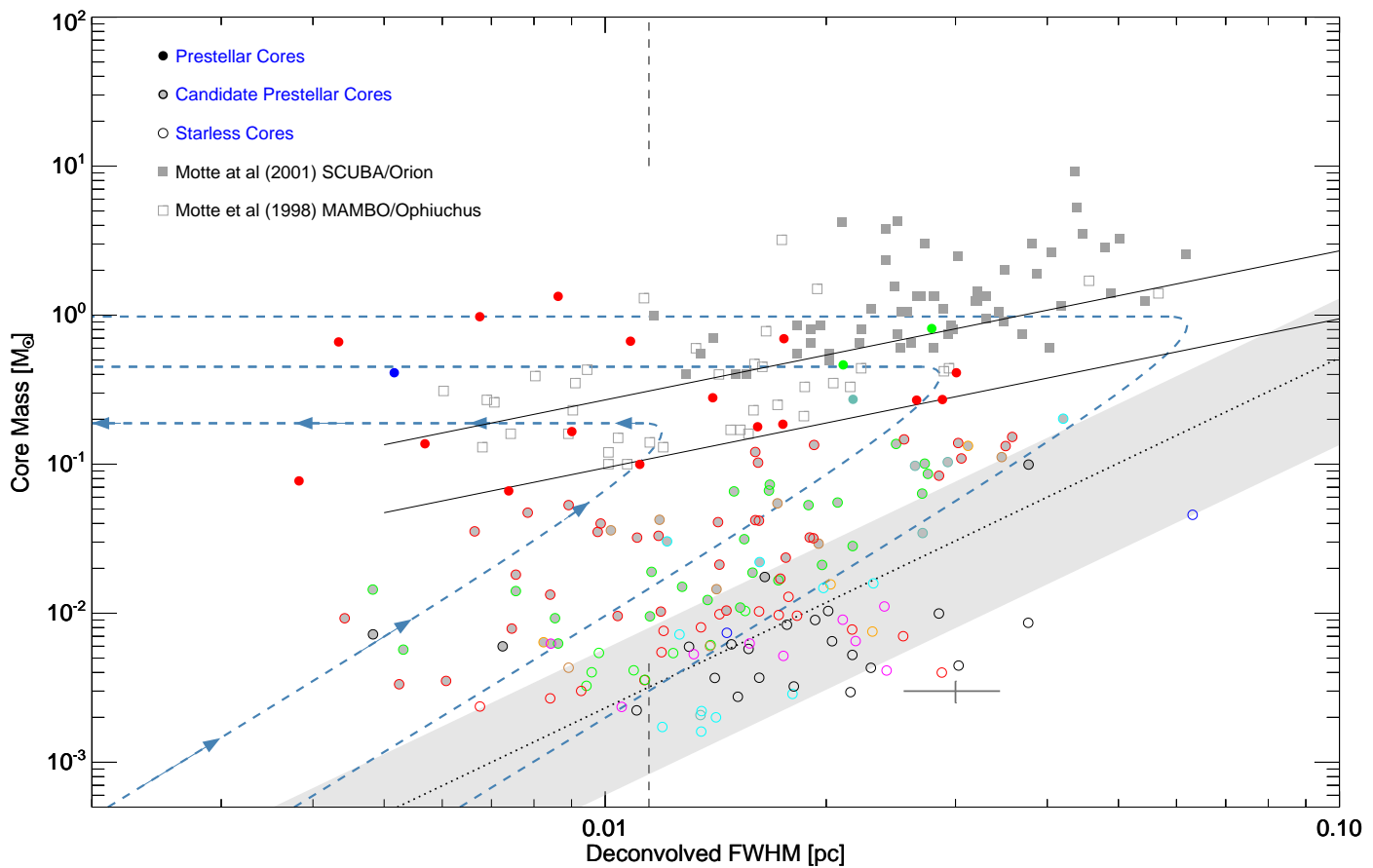}} 
\caption{The mass-size diagram for the population of 163 starless cores extracted by \textsl{getsources} from the \textit{Herschel} data. The open circles represent cores that were classified as unbound starless cores. The robust prestellar cores are indicated by circles filled using the same colour as their respective subregion. Candidate prestellar cores are filled in grey, with an outer circle colour-coded to their respective subregion (see Fig.~\ref{region_map}), or black for cores with no assigned subregion. Shaded grey squares indicate the cores found in Orion using SCUBA \citep{motte2001}, and the open squares indicate cores found in Ophiuchus using MAMBO \citep{motte2001}. The shaded grey band indicates the mass-size correlation observed for unbound CO clumps \citep{elemegreenfalgarone1996}. There are two model lines representing critical isothermal Bonnor-Ebert spheres, shown in black. The upper and lower lines represent BE spheres at $T = 20$ K and $T = 7$ K, respectively. The dashed-blue curves show the evolutionary tracks of starless cores, modelled as quasi-statically accreting isothermal Bonnor-Ebert spheres, for three ambient external pressures; 2.5, 5, and 20~$\times10^{6}$~K~cm$^{-3}$, as hypothesised by \citet{simpson2011}. The leftmost evolutionary line shows the direction of evolution. The two vertical dark grey lines indicate the physical $18.2''$ resolution in the plane of sky at the assumed 130 pc distance of CrA. The dark grey cross is the typical error in the mass and FWHM radius.}
     \label{mass_size}%
\end{figure*}

Figure~\ref{mass_size} shows a plot of the mass, $M$, versus the radius $R$, where $R$ represents the deconvolved geometric mean FWHM radius of the cores, measured using the high-resolution column density map. We plotted the cores as being unbound starless cores, bound robust-prestellar cores, and bound candidate prestellar cores using the BE criteria defined above. Results from \citet{motte1998} and \citet{motte2001} are also plotted for comparison. A typical error is shown as a dark grey cross and arises from the propagation of the uncertainty in the adopted distance ($d=129\pm11$ pc; \citealt{casey98}), and the individual errors in the masses. The uncertainty in radius for unresolved and marginally resolved cores is greater than that of resolved cores. For cores which are characterised as circular Gaussians at the limiting beam size of $18.2''$ \textit{before} deconvolution, these cores are unresolved, and are not plotted on the mass-size plane. At the adopted distance, cores below $\sim0.01$ pc are unresolved.

\section{Spatial distribution of sources}
Previous surveys, such as those by SCUBA-2 on the JCMT \citep{holland2013}, or the observations by \citet{chini1997} using SIMBA, have been limited by atmospheric filtering. Ground-based observations by \citet{yonekura1999}, \citet{chini1997}, and \citet{nutter2005} showed that the densest areas of the cloud are those adjacent to the Coronet cluster. Three dense clumps of material, which form parts of the northern tail, and the much of the material around the Coronet are well-captured by the \textit{NANTEN} C$^{18}$O ($J$=1--0) mapping by \citet{yonekura1999}.\par 
Taking advantage of the sensitivity of \textit{Herschel}, we looked for areas that are clumpy to target locations for which ground-based surveys such as those by \citet{chini1997}, \citet{yonekura1999}, and \citet{nutter2005}, were more likely to be sensitive.
Figure~\ref{region_map} shows a selection of these areas, across the high-resolution column density map, with the naming convention for the subregions of CrA continued from \citet{nutter2005}. The subregions in Fig.~\ref{region_map} all show column densities at least 3 times the rms column density of the map. Contours begin at the $\sigma_{\textrm{rms}}\sim 3\times10^{20}$~N(H$_{2}$)~cm$^{-2}$ level of the column density map. The subregion boundaries in the areas discussed below are arbitrary, with the sources having been segregated by masks made manually from the appearance of the continuum maps and the column density maps. The locations of the subregions in the molecular cloud can be viewed in Figure~\ref{fig:cra_subregs}, where we have plotted the respective colours for these subregions as outlines. Note that the subregions are also similar to those identified by \citet{andreazza1996}. 
\subsection*{CrA-A}
This part of the cloud contains the well-studied Coronet cluster and has been the focus of several previous studies over a range of scales. The Coronet was discovered by \citet{taylorandstorey1984}, who conducted a survey using the 3.9-m Anglo Australian Telescope. They identified many new members in addition to those discovered by Schmidt \citep{reynolds1916}. This area also contains the objects NGC 6726, 6727, and 6729. Variability in NGC 6729 was studied by \citet{graham1987} using optical measurements, who concluded that the variability in the visual magnitude of this object closely mimics the variability of R CrA itself. \citet{wilking1985,wilking1986,wilking1992, wilking1997} built on the observations by \citet{taylorandstorey1984}, conducting several surveys in the mid-infrared investigating spectral properties, and making comparisons with the $\rho$ Ophiuchi cloud.\par 
The largest fraction of robust prestellar cores in CrA are situated within this subregion, at 73\%. We located several prestellar cores that are interspersed with the YSO/protostellar objects that form the Coronet, including the two prestellar cores identified by \citet{nutter2005}. The prestellar core SMM 1A observed by \citet{nutter2005} is split into two prestellar cores by \textsl{getsources}. This supports the result found by \cite{chen2010}, who found that SMM 1A appears to be fragmenting. These prestellar cores have SED-derived temperatures of approximately 17 K, 15 K, respectively, from east to west across this subregion.  
\subsection*{CrA-B}
This subregion is populated by 2 starless cores, as well as 6 candidate prestellar cores, as identified by \textit{Herschel}. It also hosts a single \textit{Herschel}-identified 70-$\mu$m source, IRAS 32c (IRAS 18595-
3712; \citealt{wilking1992}), which is an embedded object (see discussion by \citealt{peterson2011}). Figure \ref{fig_cra_iras_32} shows IRAS 32c using \textit{Spitzer} 4.5-$\mu$m data and the \textit{Herschel}-derived high-resolution column density map. This well-known source has been the subject of many previous studies \citep{chen1997,chini1997,nutter2005,seale2008} and possesses a striking outflow cavity detected by \textit{Herschel} in all SPIRE bands and by PACS at 160~$\mu$m. (see Fig.~\ref{fig_cra_iras_32} for a comparison of the \textit{Herschel} and \textit{Spitzer} data).
\begin{figure}[!h]
	\begin{center}
		\begin{minipage}{1.0\linewidth}
		\resizebox{1.0\hsize}{!}{\includegraphics[angle=0]{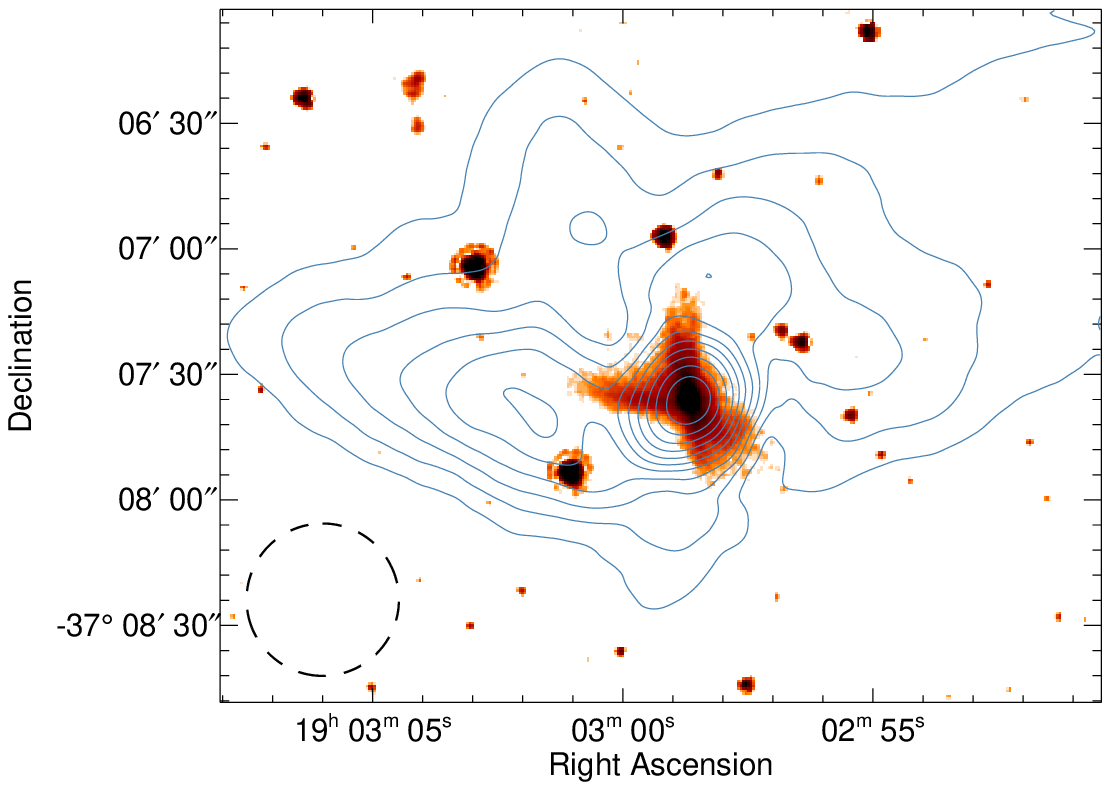}}
		\end{minipage}
	\end{center}
\caption{Image of IRAS 32c (IRAS 18595-3712 - \citep{wilking1992}) with \textit{Herschel} column density contours plotted in blue. The lowest contour is $~7\times10^{22}$~N(H$_{2}$)~cm$^{-2}$ and each sequential contour is $1.2$ times higher than the previous one. The colour image is \textit{Spitzer} 4.5-$\mu$m data showing a larger area coverage version of the same region of sky discussed and presented by \citet[Fig.~15 \& Appendix~A.44]{peterson2011}. The \textit{Herschel} column density contours form a `bay' of material to the north-east of the source. This appears to be showing the cavity that is swept out by a jet not visible within this figure. The dashed circle in the bottom left corner represents the $\sim18''$ \textit{Herschel} resolution of the column density map.}
     \label{fig_cra_iras_32}%
\end{figure}

\subsection*{CrA-C}
This subregion consists of multiple clumps of material connected by wispy striations/bridges of dust that form a triangular structure in the plane of the sky. It contains 3 robust prestellar cores, 24 candidate prestellar cores, and 8 starless cores. This structure is the last in the plane of the sky that is associated with the main Coronet region in previous studies. Much of the structure in the south-eastern and south-western sides of this has a lower column density and contains only a few starless cores. 
\subsection*{CrA-D}
This clump of material contains only two \textit{Herschel}-identified starless cores, and 2 candidate prestellar cores. The clump is positioned on the northern tail \textasciitilde$3'-4'$ to the east of CrA-C. It was previously identified by \citet{yonekura1999} as Clump 3.
\subsection*{CrA-E}
Our \textsl{getsources} extractions using the \textit{Herschel} data suggest that this subregion is fragmenting into one robust prestellar core, and 2 starless cores (one starless core is located south-east of the subregion cut-out in Figure \ref{region_map}). Although previous surveys have had no detections of any YSO within the centre of the cloud, \textit{Herschel} does have a faint detection at 70~$\mu$m, which suggests that this object could be a very low-luminosity object (VeLLO). For the purposes of our survey, this object is excluded as a YSO or a VeLLO. This subregion has been previously identified and studied in previous surveys: Cloud 42 \citep{sandqvist1976}, Condensation C \citep{andreazza1996}, Core 5 \citep{yonekura1999}, Condensation CoA7 \citep{vilas-boaes2000}, DC 000.4-19.5 \citep{Ullman2013} and CrA C \citep{suutarinen2013}. \citet{Ullman2013} studied the dynamical state of this cloud using C$^{18}$O and N$^2$H$^+$ line measurements from SEST, and find the offset in peaks to be consistent with CO depletion that has been found in other starless cores (e.g., \citealt{tafalla2002}). 
\subsection*{CrA-F}
This clump consists of a dense region at its northwestern end and a more diffuse region to the southwest. We located 8 starless cores, and 3 candidate prestellar cores within this subregion. \citet{yonekura1999} identified this as Clumps 6 \& 7. 
\subsection*{CrA-G}
This clump is the last clump along the north tail that is visible in the \textit{Herschel} data and was previously identified as Clump 8 by \citet{yonekura1999}. This subregion is made up of three distinct small clumps of material, with the largest clump containing two candidate prestellar cores. The westernmost clump also contains two candidate prestellar cores. A single starless cores is located within the easternmost clump.
\subsection*{CrA-H}
This subregion is the only region of substantial column density located on the southern tail, near the end of the very diffuse dust. It was identified by \citet{yonekura1999} as Clump 1, and contains 8 starless cores, and a single candidate prestellar core. 
\section{Discussion of properties}
Inspection of Figure \ref{mass_size} shows that the robust prestellar cores in CrA-C and CrA-G have larger radii than the majority of the prestellar cores in CrA-A, which could be indicative of these cores being at a different evolutionary stage than those in CrA-A \citep{simpson2011}.\par
On Fig. \ref{mass_size} we exemplify this by plotting evolutionary tracks of quasi-statically accreting Bonnor-Ebert spheres, according to the hypothesis of \citeauthor{simpson2011}~(\citeyear{simpson2011}, see also \citealt{johnstone2000,andre2014ppvi}). The arrow towards the lower-left of the figure shows the direction of evolution of isothermal 10~K starless cores on the mass-size plane, for three different external pressures: 2.5, 5, and 20~$\times10^{6}$~K~cm$^{-3}$. Once the Bonnor-Ebert spheres become Jeans unstable, the core undergoes a collapse, moving leftward. If the cores continue to accrete material as they collapse, then one may expect their masses to increase by a small amount as their radii decrease. Under the assumption that the core masses remain relatively constant or only increase slightly, Fig.~\ref{mass_size} suggests that the robust prestellar cores of CrA-C and CrA-G are to be at an earlier stage of evolution than some of those in CrA-A, given their relatively larger radii. Indeed, the cores in CrA-A may be influenced by environmental factors such as heating from Coronet cluster members. By similar argument, the single prestellar core in CrA-E also is at roughly the same stage of evolution as the smallest robust prestellar cores in CrA-A. \par
Figure \ref{mean_dens_temp} shows a plot of the mean core column density $N(\textrm{H}_{2}^{\textrm{ave}})=M/\pi R^{2}\mu m_{\textrm{H}}$ against the temperature for the starless cores. \citet{marsh2016} produced a similar plot for the starless and prestellar cores in Taurus. We performed a linear regression of the starless and prestellar cores with mean column densities $\leq 10^{22}$ cm$^{-2}$, as cores beyond this threshold of column density display are uncorrelated with the temperature. Cores under this mean column density show a relation of $T = -3.5\textrm{log}_{10}$~N(H$_{2}$) $+ 86.0$. \par
\begin{figure}[!h]
	\begin{center}
		\begin{minipage}{1.0\linewidth}
		\resizebox{1.0\hsize}{!}{\includegraphics[angle=0, trim={2.8cm 1.5cm 0cm 0cm}]{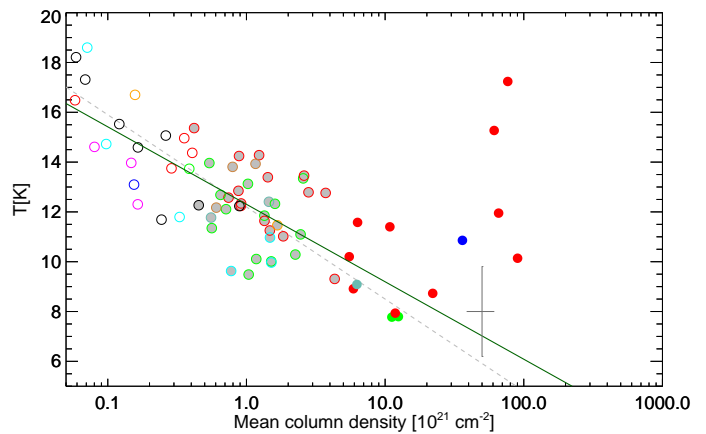}}
		\end{minipage}
	\end{center}
\caption{Mean dust temperature versus the mean column density of the respective core. Sources are colour coded by region, with solid filled cores denoting prestellar cores, and hollow circles denoting the starless cores. The cross on the figure shows the mean errors for this population of cores. The solid green line shows the relation $T = -3.5\textrm{log}_{10}$~N(H$_{2}$) $+ 86.0$. The dashed grey line is the relation $T = -3.7\textrm{log}_{10}$~N(H$_{2}) + 15.9$ found by \citet{marsh2016} for cores in Taurus L1495.}
     \label{mean_dens_temp}%
\end{figure}


Figure \ref{cra_cmf} shows the core mass function (CMF) for CrA. Here, we separated the cores into unbound starless cores, candidate prestellar cores, and robust prestellar cores. The robust prestellar cores are shown as a solid red histogram, and the additional candidate prestellar cores are shown by the dashed blue histogram. The collection of all cores that are not protostellar are shown as a dark grey histogram. The estimated 80\% completeness level at 0.1 M$_{\odot}$ is shown for prestellar cores (see Appendix~\ref{sec:completeness}). The number statistics for CrA are low. We observe fewer cores in this molecular cloud compared to Taurus \citep{marsh2016} and Aquila \citep{aquila2015HGBS}.\par 
\par
The behaviour of the CMF for all of the starless cores, which is shown as a grey dashed line on Fig.~\ref{cra_cmf}, appears consistent with a power-law, i.e. $\textrm{d}N/\textrm{dlog}M\propto M^{-\gamma}$, for M$>$0.01 M$_{\odot}$. We attempted to derive a best-fitting power-law for the cores within the aforementioned interval by using a least-squares method. We note that our core sample is dominated by candidate prestellar and unbound starless cores. The completeness limit for unbound starless cores is estimated to be $>$85\% at 0.015~M$_{\odot}$ (see Appendix~\ref{sec:completeness}, also \citealt{marsh2016}). Therefore we conducted the fit for masses greater than 0.01~M$_{\odot}$. The best-fitting value for the exponent is $\gamma=0.32\pm0.05$, which is shown as a green solid line on Fig.~\ref{cra_cmf}, accompanied by its associated $1\sigma$ error. Our value for $\gamma$ may be compared with $\gamma = 0.55$ for the HGBS result in Taurus \citep{marsh2016}.\par
The classical method used to determine the power-law index for the CMF is to conduct fits to the binned data. Fitting in this manner can be biased by the small number statistics in the higher mass bins (see, e.g., \citealt{maschberger2009}). Following \citet{kirk2015androm} and \citet{pattle2015}, we used the complementary cumulative distribution function (CCDF) of the cores to determine the index $\alpha = \gamma +1$, such that $\textrm{d}N/\textrm{d}M\propto M^{-\alpha}$. We utilised the maximum likelihood estimator for an infinite power-law distribution function \citep{koen2006,maschberger2009}. The empirical cumulative distribution function $\hat{F}$ is given by
\begin{equation}
\hat{F}(X_{i})\equiv\frac{i}{n+1},
\end{equation}
where $n$ is the number of data points of $X$. The maximum likelihood estimator $\hat{\alpha}_{\textrm{\textrm{ML}}}$ is given by
\begin{equation}
\hat{\alpha}_{\textrm{ML}} =1 + \frac{n}{\left( \sum\nolimits_{i=1}^{n} X_{i} \right) - n \textrm{ln}(\textrm{min}(X)) }.
\end{equation}
The quasi-bias-free modified maximum likelihood (MML) estimator for the exponent $\alpha$ of an infinite power-law distribution is
\begin{equation}
\hat{\alpha}_{\textrm{MML}} = \frac{n}{n-2}(\hat{\alpha}_{\textrm{ML}} - 1) + 1.
\end{equation}
\par
The power-law index found using this method was $\hat{\alpha}_{\textrm{MML}}=1.59\pm0.04$. Uncertainties in $\hat{\alpha}_{\textrm{MML}}$ were estimated by performing Monte Carlo experiments. Data were drawn randomly from the set of input masses, from which $\hat{\alpha}_{\textrm{MML}}$ was recalculated. The error given is the standard deviation of the distribution of $\hat{\alpha}_{\textrm{MML}}$ from these experiments. For comparison with the fit to the binned histogram, the equivalent power law is $\gamma =\hat{\alpha}_{\textrm{MML}}-1=0.59\pm0.04$. The cumulative distribution and fit is shown in Fig.~\ref{cra_ccdf} (see \citealt{koen2006,maschberger2009,gratier2012,kirk2015androm} and \citealt{pattle2015} for similar examples).\par
In both cases, the values for $\gamma$ are consistent with the mass function exponent, $\gamma_{\textrm{CO}}=0.7$, found for CO clumps and clouds \citep{blitz1993,kramer1998}. Our result is also close to the value $\gamma=0.5$ found by \citet{motte1998} for prestellar condensations in Ophiuchus, within the mass range 0.1-0.5 M$_{\odot}$.\par
The bulk of the prestellar core sample belongs to CrA-A. Due to the proximity of these cores to the Coronet, we investigated the properties of these cores with respect to their radial distance to the centre of the Coronet, and compared them with the rest of the prestellar cores within the molecular cloud.\par
Figure \ref{coronet_radius} shows the plane-of-sky (projected) distance of the prestellar cores against the temperatures (upper panel) and un-deconvolved geometric mean FWHM radii (lower panel) of the cores. Both quantities show a correlation with radial distance. The mean temperatures of the prestellar cores in CrA-A show a steady increase in temperature with decreasing plane-of-sky distance, especially at $\lesssim 0.1$~pc. In addition, the un-deconvolved radii of the cores decrease as the proximity to the Coronet decreases. The uncertainty in un-deconvolved radius is shown as a conservative estimate of 15\% (see Appendix~\ref{sec:completeness}). Meanwhile, the cores at distances $\gtrsim 0.1$~pc for all subregions show relatively flat distributions with respect to temperature and un-deconvolved radius. One particular exception to this is a prestellar core, which lies in CrA-E.  \par
Cores that are closer to the Coronet show higher temperatures but generally smaller radii than those found across the rest of the subregion. As mentioned above, external heating from R CrA could be causing the elevated temperatures within the local region. SED fits to the \textit{Herschel} data are also line-of-sight averaged, and temperatures could be lower towards the centre of these cores (see, e.g., \citealt{roy2014}). Furthermore, the UV flux from R CrA may be contributing to a higher external gas pressure on these prestellar cores, which could explain why cores closer to the centre of the Coronet have lower radii and higher masses (see, e.g., \citealt{lindberg2012}).\par
This correlation could be evidence that the Coronet itself is causing an increase in the local radiation field, with the prestellar cores closer to the Coronet having higher mean column densities and higher temperatures than many cores across the rest of the cloud. \citet{lindberg2012} used measurements from the Submillimeter Array (SMA) and the Atacama Pathfinder EXperiment (APEX) and concluded that the strong UV flux ($\chi_{\textrm{ISRF}}\sim 750$) from R CrA is having effects on large scales throughout the local area. The spectral classification of R CrA remains unresolved, however \citet{bibo1992} derived a B8 classification using spectroscopic data, \citet{gray2006} derived a B5 classification. They also hypothesise that this UV field will likely be influencing the evolution of the prestellar cores, being a source of external heating. the results are in agreement with this hypothesis.\par
\begin{figure}[!h]
	\begin{center}
		\begin{minipage}{1.0\linewidth}
		\resizebox{1.0\hsize}{!}{\includegraphics[angle=0, trim={2.8cm 1.5cm 1.5cm 0cm}]{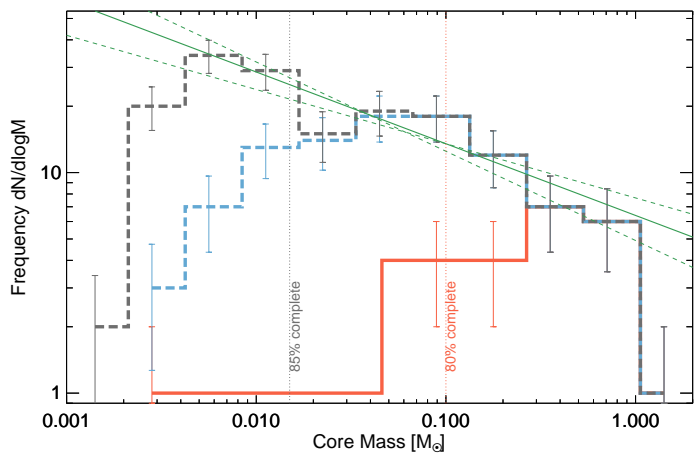}}
		\end{minipage}
	\end{center}
\caption{Core mass function (CMF) for subsets of the starless cores. The dashed grey line is the CMF for all of the starless cores regardless of their bound state. The solid red line shows robust prestellar cores, and the dashed blue line shows the additional candidate prestellar cores. The error bars are given as the errors in Poisson statistics for all three histograms. The over-plotted solid green line is the best-fitting power-law (index $\gamma=0.32\pm0.05$) for all of the starless cores with masses 0.01~M$_{\odot}<M<$1~M$_{\odot}$. The 80\% completeness limit for robust prestellar cores only is shown as a blue vertical dotted line at 0.1 M$_{\odot}$, as discussed in Appendix~B. The black vertical line is the completeness limit of all starless cores. Only bins including and above this line were used in the fitting of the green line.}
     \label{cra_cmf}%
\end{figure}
\begin{figure}[!h]
	\begin{center}
		\begin{minipage}{1.0\linewidth}
		\resizebox{1.0\hsize}{!}{\includegraphics[angle=0, trim={2.2cm 1.5cm 1.9cm 0cm}]{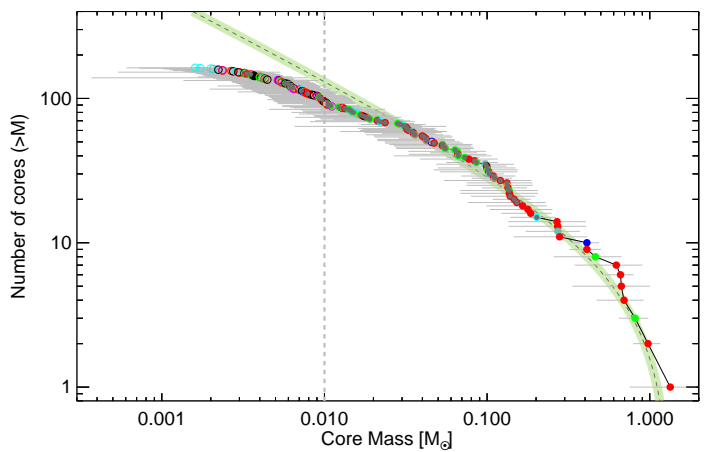}}
		\end{minipage}
	\end{center}
\caption{Complementary cumulative mass distribution for all of the starless cores. Colour coding is as in Fig.~\ref{region_map}. The quasi-bias-free maximum likelihood estimator power-law index, $\alpha_{\textrm{MML}}=1.59\pm0.04$, is plotted as a light green band with central dashed grey line, and the lower limit was taken as 0.01 M$_{\odot}$. The vertical dashed light grey line shows the lower truncation limit of 0.01M$_{\odot}$ that was used for the MML estimator. The equivalent power-law index of the $\textrm{d}N/\textrm{dlog}M$ distribution for comparison with the binned data of Fig.~\ref{cra_cmf} is $\gamma = \hat{\alpha}_{\textrm{MML}}-1=0.59\pm0.04$.}
     \label{cra_ccdf}%
\end{figure}

\begin{figure}[!h]
	\begin{center}
		\begin{minipage}{1.0\linewidth}
		\resizebox{0.98\hsize}{!}{\includegraphics[angle=0, trim={1.5cm 1.5cm 1.5cm 0cm}]{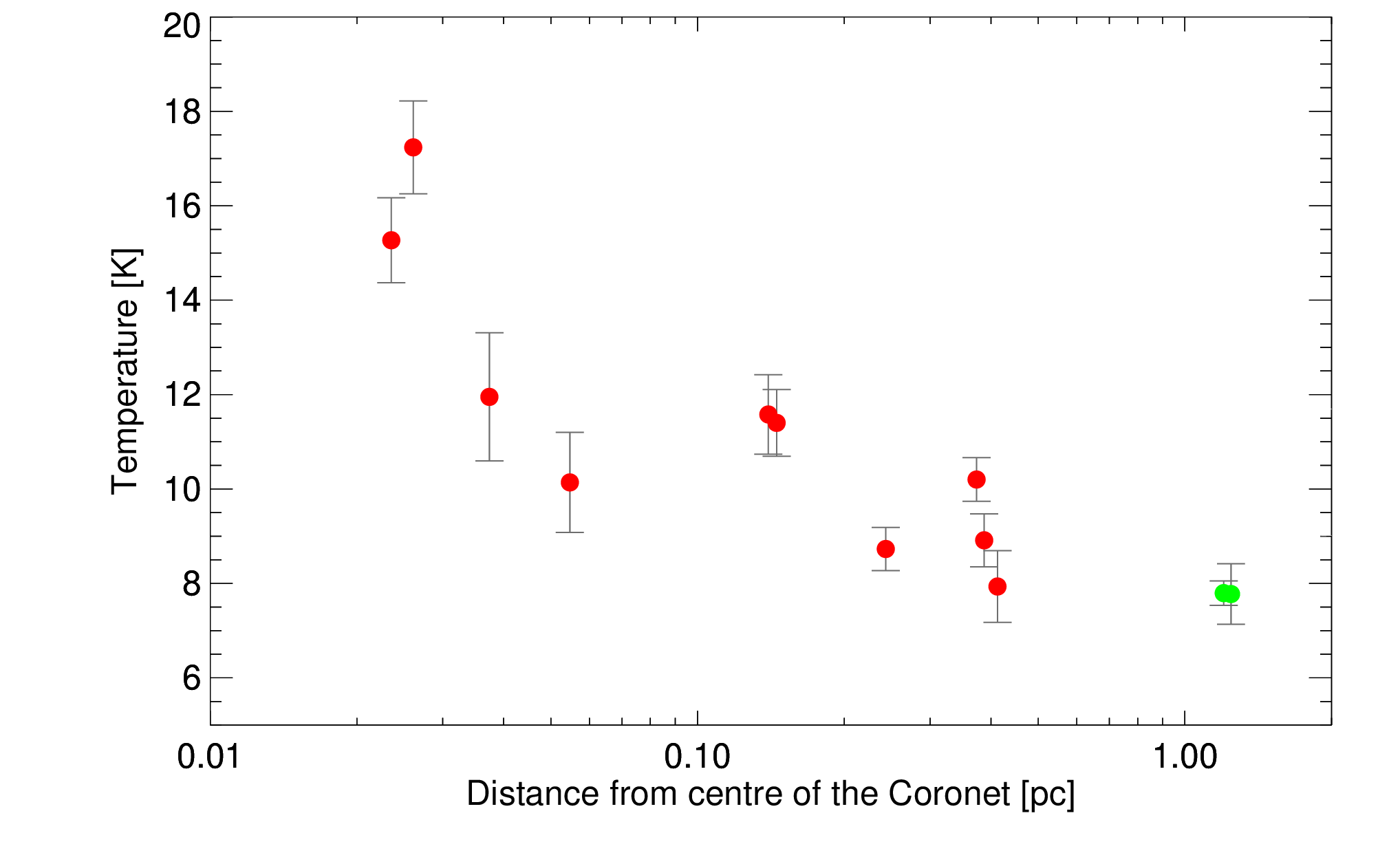}}
		\resizebox{0.98\hsize}{!}{\includegraphics[angle=0, trim={1.5cm 1.5cm 1.5cm 0cm}]{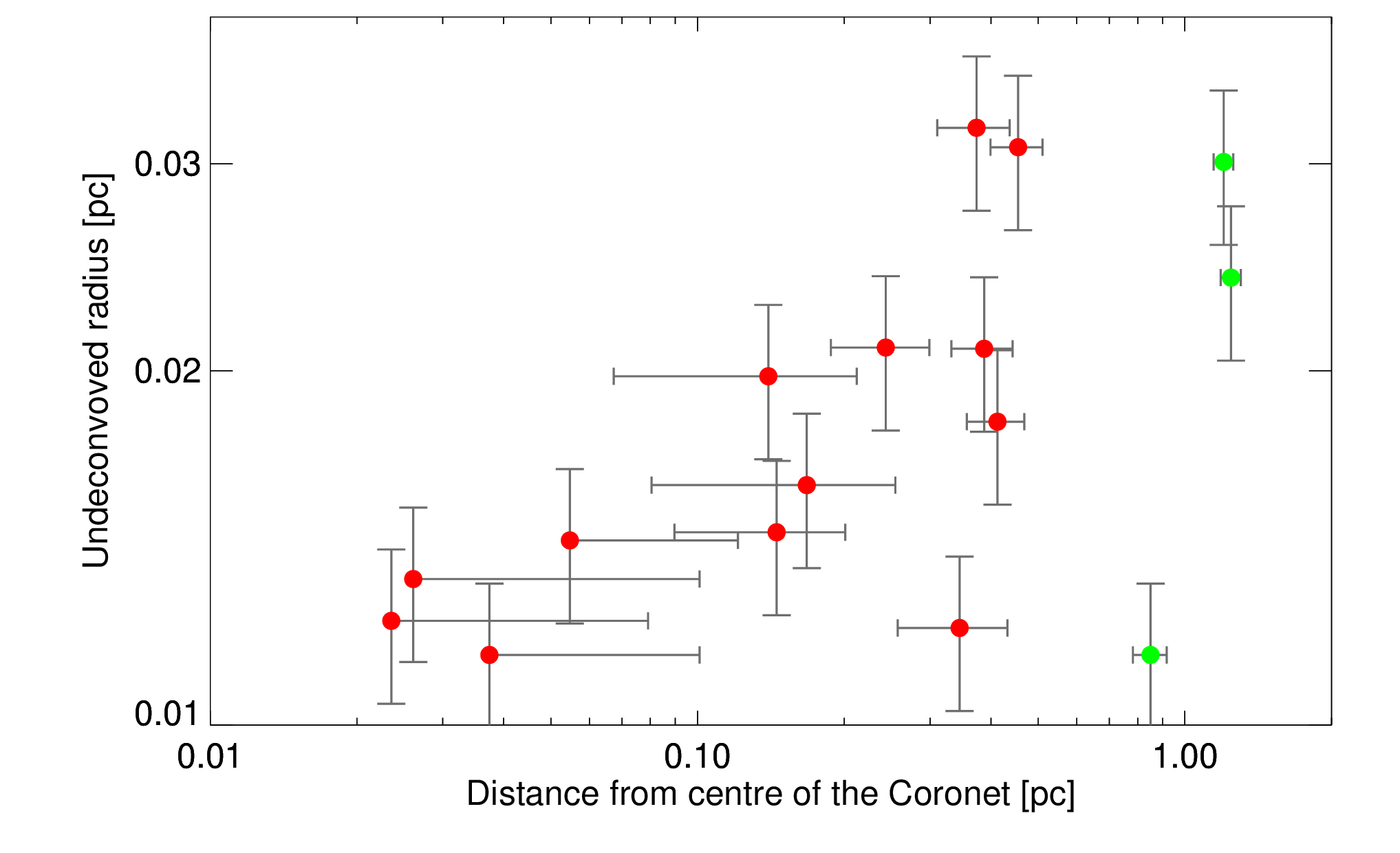}}
		\end{minipage}
	\end{center}
\caption{The upper panel shows the plane of sky distance of the robust prestellar cores (with accepted SED fits) from the Coronet against the mean temperature of those cores. The lower panel also shows the aforementioned distance, but plotted against the deconvolved geometric mean FWHM radius of the cores. The colour coding is as in Fig.~\ref{region_map}. There are no large-radius cores close to the Coronet.}
     \label{coronet_radius}
\end{figure}

\section{Filamentary properties}
A new paradigm of core formation on filaments has been previously proposed by \cite{andre2014ppvi}. Matter flows along magnetic field lines onto filamentary structures. Seeds of starless cores are formed, and the matter flows along the filaments, accreting onto these starless cores. The mass per unit length increases along this filament until a critical value is surpassed, and the cores begin collapsing. \citet{aquila2015HGBS} showed that the cores in the Aquila region (\textasciitilde$75\%$) are associated with filamentary structure, in agreement with this paradigm. They used convolved versions of the skeletons of the filaments as found by \textsl{DisPerSE}, with a Gaussian kernel that produces the typical filamentary inner-width of \textasciitilde$0.1$~pc seen first by \citet{arzoumanian2011fils}. They found that cores typically lie within the area thus defined.\par
We measured the projected distance of each core to the nearest filament found by \textsl{DisPerSE} in CrA. Figure \ref{mass_v_proxim} shows a plot of masses of starless and prestellar cores against the projected distances of those cores to the spines of the nearest filaments. The mean projected distance of the prestellar cores is within the 18.2\arcsec~beam width to their nearest filament, and is \textasciitilde0.2~pc for the starless cores. This short distance supports the paradigm that cores accrete mass from filaments, as the largest cores (\textasciitilde0.1 M$_{\odot}$) tend to be located very close to the centre of the filaments. For the starless cores, 48\% were located within 0.05~pc of their nearest filament. For candidate and robust prestellar cores, 81\% and 96\% were within 0.05~pc of their nearest filament, respectively. A deserted region can be observed towards the upper-right of the figure. There are no high-mass cores far from a filament. Again, this supports the paradigm that the majority of star formation is likely to take place along filaments.\par
We followed the techniques used by \citet{arzoumanian2011fils} who made profiles of filaments to analyse their properties. The westernmost filament (see Fig.~\ref{fig:Fig.1.}) was cleaned and repaired automatically using a custom algorithm we wrote, named \textsl{Skywalker} \citep{bresnahanphd2017}. \textsl{Skywalker} was created as a general tracing routine, primarily aimed at creating ordered, transverse profiles of morphologically thin binary skeletons, such as are created by \textsl{DisPerSE} and \textsl{getfilaments}. Most of the repairs were minor corrections and involved erasing `clusters' of pixels that would cause incorrect transverse profiles to be obtained. Though the corrections to the filament skeleton were minor, they were necessary to maintain a self-consistent transverse profile of the filament. Even single pixel errors can lead to up to 5 inaccurate profiles being taken, due to incorrect angles. Gaps in the FITS skeleton due to overlapping filaments were also automatically repaired using Skywalker,to ensure a complete profile. \par
After the filament was repaired, \textsl{Skywalker} then started at an end of the skeleton and used an implementation of Prim's algorithm \citep{prim1957} to `walk' down the filament spine pixels like a path. This process allowed us to produce images of the transverse profiles similar to those made by \citet{arzoumanian2011fils}.\par
\begin{figure}[!h]
	\begin{center}
		\begin{minipage}{1.0\linewidth}
		\resizebox{0.95\hsize}{!}{\includegraphics[angle=0, trim={2.8cm 1.5cm 1.5cm 0cm}]{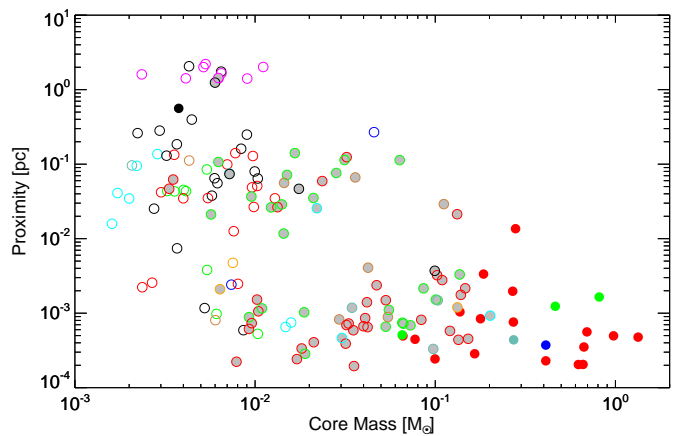}}
		\end{minipage}
	\end{center}
\caption{A plot of the core mass against the proximity of the cores to the nearest filament. Cores are colour coded as in the previous figures. Filled and hollow circles are the prestellar and starless cores respectively. No high-mass cores are seen far from any filament.}
     \label{mass_v_proxim}%
\end{figure}

Figure \ref{streamer_fil} (upper) shows a plot of the local area of the streamer to the west of the Coronet region. The bright compact source towards the north-east in the figure is S CrA. The transverse profile that was taken of this filament is shown in Fig.~\ref{streamer_fil} (lower). We manually subtracted the strong, but highly compacted column density contribution from S CrA, to prevent inflation of the column densities and dispersion on the south-east side of the filament. The mean contribution from this source is shown as a small dashed curve on top of the solid black line. The profile shows some asymmetry which we quantify in the same manner as \citet{peretto2012}. We integrated column densities above the 5$\sigma_{\textrm{rms}}$ of the high-resolution column density map, and found that the integrated column density is $\sim$15\% higher on the south-eastern side of the filament. This asymmetry cannot be explained by the stronger interstellar radiation field (ISRF) that arises from the B9V stars HD 176269 and HD 176270, which may be both associated with the cloud (see \citealt{juvela2012}). Relative to CrA, their radiation comes from the south-east, producing an anisotropic field across the filament. Though \citet{juvela2012} found that this anisotropic radiation field could slightly affect the infrared and submillimetre profiles by using radiative transfer models of the streamer, the asymmetry in the column density profile remains.\par
Our profile shows similar features to "filament 2" in the Pipe Nebula (see Fig.~2 of \citealt{peretto2012}). Despite the relatively low asymmetry in the profile, the dispersion along the south-eastern edge of the filament is high compared to that in the north west. The mean dispersion on the south-east side and north-west sides of the filament is 8.3$\times10^{20}$~N(H$_{2}$)~cm$^{-2}$, and 5.3$\times10^{20}$~N(H$_{2}$)~cm$^{-2}$, respectively. The dispersion on the south-east is over 50\% greater on the south-eastern side, than the north-western side. This could be an indicator that there is a large scale compression effect, which could be influencing the development of the region. Such a large difference in the dispersion is expected if wake turbulence from the centre of the filament were playing a significant role in the column densities behind the filament, and/or that the north-western side of the filament is being compressed. In the presence of an influence from the northwest side of the filament, this would act to strip material from the outer layers of the filament. However, the denser inner regions of the filament would act as a breaker to the incoming influence. This would form eddy-like turbulence behind the filament. This manifests in larger density dispersions, due to the non-uniformity of the stripped material, caused by the mixing effect of the turbulence. This is what we mean by `wake turbulence'.
\begin{figure}[!h]
	\begin{center}
		\begin{minipage}{1.0\linewidth}
		\resizebox{0.95\hsize}{!}{\includegraphics[angle=0, trim={1.0cm 2.0cm 0cm 0cm}]{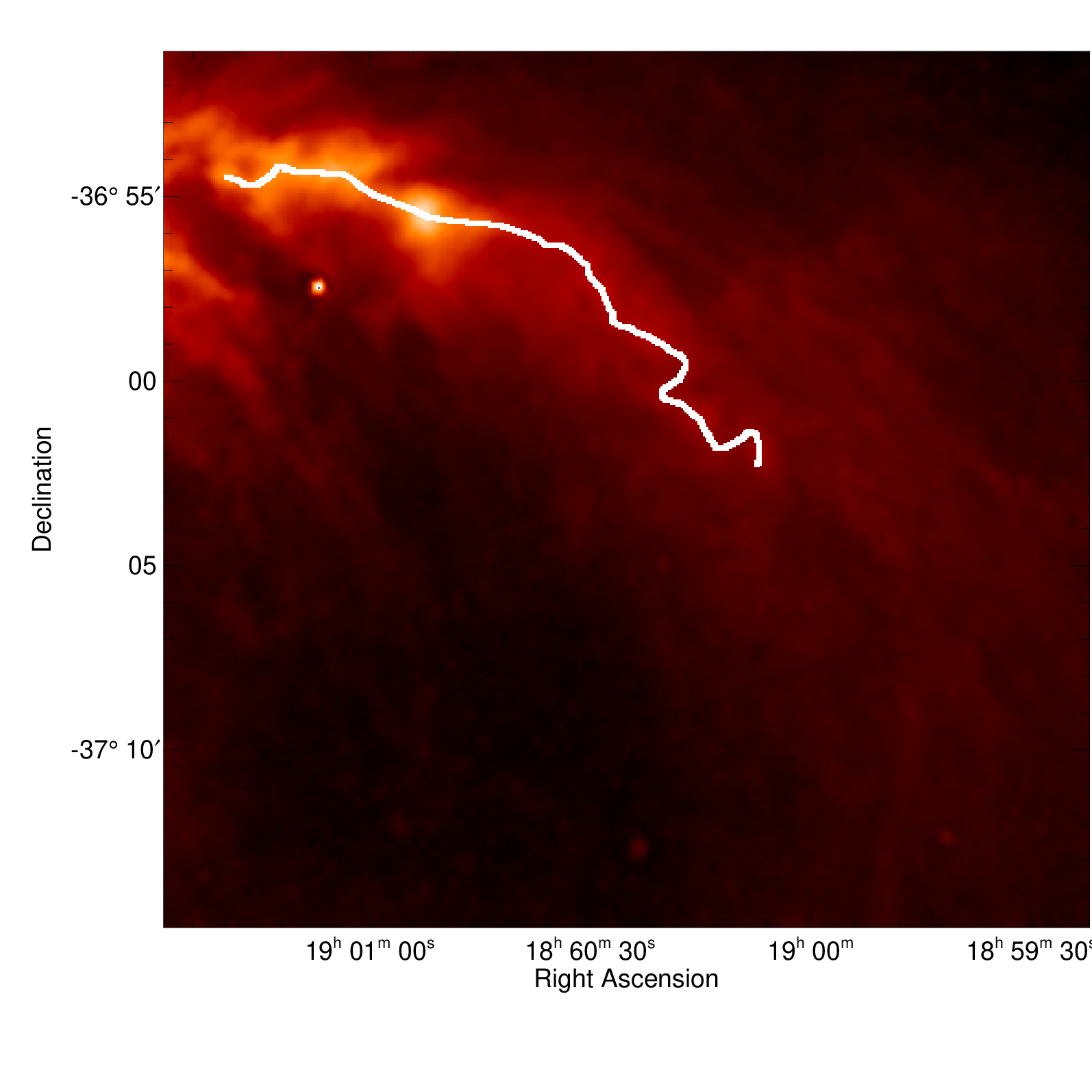}}
		\resizebox{1.0\hsize}{!}{\includegraphics[angle=0, trim={2.8cm 1.5cm 0cm 0cm}]{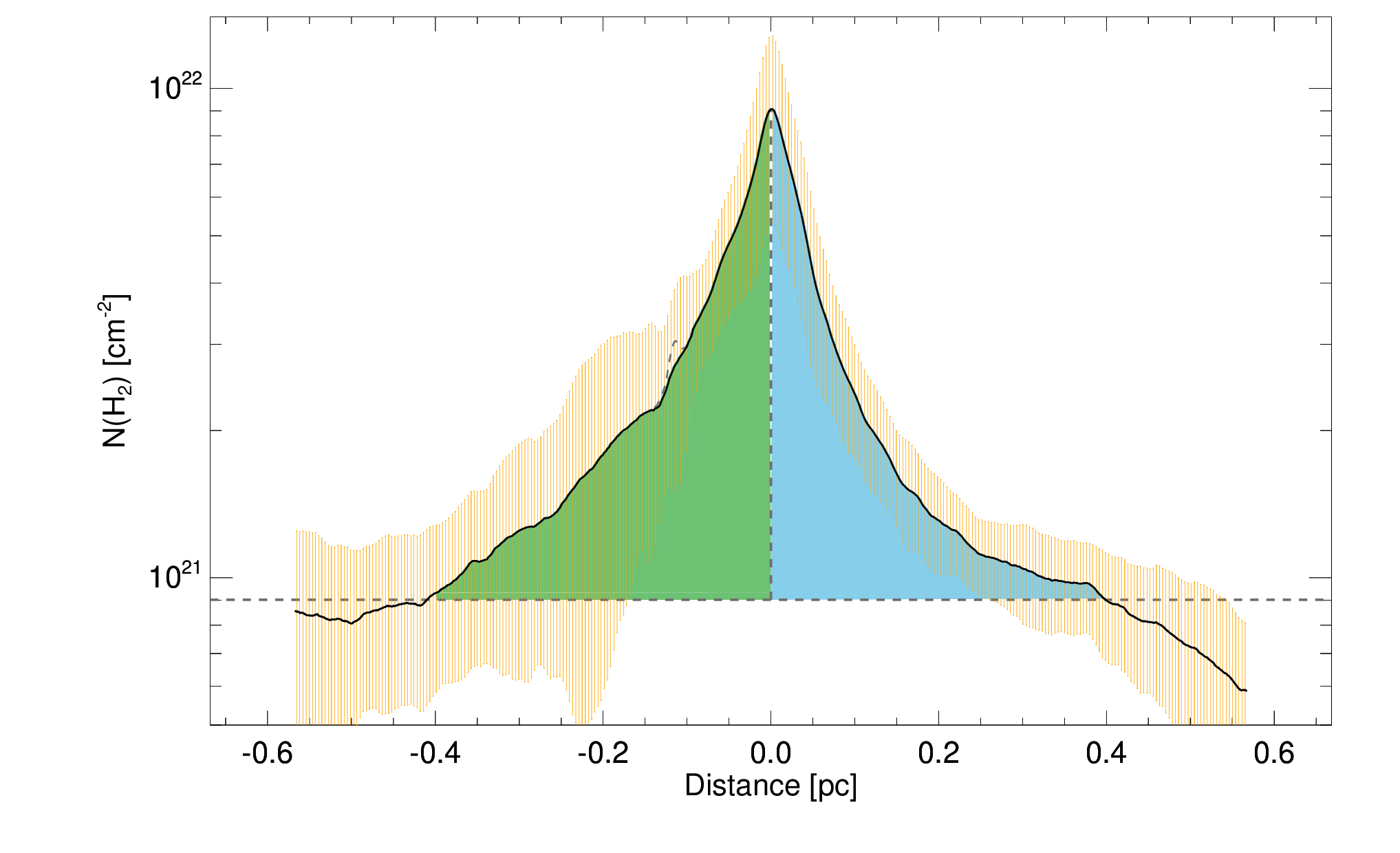}}
		\end{minipage}
	\end{center}
\caption{The upper panel shows the filament that was located using \textsl{DisPerSE}. The background image is of the high resolution column density map. The lower panel shows the profile taken from the low-resolution column density map using the skeleton shown in the upper panel in white. The \textit{x}-axis runs from south-east to north-west. The black line is the mean value of the column density at that radius. The dispersion of column densities along that radius is shown in orange. The areas which were integrated to calculate the asymmetry are shaded in green and blue. The vertical dashed line shows the crest position of the filament, and the horizontal dashed line shows the arbitrary background level of the profile. The small dashed curve on te left side of the mean column density curve is the additional column density contribution from the YSO, S CrA, which was manually subtracted from the mean column density profile.}
     \label{streamer_fil}%
\end{figure}

\section{Distribution of mass in the Corona Australis cloud}
\citet{girichidis2014} studied the evolution of a probability density function (PDF) for the mass density for self-gravitating systems. They used hydrodynamical models to show that isothermal non-self-gravitating gas goes on to develop a log-normal distribution. They suggested that despite the presence of gravity and the non-isothermality of astrophysical systems, the effects that they have are minor compared to the kinematic motions of the surrounding environment. Simulations such as those in \citet{girichidis2014} and \citet{klessen2000} showed that the high mass end of the PDF of the mass density of a molecular cloud is expected to evolve to a power-law tail under the influence of gravity. \citet{schneider2013} made use of the \textit{Herschel} data as the wide field of view combined with the resolution provided enough data to construct PDFs of the molecular clouds in the HGBS. PDFs can be used as a tool to assess the role gravity and turbulence play in the evolution of molecular clouds.\par
\citet{schneider2013} studied the PDFs for Orion B, Aquila and Polaris, and defined the PDF as
\begin{equation}
p(\eta)d\eta = (2\pi \sigma^{2}_{\eta})^{-0.5}\textrm{exp}[-(\eta - \mu)^{2} / (2\sigma^{2}_{\eta})]d\eta,
\end{equation}
where $\eta = \textrm{ln}(\textrm{N(H}_{2})/\langle \textrm{N(H}_{2}\textrm )\rangle)$ and $\sigma_{\eta}$ is the dimensionless dispersion of the logarithmic field and $\mu$ is the mean. The power-law slope, $p(\eta) = p_{0}(\eta/\eta_{0})^{s}$ was fitted using a regression to the tail.\par
\citet{federrath2013} created mass PDFs from spherical radial density distributions characterised as
\begin{equation}
\rho\propto r^{-\alpha},
\end{equation}
where $\rho$ is the volume density of the cloud. In reality, the 3D distribution of the cloud is unknown, and we have access only to the column density along the line of sight. The equivalent relation for the column density was shown by \citet{federrath2013} to be,
\begin{equation}
\Sigma \propto r^{-\alpha + 1},
\end{equation}
where $\Sigma$ is the surface density, and the quantity $\alpha$ is related to the slope of a column density PDF as:
\begin{equation}
\alpha = -\frac{2}{s} + 1,
\end{equation}
where $s$ is the slope of the power-law tail of the PDF obtained from the column density map.
\par
We created a PDF for CrA using the high resolution column density map, which is shown in Fig.~\ref{cra_global_pdf}. We excluded noise spikes and artefacts created during the creation of the high resolution map by applying a custom mask to this map. We adopted a bin-size of 0.1~$\eta$. \citet{schneider2015} found that changes in PDF derived parameters, with respect to the bin-size used, are negligible. Features of PDFs can be smoothed out by increasing the bin-size, and the statistical noise will increase for a sampling that is too small. However, tests conducted on the Auriga column density map showed that the parametrisations of the power-law tail, and that of the log-normal fitting, remain almost identical (see Appendix A of \citealt{schneider2015}). \par
The parameter $\sigma_{\eta}$ was determined using a $\chi ^2$ minimisation routine. We created a grid of PDF models, which were parametrised by $\sigma_{\eta}$ and $\mu$ (see equation 6), and minimised the variance between the observed data and the PDF models. We found a value of  $\sigma=0.43$, which is similar to the values found in subregions called 1 and 3 within the Rosette molecular cloud \citep{schneider2012rosette} as well as Orion B \citep{schneider2013}.
We conducted three separate fits to the slope of the PDF beyond the quantity $N$(H$_{2}$)$_{\textrm{break}}$. The first involved fitting the data within the boundaries of column density threshold of $N$(H$_{2}$)$_{\textrm{break}}\leq$ $N$(H$_{2}$)$\leq1\times10^{21}$~N(H$_{2}$)~cm$^{-2}$. The second, involved fitting the power-law for the boundary $N$(H$_{2}$)$\geq1\times10^{21}$ cm$^{-2}$. The third fit involved fitting the whole of the power-law tail beyond $N$(H$_{2}$)$_{\textrm{break}}$. The results of the fits are given in Table \ref{table:cra_pdf}.\par
In Fig.~\ref{cra_global_pdf}, the black histogram shows the binned column densities, with the Gaussian fit over-plotted in green and the power-law tail plotted blue and green solid lines, and a dashed dark grey line. The two vertical dashed lines indicate the limits for which the Gaussian fit was conducted. The left hand dashed line is the location where the fitted PDF breaks away from the very low mass material. Most of this low mass material lies below the noise level of the high-resolution map. The right-hand dashed line is the location of $N$(H$_{2}$)$_{\textrm{break}}$. The very high mass end of the PDF in Fig.~\ref{cra_global_pdf} shows significant deviation away from the power-law behaviour. This can be partially explained by effects along the line of sight, as well as the limited angular resolution of the maps. However, physical effects could also play a significant role in the production of this behaviour beyond $N$(H$_{2}$)$_{\textrm{break}}$.\par
For two of the fits to the tail of the PDF in CrA, we found that the parameter $\alpha$ lies around the top end of the expected range (between 1.5 and 2) for dense cores for intermediate-mass \citep{schneider2012rosette,schneider2013,federrath2013,girichidis2014}. The high-mass end of the PDF shows a value $\alpha=2.4$. For such a high value of $\alpha$, one can expect that some other compressive process is occurring \citep{schneider2012rosette,tremblin2014}. Figure \ref{cra_global_pdf} shows that the CrA-A subregion has a second peak at $\sim10^{22}$~N(H$_{2}$)~cm$^{-2}$. \citet{schneider2012rosette} also found that two subregions within the Rosette molecular cloud showed a second peak in their column density PDFs. The Rosette molecular cloud is a region of high mass star formation, and the feedback processes generated by OB stars include gas compression due to the expanding ionisation front, thereby causing a second peak.\par 
Though \citet{schneider2012rosette} noted that the presence of a UV-illuminating source does not necessarily imply that a double peaked PDF should be expected for all regions, compression effects may be a contributing factor on a case-by-case basis. The study of irradiation of protostellar cores in the Coronet by \cite{lindberg2012}, led to R CrA being a primary suspect for the source of localised heating. The secondary peak in the PDF of the molecular cloud supports the possibility that the UV radiation from R CrA is causing, or significantly contributing to, the development of this secondary peak.
\begin{table}
	\caption{Results of the three fits to the power-law tail shown in Fig.~\ref{cra_global_pdf}. }
\centering
\begin{tabular}{ l cccc }
\hline\hline
& Full tail & High-mass & Intermediate-mass \\
& (Dashed grey) &	(Orange)	&	(Blue) \\
\hline
$s$ &  -1.61 &  -1.46 &  -2.15 \\
$\alpha$ & 2.2 & 2.4 & 1.9	\\ \hline
\end{tabular}
\label{table:cra_pdf}
\end{table}
\begin{figure}[!h]
	\begin{center}
		\begin{minipage}{1.0\linewidth}
		\resizebox{1.0\hsize}{!}{\includegraphics[angle=0, trim={0.5cm 0cm 0cm 0cm}]{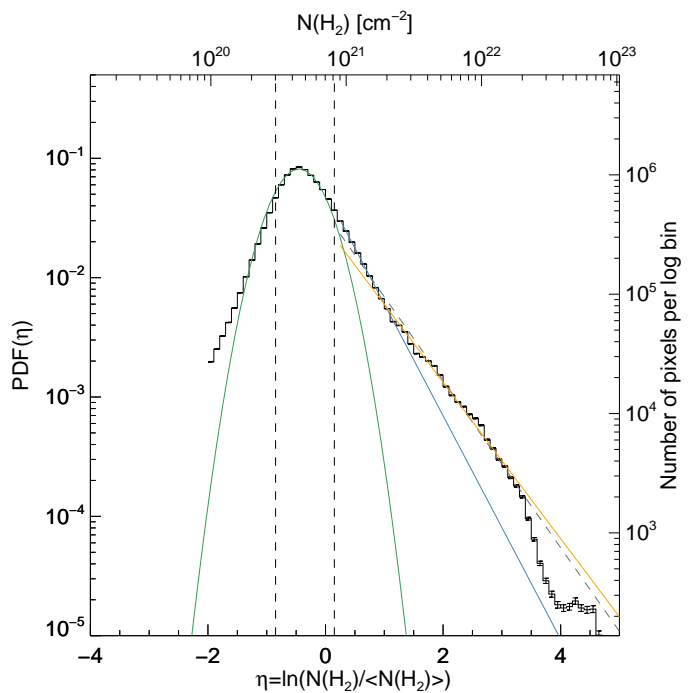}}
		\end{minipage}
	\end{center}
\caption{The probability distribution functions of column density for Corona Australis, at an angular resolution of $18''$. The black histogram shows the global PDF for the molecular cloud. The two histograms in red and green show the regional PDFs for CrA-A, and CrA-C, respectively.The left y axis is the normalised probability $p(\eta)$. The pixel size of $3''$ means that the errors calculated using Poisson statistics are very small, however the characteristics of the PDF do not change on lower resolution grids (see also \citealt{schneider2013}). The green curve is the Gaussian fit to this PDF between the two vertical dashed lines.The right hand dashed line represents the point at which the power-law tail was determined to develop, departing the log-normal distribution. The dashed grey (full tail), blue (intermediate-mass) and orange (high-mass) lines are the regression fits to the column densities higher than right hand dashed line, and represent the more dense material on the column density map.}
     \label{cra_global_pdf}%
\end{figure}
\section{Conclusions}
We have used data from the \textit{Herschel} Gould Belt Survey to create the first highly sensitive wide field survey of cores in Corona Australis. We identified a total of 163 starless cores, including 99 candidate prestellar cores, of which are 23 robust prestellar cores. We also identified 14 protostellar cores based on extractions conducted on the 70-$\mu$m images.\par
We found that the mass-size relation for these cores provides evidence that the evolution of star formation throughout the molecular cloud could be primarily being influenced by members of the Coronet cluster, especially R CrA.\par
We determined the masses and temperatures of the starless dense cores using SED fits to the \textsl{getsources} fluxes. The distribution of the CMF is consistent with the expected shape of the core mass function of unbound cores. We determined the power-law index $\gamma$ for the cores using two different methods. The maximum likelihood estimate of the power-law index of the $\textrm{d}N/\textrm{dLog}M$ distribution for starless cores is 0.59$\pm$0.04, which is consistent with the slope found for CO clumps, as found by \citet{kramer1998}, and the census of (mostly unbound) starless dense cores in Taurus L1495 \citep{marsh2016}. \par
We looked for evidence that the Coronet members could be influencing the local prestellar cores and found that there is a substantial increase in temperature for prestellar cores that are located within \textasciitilde0.1~pc of the Coronet cluster. The radius of these cores also appears to decrease as the distance from the Coronet decreases. This could be evidence that the Coronet is heating the local gas, which then exerts a higher external pressure on the cores that are within the local area. There is a temperature-column density relation throughout many of the regions in CrA. However this breaks down for most of the prestellar cores in CrA-A, the region containing the Coronet cluster. \par
We found that prestellar cores are closely associated with filaments, lending support to the hypothesis of \citet{andre2010HGBS}, that compact dense cores are primarily forming within the filamentary structure of molecular clouds. All of the BE defined prestellar cores lie well within the 0.1~pc quasi-universal filament width found by \citet{arzoumanian2011fils}.\par
We extracted the filamentary structure of the cloud and found that the streamer to the west of the Coronet appears to have a slightly asymmetric profile, with high and variable dispersion of column densities on the eastern side of the filament, but lower and uniform dispersion to the west. This could be evidence that the entire cloud could be under an external influence such as that of the Upper Centaurus Lupus association, or another unknown source. \citet{harju1993} observed that the morphology of CrA is very similar to that of the Orion A molecular cloud. In both of these molecular clouds, there is a nucleus of material, with tail-like structure, pointing away from OB associations. In the case of Orion A, the head points to the Ori I association. In CrA, the head points towards the UCL association. \citet{harju1993} concluded that the Orin A cloud, and CrA, were likely formed under the same physical process. Our observation of a slightly asymmetric profile, and the disparity in dispersion in the westernmost filament appears to support these claims.\par
We investigated the mass distribution across the molecular cloud. We used the column density map as an empirical probability density function, and analysed the global and regional PDF properties of CrA. We observed that the PDF of CrA is dominated by the lognormal component of the low density cirrus-like material, with contributions from the dense clumps forming two different power-law tails. The densest regions within CrA, which we referred to as CrA-A and CrA-C, dominate the high mass end of the PDF. This appears to be further evidence that the CrA-A, CrA-C and possibly CrA-B regions of the complex are further evolved than the other regions. These regions also contain the highest number of `bound' cores, which would support this claim.\par

\label{lastpage}
\begin{acknowledgements} 
D.W.B wishes to thank the University of Central Lancashire for studentship support while this research was carried out. GJW gratefully acknowledges support of an Emeritus Fellowship from the Leverhulme trust. SPIRE was developed by a consortium of institutes led by Cardiff Univ. (UK) and including: Univ. Lethbridge (Canada); NAOC
(China); CEA, LAM(France); IFSI, Univ. Padua (Italy); IAC (Spain); Stockholm
Observatory (Sweden); Imperial College London, RAL, UCL-MSSL, UKATC,
Univ. Sussex (UK); and Caltech, JPL, NHSC, Univ. Colorado (USA). This development
was supported by national funding agencies: CSA (Canada); NAOC
(China); CEA, CNES, CNRS (France); ASI (Italy); MCINN (Spain); SNSB
(Sweden); STFC, UKSA (UK); and NASA (USA). PACS was developed
by a consortium of institutes led by MPE (Germany) and including UVIE (Austria);
KUL, CSL, IMEC (Belgium); CEA, OAMP (France); MPIA (Germany);
IFSI, OAP/AOT, OAA/CAISMI, LENS, SISSA (Italy); IAC (Spain). This development
was supported by the funding agencies BMVIT (Austria), ESAPRODEX
(Belgium), CEA/CNES (France), DLR (Germany), ASI (Italy), and
CICT/MCT (Spain). This work received support from the European Research
Council under the European Union’s Seventh Framework Programme (ERC Advanced
Grant Agreement no. 291294 – ’ORISTARS’) and from the French National
Research Agency (Grant no. ANR–11–BS56–0010 – ‘STARFICH’). We thank an anonymous referee for the comments that improved the paper.
\end{acknowledgements}

\appendix
\section{A catalogue of dense cores identifed with \textit{Herschel} in the Corona Australis molecular cloud}
\label{sec:app.catalogue}
We identified 163 starless cores and 14 protostellar cores. A template of the catalogue which is available in complete online, is provided below to illustrate its content and form. Of the 163 starless cores, 122 were located within the western extracted field, with the remaining 41 within the eastern extraction field. \par
In addition to starless cores, both tables contain entries for dense cores with embedded protostars. The derived properties for these cores is more uncertain, with their masses underestimated by a factor \textasciitilde2. We detect 14 protostellar cores within Corona Australis, and their properties are not discussed within this work. We also include 62 additional sources not included within the scientific discussion. These additional sources are given a `core type' flag of -1, as their nature is uncertain (see Sect. \ref{sec:completeness}).\par
Most previous surveys have been carried out at a lower resolution than that of the \textit{Herschel} data. Disparities within the matches for SIMBAD objects are to be expected. One such core is the prestellar core in CrA-E. \citet{Ullman2013} estimated the mass of this core to be 21 M$_{\odot}$ at 130~pc. We estimate that this object has a mass of 0.4 M$_{\odot}$. This particular core was found to be structured such that there is a highly compact object at the centre of the larger scale clump. In effect, this may be treated as a single large scale object, or as a highly compact object, embedded within the larger scale structure, which is the case for the HGBS source extractions.\par
Comments in the far right of both tables contain the same information. Rejected SED fits are indicated, along with low 250-$\mu$m flux sources, two sources within reflection nebula, and sources that are spatially unresolved within the high-resolution column density map.\par

\begin{sidewaystable*}[htb]\tiny\setlength{\tabcolsep}{2.5pt}
\caption{Catalogue of  dense cores identified in the HGBS maps of the Corona Australis molecular cloud (template, full catalog only provided online).} 
\label{tab_obs_cat}
\renewcommand{\arraystretch}{1.2}
\begin{tabular}{ | c | c | r@{:}c@{:}l r@{:}c@{:}l | c r l c c l r c c c }
\hline\hline 
 C.No. & Source name & \multicolumn{3}{c}{R.A. (2000)} & \multicolumn{3}{c}{Dec. (2000)} & Sig$_{70}$ & \multicolumn{2}{c}{$S^\mathrm{peak}_{70}$} & $S^\mathrm{peak}_{70}/S_\mathrm{bg}$ & $S^{\textrm{conv,}500}_{70}$ & \multicolumn{2}{c}{$S^\mathrm{tot}_{70}$} & a$_{70}$ & b$_{70}$ & PA$_{70}$  \\ 
 & & \multicolumn{3}{c}{(h m s)} & \multicolumn{3}{c}{($^{\circ}$~\arcmin~\arcsec)} &  & \multicolumn{2}{c}{(Jy~beam$^{-1}$)} &  & (Jy~beam$^{-1}_{500}$) & \multicolumn{2}{c}{(Jy)} & (\arcsec) & (\arcsec) & ($^{\circ}$) \\ 
 (1) & (2) & \multicolumn{3}{c}{(3)} & \multicolumn{3}{c}{(4)} & (5) & \multicolumn{1}{c}{(6)} & \multicolumn{1}{c}{(7)} & (8) & (9) & \multicolumn{1}{c}{(10)} & \multicolumn{1}{c}{(11)} & (12) & (13) & (14) \\ 
\hline
  5 & 190008.1-370137 & 19 & 00 & 08.16 & -37 & 01 & 37.9 & 8 &   -2.30e-02&   2.91e-03 &     -4.47 &   -3.03e+00 &   -1.17e-01&   1.49e-02 &  59 &  10 & 100 \\
 36 & 190141.7-365832 & 19 & 01 & 41.71 & -36 & 58 & 32.8 & 2708 &    5.37e+01&   3.26e-02 &    317.73 &    1.53e+03 &    7.95e+01&   4.82e-02 &   8 &   8 &  20 \\
151 & 191020.1-370826 & 19 & 10 & 20.17 & -37 & 08 & 27.0 & 6 &    8.14e-02&   1.58e-02 &      1.26 &    4.90e+00 &    1.49e-01&   2.88e-02 &  17 &   9 & 142 \\
\hline\hline
\end{tabular}
\scalebox{1.2}{$\sim$}
\vspace{0.2cm}

\scalebox{1.2}{$\sim$}
{\renewcommand{\arraystretch}{1.2}
\begin{tabular}{ c r l c c l r c c c c r l c c l r c c c }
\hline\hline 
  Sig$_{160}$ & \multicolumn{2}{c}{$S^\mathrm{peak}_{160}$} & $S^\mathrm{peak}_{160}/S_\mathrm{bg}$ & $S^{\textrm{conv,}500}_{160}$ & \multicolumn{2}{c}{$S^\mathrm{tot}_{160}$} & a$_{160}$ & b$_{160}$ & PA$_{160}$ &   Sig$_{250}$ & \multicolumn{2}{c}{$S^\mathrm{peak}_{250}$} & $S^\mathrm{peak}_{250}/S_\mathrm{bg}$ & $S^{\textrm{conv,}500}_{250}$ & \multicolumn{2}{c}{$S^\mathrm{tot}_{250}$} & a$_{250}$ & b$_{250}$ & PA$_{250}$ \\ 
  & \multicolumn{2}{c}{(Jy~beam$^{-1}$)} &  & (Jy~beam$^{-1}_{500}$) & \multicolumn{2}{c}{(Jy)} & (\arcsec) & (\arcsec) & ($^{\circ}$) &  & \multicolumn{2}{c}{(Jy~beam$^{-1}$)} &  & (Jy~beam$^{-1}_{500}$) & \multicolumn{2}{c}{(Jy)} & (\arcsec) & (\arcsec) & ($^{\circ}$) \\
 (15) & \multicolumn{1}{c}{(16)} & \multicolumn{1}{c}{(17)} & (18) & (19) & \multicolumn{1}{c}{(20)} & \multicolumn{1}{c}{(21)} & (22) & (23) & (24) & (25) & \multicolumn{1}{c}{(26)} & \multicolumn{1}{c}{(27)} & (28) & (29) & \multicolumn{1}{c}{(30)} & \multicolumn{1}{c}{(31)} & (32) & (33) & (34) \\
\hline
12 &    1.70e-01&   2.56e-02 &      0.28 &    1.77e+01 &    1.24e+00&   1.86e-01 &  42 &  28 &  64 & 25 &    3.58e-01&   2.92e-02 &      0.32 &    1.68e+01 &    1.17e+00&   9.55e-02 &  37 &  24 &  54 \\
2138 &    5.76e+01&   7.29e-01 &     12.02 &    1.64e+03 &    5.72e+01&   7.23e-01 &  13 &  13 & 150 & 2264 &    3.42e+01&   9.75e-01 &      6.22 &    8.59e+02 &    3.05e+01&   8.70e-01 &  18 &  18 &  42 \\
216 &    5.49e+00&   5.00e-02 &      8.97 &    1.56e+02 &    5.10e+00&   4.65e-02 &  13 &  13 & 115 & 655 &    8.89e+00&   2.43e-01 &      3.94 &    2.23e+02 &    7.49e+00&   2.05e-01 &  18 &  18 &  40 \\
\hline\hline
\end{tabular}
}
\scalebox{1.2}{$\sim$}
\vspace{0.2cm}

\scalebox{1.2}{$\sim$}
{\renewcommand{\arraystretch}{1.2}
\begin{tabular}{ c r l c c l r c c c c r l c c l r c c c }
\hline\hline 
  Sig$_{350}$ & \multicolumn{2}{c}{$S^\mathrm{peak}_{350}$} & $S^\mathrm{peak}_{350}/S_\mathrm{bg}$ & $S^{\textrm{conv,}500}_{350}$ & \multicolumn{2}{c}{$S^\mathrm{tot}_{350}$} & a$_{350}$ & b$_{350}$ & PA$_{350}$ &   Sig$_{500}$ & \multicolumn{2}{c}{$S^\mathrm{peak}_{500}$} & $S^\mathrm{peak}_{500}/S_\mathrm{bg}$ & \multicolumn{2}{c}{$S^\mathrm{tot}_{500}$} & a$_{500}$ & b$_{500}$ & PA$_{500}$ &\\ 
  & \multicolumn{2}{c}{(Jy~beam$^{-1}$)} &  & (Jy~beam$^{-1}_{500}$) & \multicolumn{2}{c}{(Jy)} & (\arcsec) & (\arcsec) & ($^{\circ}$) &  & \multicolumn{2}{c}{(Jy~beam$^{-1}$)} &  & \multicolumn{2}{c}{(Jy)} & (\arcsec) & (\arcsec) & ($^{\circ}$) \\
 (35) & \multicolumn{1}{c}{(36)} & \multicolumn{1}{c}{(37)} & (38) & (39) & \multicolumn{1}{c}{(40)} & \multicolumn{1}{c}{(41)} & (42) & (43) & (44) & (45) & \multicolumn{1}{c}{(46)} & \multicolumn{1}{c}{(47)} & (48) & \multicolumn{1}{c}{(49)} & \multicolumn{1}{c}{(50)} & (51) & (52) & (53) &\\
\hline
22 &    3.55e-01&   2.91e-02 &      0.32 &    1.11e+01 &    6.59e-01&   5.40e-02 &  36 &  25 &  64 & 14 &    2.36e-01&   4.76e-02 &      0.23 &    2.57e-01&   5.18e-02 &  42 &  36 &  67 &\\
988 &    1.67e+01&   1.71e+00 &      3.69 &    4.34e+02 &    1.57e+01&   1.61e+00 &  24 &  24 &  60 & 486 &    8.18e+00&   1.78e+00 &      1.94 &    7.20e+00&   1.57e+00 &  36 &  36 & 177 &\\
534 &    7.99e+00&   3.43e-01 &      2.72 &    2.08e+02 &    8.34e+00&   3.58e-01 &  24 &  24 &  21 & 291 &    6.54e+00&   4.53e-01 &      2.21 &    6.77e+00&   4.69e-01 &  36 &  36 &   8 &\\
\hline\hline
\end{tabular}
}
\scalebox{1.2}{$\sim$}
\vspace{0.2cm}

\scalebox{1.2}{$\sim$}
{\renewcommand{\arraystretch}{1.2}
\begin{tabular}{ c c c c c c c c c c c c c c c c | }
\hline\hline 
  Sig$_{\textrm{N}(\textrm{H}_{2})}$ & $N^\mathrm{peak}_{\textrm{H}_{2}}$ & $N^\mathrm{peak}_{\textrm{H}_{2}}/N_\mathrm{bg}$ & $N^{\textrm{conv,}500}_{\textrm{H}_{2}}$ & $N^{\textrm{bg}}_{\textrm{H}_{2}}$ & a$_{\textrm{N}(\textrm{H}_{2})}$ & b$_{\textrm{N}(\textrm{H}_{2})}$ & PA$_{\textrm{N}(\textrm{H}_{2})}$ &   N$_{\textrm{SED}}$ & \textsc{csar} & CuTEx & Core type & SIMBAD & WISE & \textit{Spitzer} & Comments\\
  & (10$^{21}$~cm$^{-2}$) &   & (10$^{21}$~cm$^{-2}$) & (\arcsec) & (\arcsec) & ($^{\circ}$) & & & & & & & & &\\
 (54) & (55) & (56) & (57) & (58) & (59) & (60) & (61) & (62) & (63) & (64) & (65) & (66) & (67) & (68) & (69) \\
\hline
  19 &   1.28 &     0.41 &   0.59 &   3.15 &  35 &  24 &  53 &        4 &    2 &    1 &  3 &   &  &  &  \\
 934 &  69.27 &     8.77 &  17.41 &   7.90 &  18 &  18 &  53 &        5 &    2 &    2 &  4 & 2MASS J19014156-3658312 & J190141.58-365831.6 & IRS 2 &  \\
 550 &  48.11 &     1.83 &  13.83 &  26.35 &  21 &  18 &  67 &        5 &    2 &    2 &  2 & [YMS99b] 5 &  &  &  \\
\hline\hline
\end{tabular}
\tablefoot{Catalogue entries are as follows: 
{\bf(1)} Core number;
{\bf(2)} Core name $=$ HGBS\_J prefix directly followed by a tag created from the J2000 sexagesimal coordinates; 
{\bf(3)} and {\bf(4)}: Right ascension and declination of core center; 
{\bf(5)}, {\bf(15)}, {\bf(25)}, {\bf(35)}, and {\bf(45)}: Detection significance from monochromatic single scales, in the 70-, 160-, 250-, 350-, and 500-$\mu$m maps, respectively. 
(NB: the detection significance has the special value of -999 when the core is not visible in clean single scales. A few sources within Table \ref{tab_obs_cat} have the flux quantities $S$ of 0.00e+00 within their PACS measurements, as these sources were outside of the masks used for those particular wavelengths of the extractions); 
{\bf(6)}$\pm${\bf(7)}, {\bf(16)}$\pm${\bf(17)} {\bf(26)}$\pm${\bf(27)} {\bf(36)}$\pm${\bf(37)} {\bf(46)}$\pm${\bf(47)}: Peak flux density and its error in Jy/beam as estimated by \textsl{getsources};
{\bf(8)}, {\bf(18)}, {\bf(28)}, {\bf(38)}, {\bf(48)}: Contrast over the local background, defined as the ratio of the background-subtracted peak intensity to the local background intensity ($S^{\rm peak}_{\rm \lambda}$/$S_{\rm bg}$); 
{\bf(9)}, {\bf(19)}, {\bf(29)}, {\bf(39)}: Peak flux density measured after smoothing to a 36.3$\arcsec$ beam; 
{\bf(10)}$\pm${\bf(11)}, {\bf(20)}$\pm${\bf(21)}, {\bf(30)}$\pm${\bf(31)}, {\bf(40)}$\pm${\bf(41)}, {\bf(49)}$\pm${\bf(50)}: Integrated flux density and its error in Jy as estimated by \textsl{getsources}; 
{\bf(12)}--{\bf(13)}, {\bf(22)}--{\bf(23)}, {\bf(32)}--{\bf(33)}, {\bf(42)}--{\bf(43)}, {\bf(51)}--{\bf(52)}: Major \& minor FWHM diameters of the core (in arcsec), respectively, 
as estimated by \textsl{getsources}. (NB: the special value of $-1$ means that no size measurement was possible); 
{\bf(14)}, {\bf(24)}, {\bf(34)}, {\bf(44)}, {\bf(53)}: Position angle of the core major axis, measured east of north, in degrees; 
{\bf(54)} Detection significance in the high-resolution column density image;  
{\bf(55)} Peak H$_{2}$ column density in units of $10^{21}$ cm$^{-2}$ as estimated by \textsl{getsources} in the high-resolution column density image; 
{\bf(56)} Column density contrast over the local background, as estimated by \textsl{getsources} in the high-resolution column density image;
{\bf(57)} Peak column density measured in a 36.3$\arcsec$ beam; 
{\bf(58)} Local background H$_{2}$ column density as estimated by \textsl{getsources} in the high-resolution column density image; 
{\bf(59)}--{\bf(60)}--{\bf(61)}: Major \& minor FWHM diameters of the core, and position angle of the major axis, respectively, as measured in the high-resolution column density image; 
{\bf(62)} Number of {\it Herschel} bands in which the core is significant (Sig$_{\rm \lambda} >$ 5) and has a positive flux density, excluding the column density plane; 
{\bf(63)} `\textsl{CSAR}' flag: 2 if the \textsl{getsources} core has a counterpart detected by the \textsl{CSAR} source-finding algorithm \citep{kirk2013} within 6$\arcsec$ of its peak position,
equal to 1 if source found independently by \textsl{CSAR} within 6$\arcsec$ of \textsl{getsources} source, 2 if source found within source 50\% elliptical contour of \textsl{getsources} source ,and 3 if the source is not already idenfied using the former criteria. However is located using the mask image produced by \textsl{CSAR}
0 otherwise;
{\bf(63)} `\textsl{CuTEx}' flag: 2 if the \textsl{getsources} core has a counterpart detected by the \textsl{CuTEx} source-finding algorithm \citep{molinari2011} within 6$\arcsec$ of its peak position,
1 if no close \textsl{CuTEx} counterpart exists but the peak position of a \textsl{CuTEx} source lies within the FWHM contour of the \textsl{getsources} core in the high-resolution column density map, 
0 otherwise;
{\bf(65)} Core type: 1=unbound starless, 2=prestellar, 3=candidate prestellar (non-robust), 4=dense core with embedded protostar, or -1=tentative additional candidate core; 
{\bf(66)} Closest counterpart found in SIMBAD, if any, up to 1$\arcmin$ from the {\it Herschel} peak position;
{\bf(67)} Closest WISE-identified YSO from the WISE all-sky YSO catalogue given by \citep{marton2015} within 6$\arcsec$ of the {\it Herschel} peak position, if any. When present, the WISE source name contained within the aforementioned catalogue is given;
{\bf(68)} Closest {\it Spitzer}-identified YSO from the \textit{Spitzer} Gould Belt Survey \citep{peterson2011} within 6$\arcsec$ of the {\it Herschel} peak position, if any. When present, the {\it Spitzer} source name contained within the aforementioned catalogue is given;
{\bf(69)} Comments may be \textit{no SED fit}, \textit{spatially unresolved in col. dens}, \textit{low 250 micron flux}, \textit{low mass}, or \textit{reflection nebula?} (see text for details).}
}
\end{sidewaystable*}
\clearpage

\begin{sidewaystable*}[htb]\tiny\setlength{\tabcolsep}{4.0pt}
\caption{Derived properties of the dense cores identified in the HGBS maps of the Corona Australis molecular cloud (template, full table only provided online).}
\label{tab_der_cat_cores}
{\renewcommand{\arraystretch}{1.2}
\begin{tabular}{ | c | c | r@{:}c@{:}l r@{:}c@{:}l | c c c c c c c c c c c c c c c c|}
\hline\hline 
 C.No. & Source name & \multicolumn{3}{c}{R.A. (2000)} & \multicolumn{3}{c}{Dec. (2000)} & \multicolumn{2}{c}{$R_{\textrm{core}}$} & \multicolumn{2}{c}{$M_{\textrm{core}}$} & \multicolumn{2}{c}{$T_{\textrm{core}}$} & $N^\mathrm{\textrm{peak}}_{\textrm{H}_{2}}$ & \multicolumn{2}{c}{$N^\mathrm{ave}_{\textrm{H}_{2}}$} & $n^\mathrm{peak}_{\textrm{H}_{2}}$ & \multicolumn{2}{c}{$n^\mathrm{ave}_{\textrm{H}_{2}}$} & $\alpha_{\textrm{BE}}$ & Core type & Subregion & comments \\ 
 & & \multicolumn{3}{c}{(h m s)} & \multicolumn{3}{c}{($^{\circ}$~\arcmin~\arcsec)} & \multicolumn{2}{c}{(pc)} & \multicolumn{2}{c}{(M$_{\odot}$)} & \multicolumn{2}{c}{(K)} & (10$^{21}$~cm$^{-2}$) & \multicolumn{2}{c}{(10$^{21}$~cm$^{-2}$)} & (10$^{4}$~cm$^{-3}$) & \multicolumn{2}{c}{(10$^{4}$~cm$^{-3}$)} & & & &\\ 
 (1) & (2) & \multicolumn{3}{c}{(3)} & \multicolumn{3}{c}{(4)} & (5) & \multicolumn{1}{c}{(6)} & \multicolumn{1}{c}{(7)} & (8) & (9) & \multicolumn{1}{c}{(10)} & \multicolumn{1}{c}{(11)} & (12) & (13) & (14) &(15)&(16)&(17)&(18)&(19)&20\\ 
\hline
  5 & 190008.1-370137 & 19 & 00 & 08.16 & -37 & 01 & 37.9 & 0.019 & 0.015 &  0.010& 0.003 &    15.4&1.0 &   0.61 &     0.42 &     0.68 &   1.00 &     0.55 &     1.13 &   22.58 &  3 & CrA-A& \\
 36 & 190141.7-365832 & 19 & 01 & 41.71 & -36 & 58 & 32.8 & 0.011 & 0.011 &  0.134& 0.033 &    19.3&1.3 &   9.43 &    14.21 &    14.21 &  25.04 &    30.10 &    30.10 &    0.00 &  4 & CrA-A& \\
151 & 191020.1-370826 & 19 & 10 & 20.17 & -37 & 08 & 27.0 & 0.013 & 0.005 &  0.411& 0.078 &    10.9&0.4 &  24.07 &    36.10 &   214.39 &  58.26 &    69.76 &  1009.58 &    0.20 &  2 & CrA-E& \\
\hline\hline
\end{tabular}
}
\tablefoot{Table entries are as follows: {\bf(1)} Core running number; {\bf(2)} Core name $=$ HGBS\_J prefix directly followed by a tag created from the J2000 sexagesimal coordinates; 
{\bf(3)} and {\bf(4)}: Right ascension and declination of core center; 
{\bf(5)} and {\bf(6)}: Geometrical average between the major and minor FWHM sizes of the core (in pc), as measured in the high-resolution column density map 
before deconvolution, and after deconvolution from the 18.2$\arcsec$ HPBW resolution of the map, respectively.
(NB: Both values provide estimates of the object's outer {\it radius} when the core can be approximately described by a Gaussian distribution, as is the case 
for a critical Bonnor-Ebert spheroid); 
{\bf(7)} Estimated core mass ($M_\odot$) assuming the dust opacity law advocated by \citet{roy2014}; 
{\bf(9)} SED dust temperature (K); {\bf(8)} \& {\bf(10)} Statistical errors on the mass and temperature, respectively, including calibration uncertainties, but excluding dust opacity uncertainties; 
{\bf(11)} Peak H$_2$ column density, at the resolution of the 500$~\mu$m data, derived from a graybody SED fit to the core peak flux densities measured in a common 36.3$\arcsec$ beam at all wavelengths; 
{\bf(12)} Average column density, calculated as $N^{\rm ave}_{\rm H_2} = \frac{M_{\rm core}}{\pi R_{\rm core}^2} \frac{1}{\mu m_{\rm H}}$, 
          where $M_{\rm core}$ is the estimated core mass (col. {\bf 7}), $R_{\rm core}$ the estimated core radius prior to deconvolution (col. {\bf 6}), and $\mu = 2.86$;
{\bf(13)} Average column density calculated in the same way as for col. {\bf 12} but using the deconvolved core radius (col. {\bf 5}) instead of the core radius measured prior to deconvolution;  
{\bf(14)} Beam-averaged peak volume density at the resolution of the 500-$\mu$m data, derived from the peak column density (col. {\bf 11}) assuming a Gaussian spherical distribution: 
          $n^{\rm peak}_{\rm H_2} = \sqrt{\frac{4 \ln2}{\pi}} \frac{N^{\rm peak}_{\rm H_2}}{\overline{FWHM}_{\rm 500}}$; 
{\bf(15)} Average volume density, calculated as
          $n^{\rm ave}_{\rm H_2} = \frac{M_{\rm core}}{4/3 \pi R_{\rm core}^3} \frac{1}{\mu m_{\rm H}}$, using the estimated core radius prior to deconvolution; 
{\bf(16)} Average volume density, calculated in the same way as for col. {\bf 15} but using the deconvolved core radius (col. {\bf 5}) instead of the core radius measured prior to deconvolution; 
{\bf(17)} Bonnor-Ebert mass ratio: $\alpha_{\rm BE} = M_{\rm BE,crit} / M_{\rm obs} $ (see text for details); 
{\bf(18)} Core type: 1=unbound starless, 2=prestellar, 3=candidate prestellar (non-robust), 4=dense core with embedded protostar, or -1=tentative additional candidate core;
{\bf(19)} Subregion using an arbitrary mask;
{\bf(20)} Comments may be \textit{no SED fit}, \textit{spatially unresolved in col. dens}, \textit{low 250 micron flux}, \textit{low mass}, or \textit{reflection nebula?} (see text for details).. 
}
\end{sidewaystable*}
\clearpage
\newpage
   \begin{figure*}[!!!!!!!bbbb]
   \begin{center}
    \resizebox{0.6\hsize}{!}{\includegraphics[angle=0, trim={1cm 2cm 1cm 0cm}]{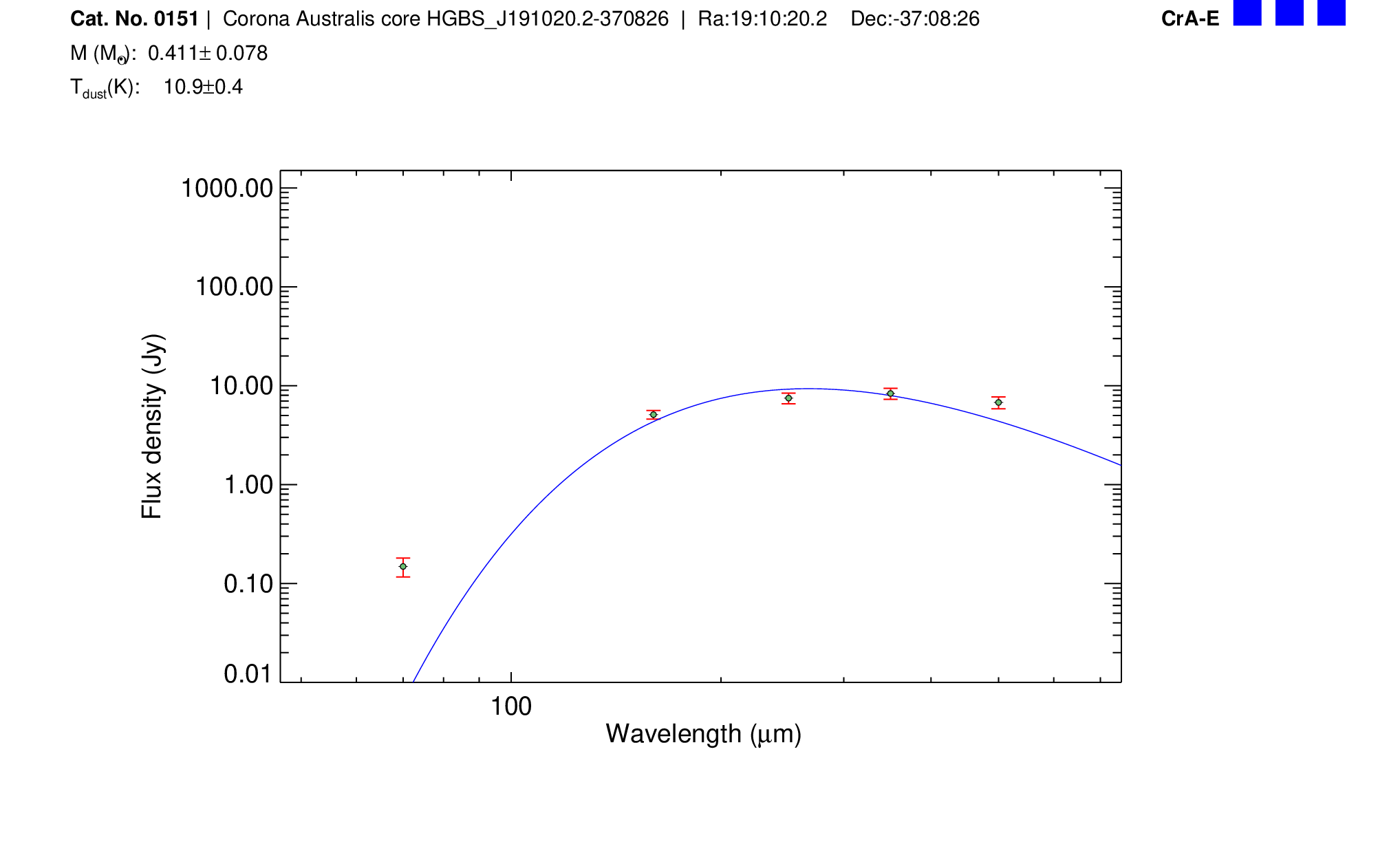}}
 \hspace{2mm}
    \resizebox{0.6\hsize}{!}{\includegraphics[angle=0, trim={1cm 4cm 1cm 0cm}]{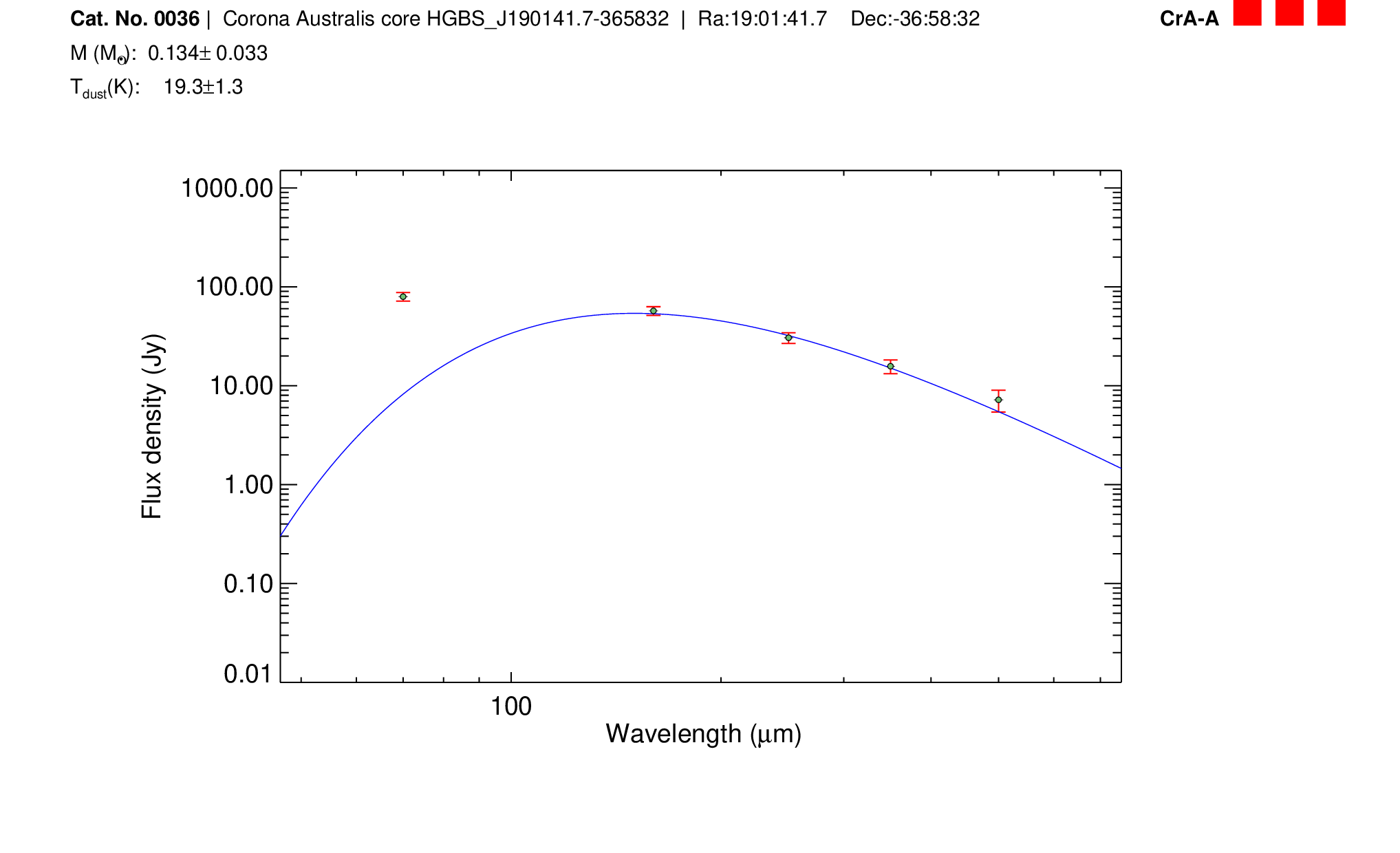}}
   \end{center}
   \caption{
	Examples of \textit{Herschel} spectral energy distributions (SEDs). The upper panel shows an SED for a prestellar core, and a protostellar core SED is shown on the lower panel. The corresponding source card images are shown in Figs. \ref{cardexamplecore}, and \ref{cardexampleproto}, respectively.
   }
    \label{fig_SED}%
    \end{figure*}
    
\begin{figure*}[!]
		\centering
		\resizebox{0.95\hsize}{!}{\includegraphics[angle=0, trim={0cm 2.5cm 0cm 1cm}]{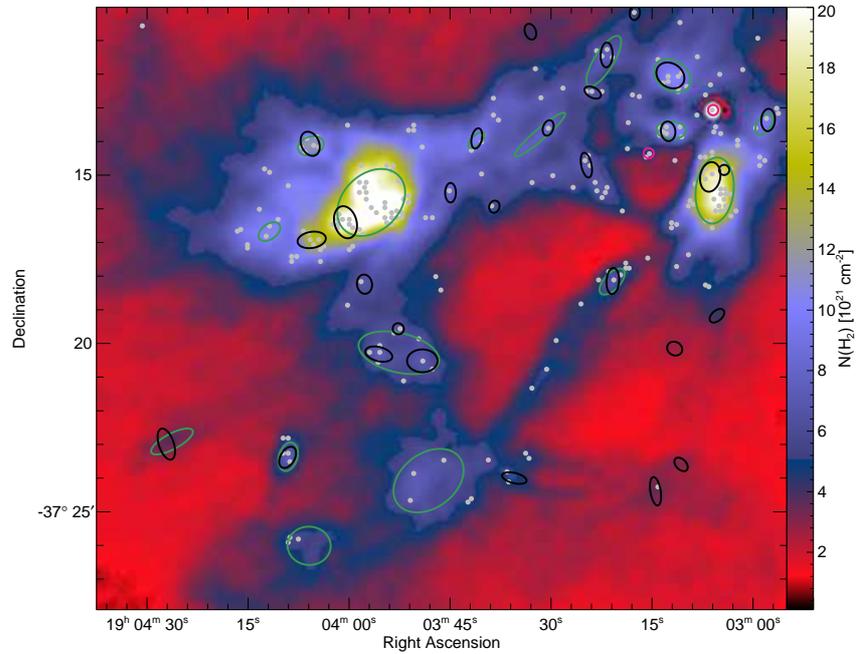}} 
\caption{A magnified view of an arbitrary sub-field within CrA. The background map is the high-resolution column density map. Black and pink ellipses mark the FWHM sizes of the dense cores, and protostellar cores, respectively. The green ellipses show the \textsl{CSAR} \citep{kirk2013} identified sources. The grey points show sources identified using \textsl{CuTEx} \citep{molinari2011}. Both \textsl{CSAR} and \textsl{CuTEx} were used on the high-resolution column density image.}
     \label{subfield}%
\end{figure*}   

\clearpage
\begin{figure*}[!]
		\centering
		\resizebox{0.95\hsize}{!}{\includegraphics[angle=0, trim={0cm 0cm 0cm 0cm}]{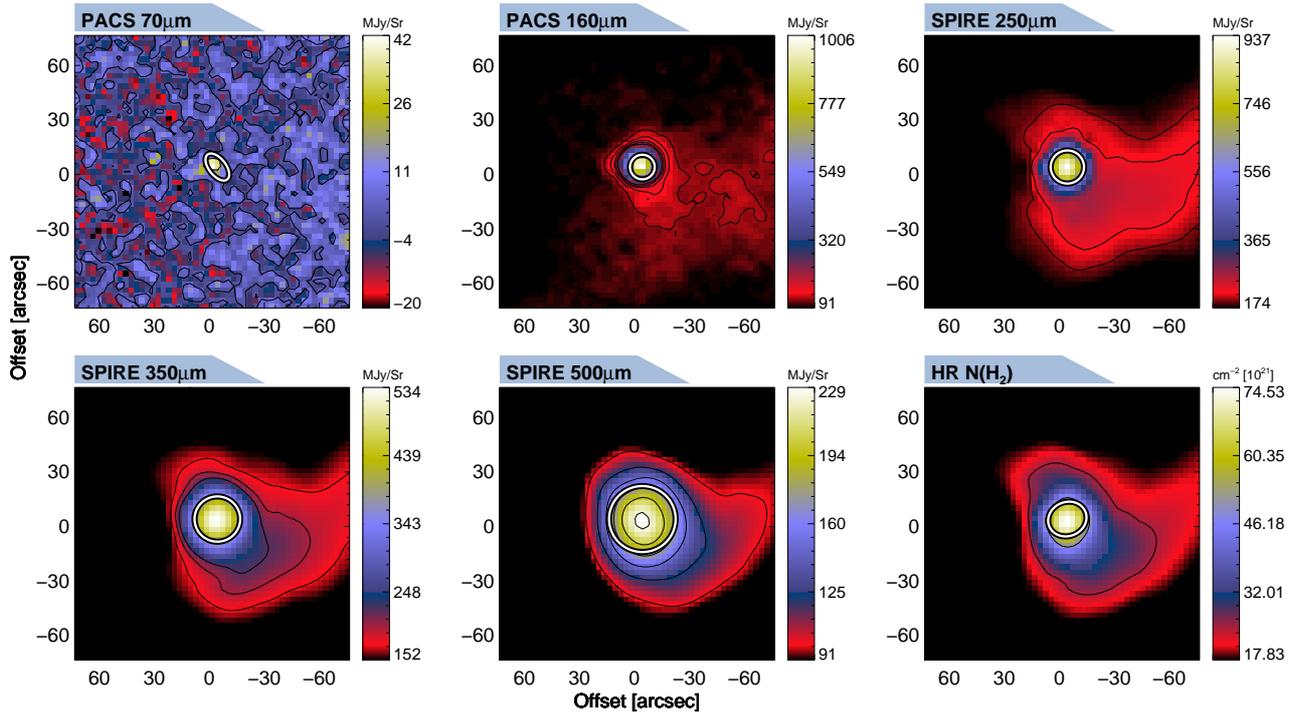}} 
\caption{Example `card' for a (bound) prestellar core. We show \textit{Herschel} images at 70~$\mu$m, 160~$\mu$m, 250~$\mu$m, 350~$\mu$m, and 500~$\mu$m. We also show the high-resolution column density image of the core. Ellipses represent the estimated major and minor FWHM sizes of the core at each wavelength. If a core is significantly detected at the respective wavelength, the line is solid, and is dashed otherwise. We provide a complete set of these images for the dense cores. Each card cut-out is \textasciitilde2.5\arcmin~across, on a 3\arcsec~pixel grid.}
     \label{cardexamplecore}%
\end{figure*}
\begin{figure*}[!]
		\centering
		\resizebox{0.95\hsize}{!}{\includegraphics[angle=0, trim={0cm 0cm 0cm 0cm}]{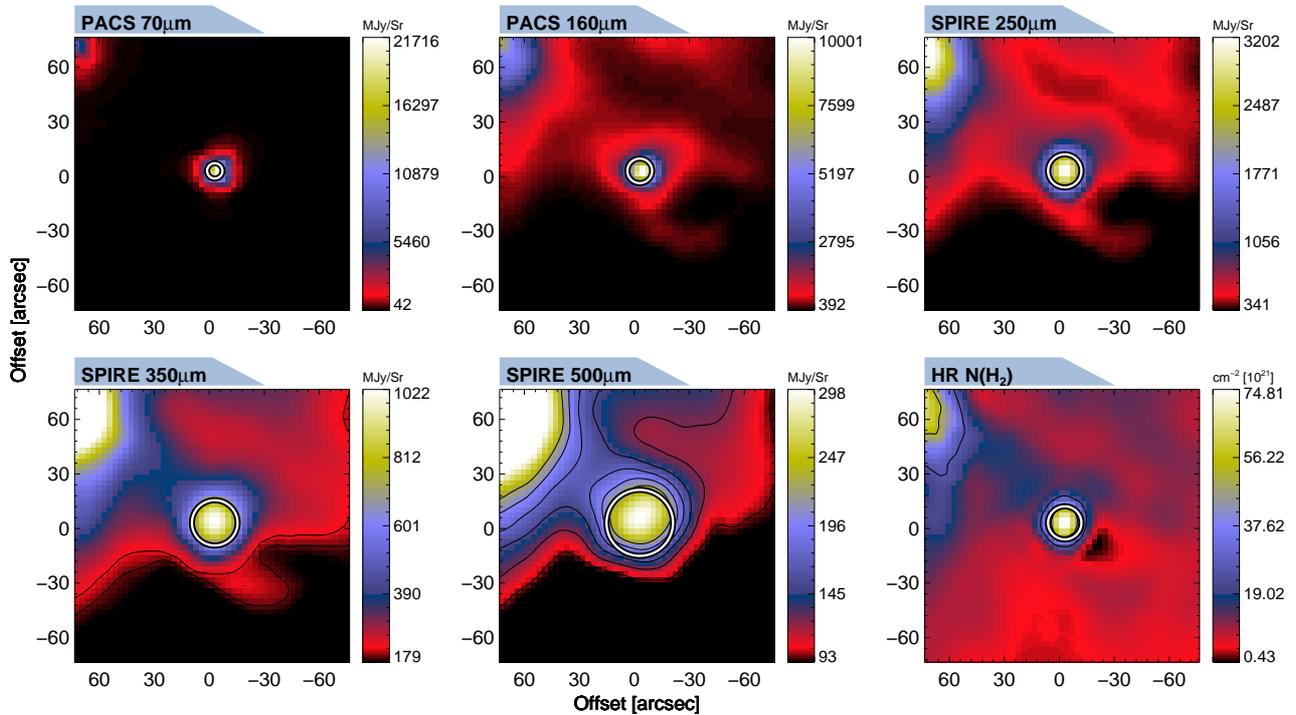}} 
\caption{Same as Fig.~\ref{cardexamplecore} for a protostellar core. The same images cut-outs are shown.}
     \label{cardexampleproto}%
\end{figure*}

\clearpage
\newpage

\section{Estimating the completeness of the prestellar core extractions in Corona Australis} \label{sec:completeness}
We estimated the completeness of the survey by conducting full source extractions using synthesised critical BE spheres. We used clean maps produced using \textsl{getsources}, which consisted of the remaining emission across all \textit{Herschel} wavelengths when all sources located using \textsl{getsources} were subtracted from the observed maps. Following \citet{aquila2015HGBS}, several sets of critical BE spheres were then injected into the clean-background images to produce synthetic \textit{Herschel} and column density maps. A population of 231 model starless cores with the flat input mass distribution $dN/dlogM\propto M^{-1.0}$, which is similar to the \citet{kroupa2001} IMF for the mass interval $0.5M_{\odot} \gtrsim M \gtrsim 0.08M_{\odot}$.\par
The dust continuum emission from the synthetic BE cores was constructed using dust radiative transfer models using the MODUST code \citep{bouwman2000,bouwman2001}, with a temperature drop towards the centre of these cores. The cores were distributed across the two tiles we used for the extractions. For the western extraction field these cores were distributed on the background column density map where $N_{\textrm{H}_{2}}^{bg}\geq 3.8 \times 10^{21}$ cm$^{-2}$. While for the eastern extraction field, these cores were distributed where $N_{H_{2}}^{bg}\geq 1.9 \times 10^{21}$ cm$^{-2}$ (see Fig.~\ref{getsources_tiles}). After constructing these synthetic \textit{Herschel} images across all wavelengths, we conducted the source extraction with \textsl{getsources} and the classification process outlined in Sect. 5.1.\par
To gain a more representative completeness limit, we sampled the the model starless cores with a background column density $>$5$\times 10^{21}$~N(H$_{2}$)~cm$^-2$. The completeness limit for starless cores is background dependent (see Appendix B.2 of \citealt{aquila2015HGBS}). 90\% of the prestellar cores lie above a column density of 5$\times 10^{21}$~N(H$_{2}$)~cm$^-2$. Figure \ref{complete_ratio} shows the completeness as a function of mass for the simulated data for bound cores. We estimate that the completeness of the survey is falls below 80\% for prestellar cores of less than 0.1 M$_{\odot}$.\par
The completeness limit for \textit{unbound} cores has been studied by \citet{marsh2016}. They found that the completeness of unbound cores is higher at lower masses in low column density regions (A$_{\textrm{v}}<5$), which is to be expected owing to the higher temperatures of unbound starless cores. They estimate that the completeness for unbound cores is $\simeq$85 percent complete for $M/M_{\odot}>0.015$ at $A_{\text{V}}\lesssim5$ mag. We do not make an attempt here to estimate the completeness for the unbound starless cores, but the completeness limit can be expected to be similar for these objects in CrA, given its similar distance.\par
Figure \ref{complete_ratio_radius} shows a plot of the masses of the simulated cores, against the median ratio of the derived core size to the true core size. The same mass bins as the completeness testing are used. The figure shows that the derived core sizes remain within 5\% of the true core sizes. Our results are consistent with the results of \citet{aquila2015HGBS}, who conducted a similar study towards Aquila. We estimate a conservative error of 15\% in the un-deconvolved radius of our cores.

\begin{figure}[!h]
	\begin{center}
		\begin{minipage}{1.0\linewidth}
		\resizebox{1.0\hsize}{!}{\includegraphics[angle=0, trim={0.5cm 1.5cm 0.2cm 0.5cm}]{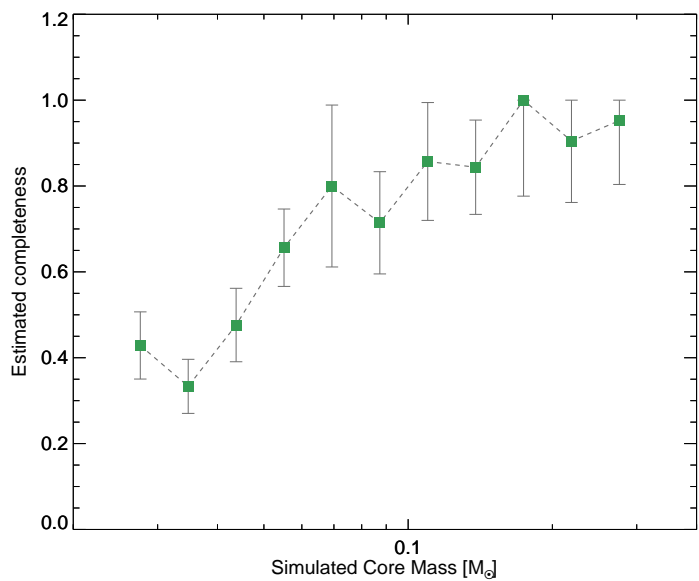}}
		\end{minipage}
	\end{center}
\caption{Catalogue completeness as a function of mass based on simulated data for bound (prestellar) cores. The error bars represent Poisson statistical errors only.}
     \label{complete_ratio}%
\end{figure}
\begin{figure}[!h]
	\begin{center}
		\begin{minipage}{1.0\linewidth}
		\resizebox{1.0\hsize}{!}{\includegraphics[angle=0, trim={2.5cm 1.5cm 0.2cm 0.5cm}]{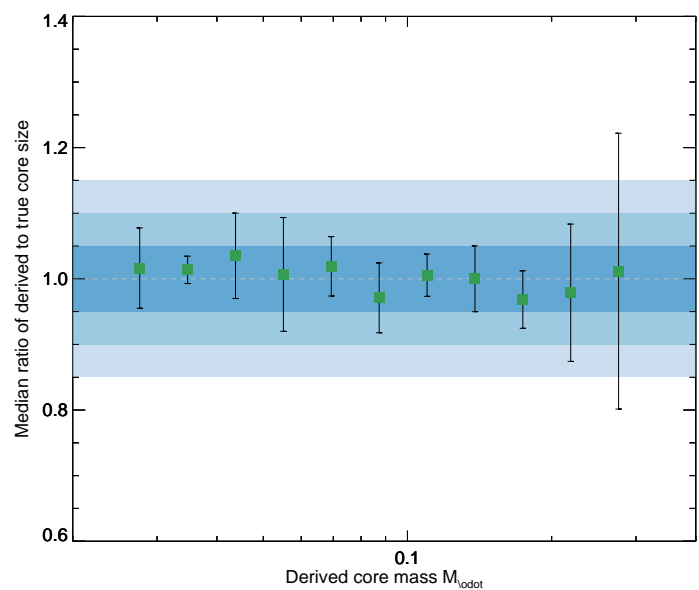}}
		\end{minipage}
	\end{center}
\caption{True core mass against the median ratio of derived to true core size. The error bars is the standard deviation of the sizes in each core bin. The horizontal dashed line marks a size ratio of 1. The three shaded blue bands from darkest to lightest, represent 5\%, 10\%, and 15\% differences in mass ratio, respectively. The median core size in each bin remains within 5\% of the true core size.}
     \label{complete_ratio_radius}%
\end{figure}
\subsection*{Extragalactic contaminants}
The environment within the eastern field (blue outline, Fig.~\ref{getsources_tiles}) is much more diffuse than that found within the western field (orange outline, Fig.~\ref{getsources_tiles}). For the purposes of source extraction, the estimated background level within the eastern sub-field is lower than that of the western field (see previous section). As a result, \textsl{getsources} is able to go much deeper into the eastern field. This increases the chances of including extragalactic contaminants within the extraction results. To asses the contamination with the results, we used the 250-$\mu$m data to estimate the number of extragalactic contaminants within the results.\par
\citet{oliver2010} studied the number counts of galaxies within SPIRE images as part of the The Herschel Multi-tiered Extragalactic Survey (HerMES - \citealt{oliver2012}). They found that the number density of extragalactic sources with $S_{250}>100$ mJy is 12.8$\pm$3.5 deg$^{-2}$. Therefore, for the 5.3 deg$^2$ eastern tile, and 5.7 deg$^2$ western tile, we expect 67$\pm$19 and 72$\pm$20 extragalactic sources, respectively. In reality, the contamination by extragalactic sources is substantially lower within the western field. The extraction is dominated by the bright region surrounding the Coronet. As a result of this, \textsl{getsources} does not go as deeply into the background within the western tiles as within the eastern tile. \par 
Figure \ref{mass_250flux} shows a plot of the masses of the sources, against their respective 250-$\mu$m integrated fluxes. The symbols are as in Fig.~\ref{region_map}, where the protostellar candidates are shown as stars. The vertical dashed line shows the 0.001 M$_{\odot}$ limit, below which we exclude cores. The lower horizontal dashed line shows the \textit{Herschel} 250-$\mu$m contamination limit, below which extragalactic sources are expected to dominate. The upper horizontal dashed line shows the 150 mJy integrated flux density level, below which we exclude sources from the scientific discussion, but are included in the catalogue. We identified 43 sources with 100 mJy $<S^{\textrm{tot}}_{250}<$150 mJy, and M$>$0.001 M$_{\odot}$, after the visual checks have been completed across both extraction fields. These sources are marked as having a low 250-$\mu$m flux within the catalogue. Within the western field, only one source is detected below 100 mJy. The parameter range of sources within the western field compares well with results within Taurus L1495 (see Figure A1 of \citealt{marsh2016} for a similar plot).\par 
We include 19 sources that posses a mass $<0.001$ M$_{\odot}$ within the catalogue, but exclude them from the scientific discussion. Of these sources, 12 have integrated flux densities above 150 mJy. Approximately 58\% of these sources have no SED fit. The sources with a valid SED fit within this group have a median temperature of \textasciitilde22~K, which is higher than the typical temperature for starless, unbound cores (see, e.g., \citet{marsh2016,dwt2016}). As these sources are faint at 250~$\mu$m (despite having integrated flux densities greater than the 150 mJy contamination limit), their derived masses would be low, even under the assumption of low dust temperatures typically found within prestellar cores (\textasciitilde10 K). The majority of these sources are also unresolved, or marginally resolved, with 63\% having an undeconolved radius lower than 18.5\arcsec. We therefore include 19 sources which have $S^{\textrm{tot}}_{250}>$100 mJy, but $M<$0.001 M$_{\odot}$, in addition to the 43 sources already discussed. Within the catalogue, we round these sources to 0.001 M$_{\odot}$, and note them as `low mass' within both catalogue tables. All 62 sources which were excluded from the scientific discussion, but included within the catalogue, are given a negative `core type' flag of -1, as their nature is uncertain. We note that many of the sources with this flag have 2MASS detections within the SIMBAD database, which supports the argument that these sources are not likely to be starless cores.
\begin{figure}[!h]
	\begin{center}
		\begin{minipage}{1.0\linewidth}
		\resizebox{1.0\hsize}{!}{\includegraphics[angle=0, trim={0.5cm 1.5cm 0.5cm 00cm}]{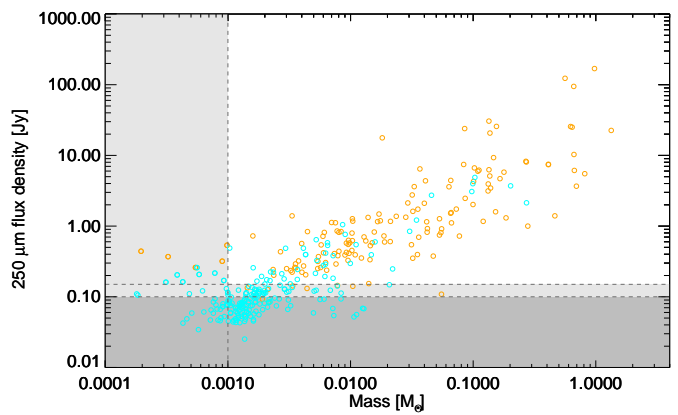}}
		\end{minipage}
	\end{center}
\caption{Mass against integrated flux density at 250~$\mu$m for all sources, regardless of their bound state or subregion. The orange circles and cyan circles show sources that lie within the western and eastern tiles, respectively. The lower dashed line shows the approximate 100 mJy 250-$\mu$m contamination limit, below which we exclude sources from the catalogue on the ground that they are most likely are extragalactic. The upper dashed line shows the 150 mJy cut applied to the sources, below which sources are included in the catalogue, but excluded from the scientific discussion because a high fraction of them may be extragalactic. The vertical dashed line shows the 0.001 M$_{\odot}$ mass limit, below which we exclude cores from the scientific discussion. In general, sources within the dark grey area are excluded from the catalogue and discussion. Sources within the light-grey area are included within the catalogue, but excluded from the scientific discussion, and given a `core type' of -1. Sources within the white area are included within the catalogue and scientific discussion.}
     \label{mass_250flux}%
\end{figure}

\section{Comparison with optical extinction}
\label{sec:app.dobashi}
We compared the \textit{Herschel} derived low resolution column density map with the optical extinction maps produced using Digitized Sky Survey I (DSS1) maps \citep{dobashi2005}. The resolution of the optical extinction maps is \textasciitilde$6'$, compared to the lowest resolution of ~$36''$. We convolved the low-resolution column density map to the same grid as the DSS1 optical extinction map, and regridded the data onto that of the DSS1 map. We then extracted the area for which both maps shared a common area. We then compared each pixel of the column density map to the respective $A_{\text{V}}$ extinction value found by \citet{dobashi2005}. Figure \ref{dobashi} shows the DSS $A_{\text{V}}$ extinction map (see Figs. 9 and 18-1-1 of \citealt{dobashi2005} for extinction maps around CrA) with overlaid contours from the convolved \textit{Herschel} low-resolution column density map. The maps are in good agreement, with only minor exceptions. The lower panel of the same figure shows a plot of the DSS1 optical extinction magnitudes against the \textit{Herschel} column density values along the line of sight. Again, there is good agreement between these quantities below N(H$_{2}$) column densities of ~$7.5\times10^{21}$~cm$^{-2}$. For column densities below this value, we derived a relation of $N(\textrm{H}_{2})/A_{\text{V}}= 1\times10^{21}$~cm$^{-2}$~mag$^{-1}$, in agreement of the relation, $N(\textrm{H}_{2})/A_{\text{V}}= 9.4\times10^{20}$ cm$^{-2}$~mag$^{-1}$, found by \citet{bohlin1978}.
\begin{figure}[!h]
	\begin{center}
		\begin{minipage}{1.0\linewidth}
		\resizebox{1.0\hsize}{!}{\includegraphics[angle=0, trim={0cm 2cm 0cm 0.5cm}]{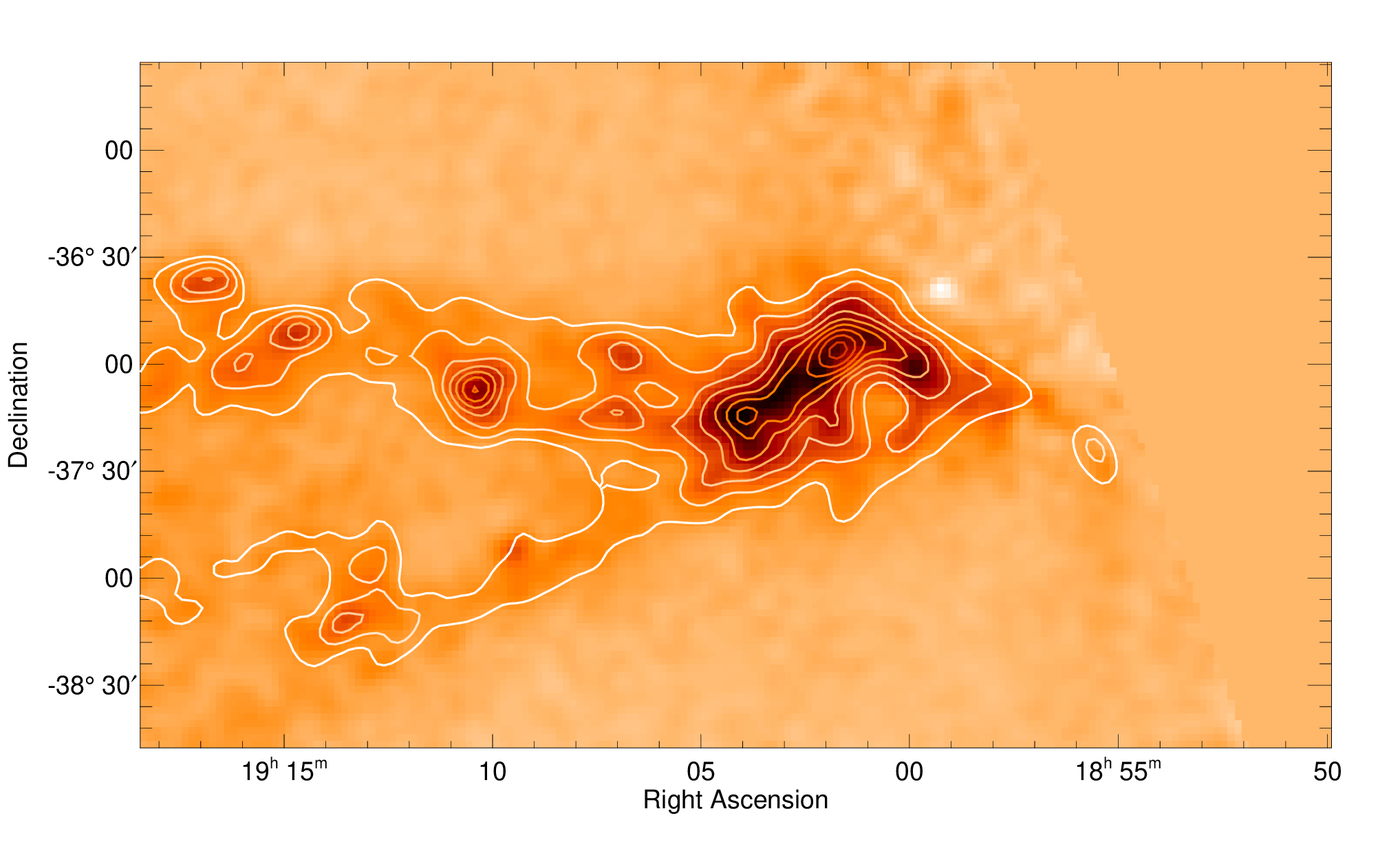}}
		\resizebox{1.0\hsize}{!}{\includegraphics[angle=0, trim={0cm 2cm 0cm 0.5cm}]{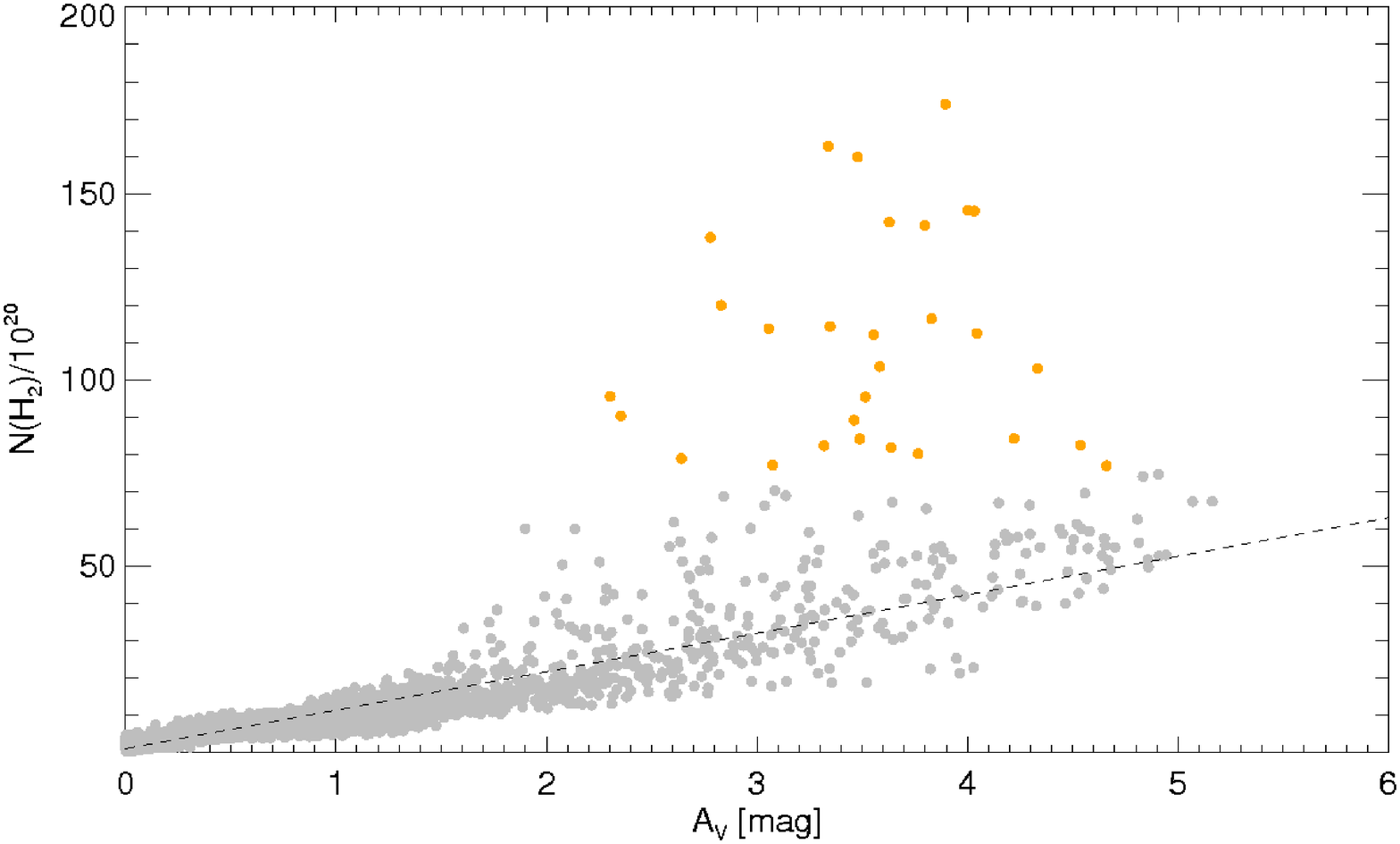}}
		\end{minipage}
	\end{center}
\caption{The top panel shows the $A_{\text{V}}$ extinction map derived by \citet{dobashi2005}, with the \textit{Herschel} low column density map contours overlaid in white. The contours start from 3$\sigma$, with each consecutive contour being 1.5 times the previous level. The lower panel is a plot of $A_{\text{V}}$ against the column density along the line of sight. The grey dashed line is the fit $N(H_{2})/A_{\text{V}}= ~1\times10^{21}$~cm$^{-2}$~mag$^{-1}$.The orange points are those that were excluded from the fitting process, as they are in very high extinction areas, where the optical extinction relation breaks down.}
     \label{dobashi}%
\end{figure}
\clearpage
\newpage
\section{Getsources extraction areas and selected subregions}
\label{sec:app.tiles}
Figure \ref{getsources_tiles} shows the two areas that were selected for extraction from CrA, outlined in orange and blue. The background image is the high resolution column density map. A small overlap is visible, though this did not add any complications to the source extraction process. 
\begin{figure*}[!h]
	\centering
	\includegraphics[ width=1.0\hsize, angle=0, trim={2.5cm 2.5cm 2.5cm 2.5cm}]{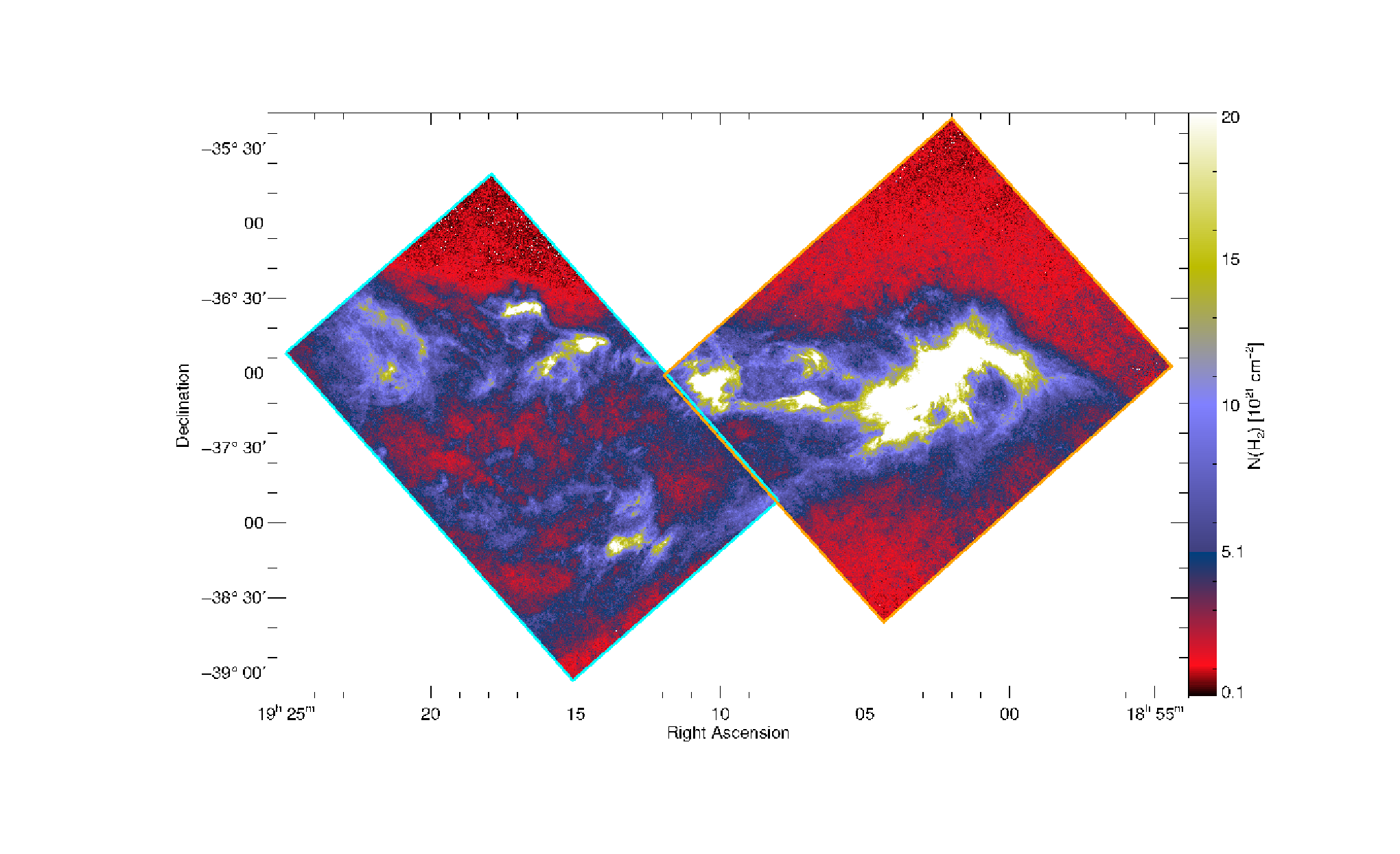}
\caption{The two areas we used for extraction with \textsl{getsources} are shown in blue and orange. The background is the high resolution column density map.}
 \label{getsources_tiles}
\end{figure*}
\begin{figure*}[!h]
	\centering
	\includegraphics[ width=1.0\hsize, angle=0, trim={2.5cm 2.5cm 2.5cm 2.5cm}]{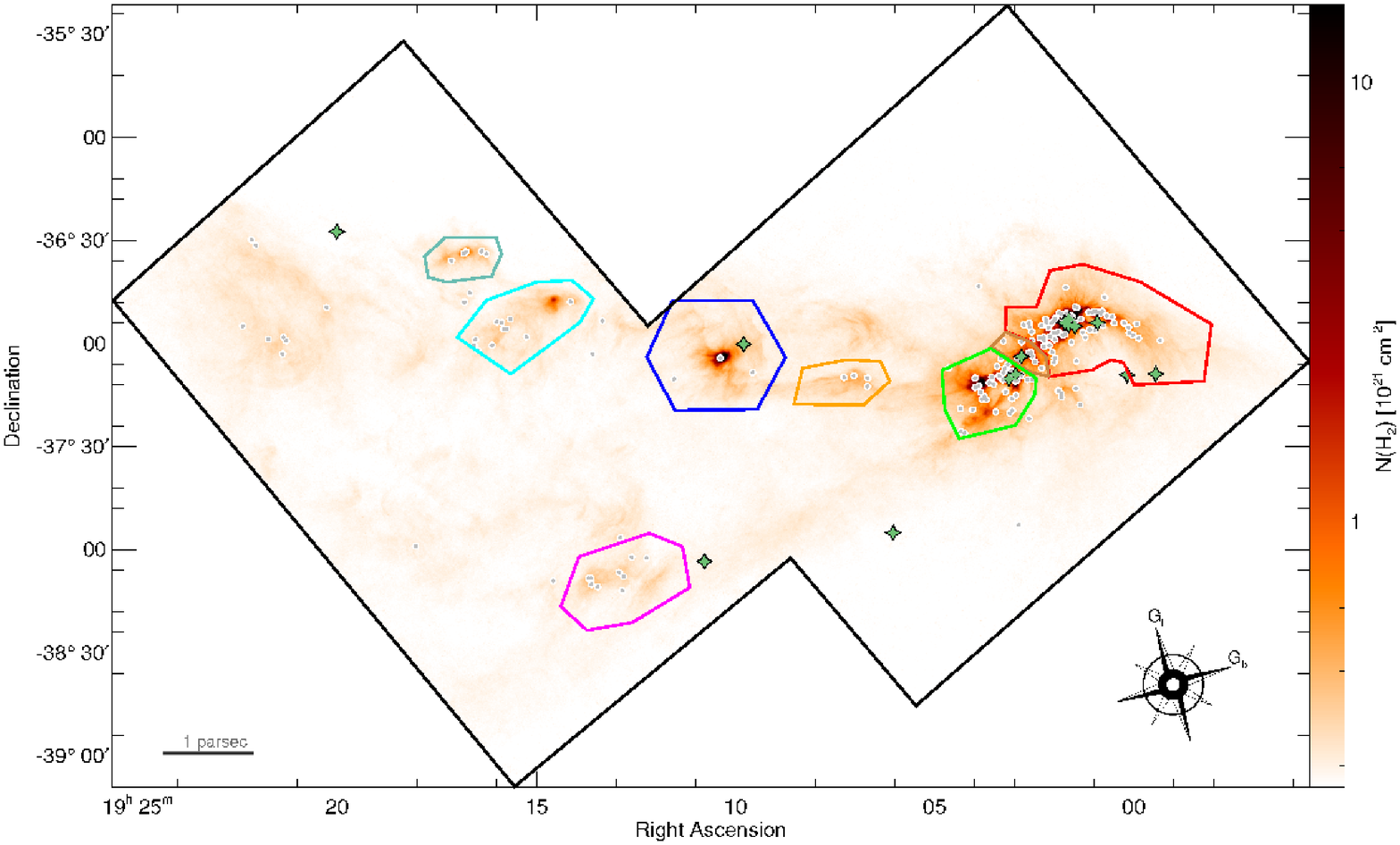}
\caption{The following colours are used for the subregions; red for CrA-A; brown for CrA-B; green for CrA-C, orange for CrA-D; dark blue for CrA-E; cyan for CrA-F; sea green for CrA-G and magenta for CrA-H. The background is the high resolution column density map.}
 \label{fig:cra_subregs}
\end{figure*}
\clearpage
\newpage
\section{Coronet region filaments}
\label{sec:app.coronetfils}
Figure \ref{coronet_fil_zoom} shows a zoom-in of the filaments around the Coronet region. Both \textsl{DisPerSE} and \textsl{getfilaments} identified filaments are shown. There is a good agreement between the two algorithms. The background \textsl{getfilaments} map is shown as a map of column density, for which the scale is given on the right of the figure.
\begin{figure*}[!h]
	\centering
	\includegraphics[ width=0.95\hsize, angle=0, trim={0.0cm 0.0cm 0.0cm 0.0cm}]{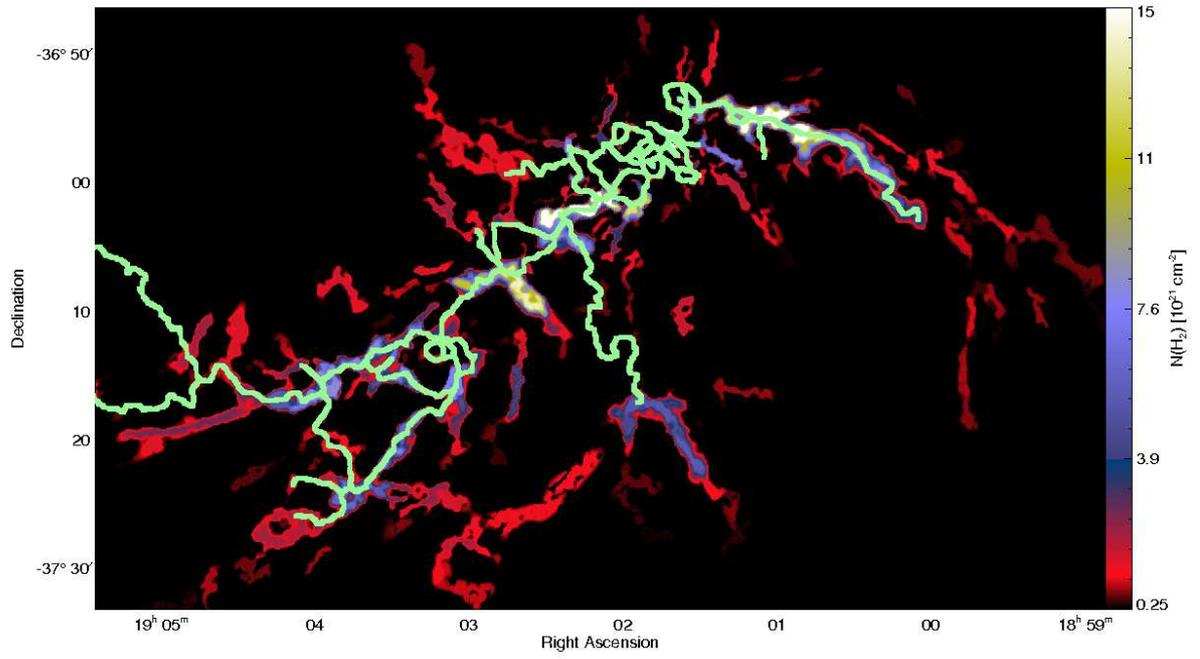}
\caption{A zoom-in of the filamentary structure around the Coronet. Filaments found by \textsl{DisPerSE} are shown in green. The background image shows filaments identified by \textsl{getfilaments}, where angular scales up to $145\arcsec$ are shown.}
 \label{coronet_fil_zoom}
\end{figure*}
\clearpage
\newpage
\section{Data}
\label{sec:app.data}
We show here the regions that have been observed by the \textit{Herschel} PACS and SPIRE instruments. Figure A1-5 show the 70-$\mu$m, 160-$\mu$m, 250-$\mu$m, 350-$\mu$m, and 500-$\mu$m data, respectively. All images share a common $3''$ pixel grid for comparison.
\newpage
\renewcommand\thefigure{\thesection.\arabic{figure}}
\setcounter{figure}{0}
\begin{figure*}
	\centering
	\includegraphics[ width=\vsize, angle=90, trim={4cm 0cm 1cm 0cm}]{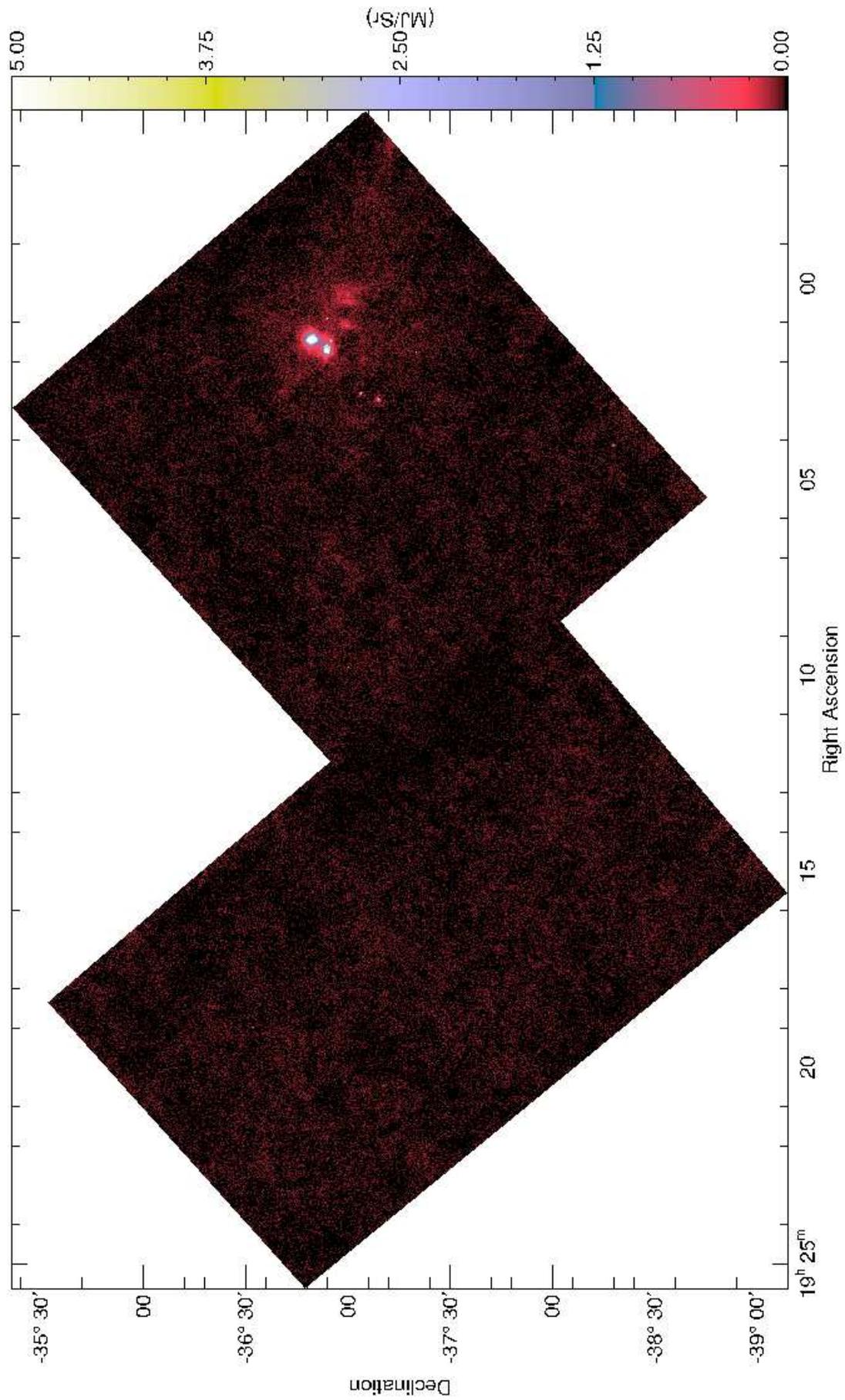}
\caption{The 70-$\mu$m flux density measured in Corona Australis with PACS.}
\end{figure*}

\newpage
\begin{figure*}
	\centering
	\includegraphics[ width=\vsize, angle=90, trim={4cm 0cm 1cm 0cm}]{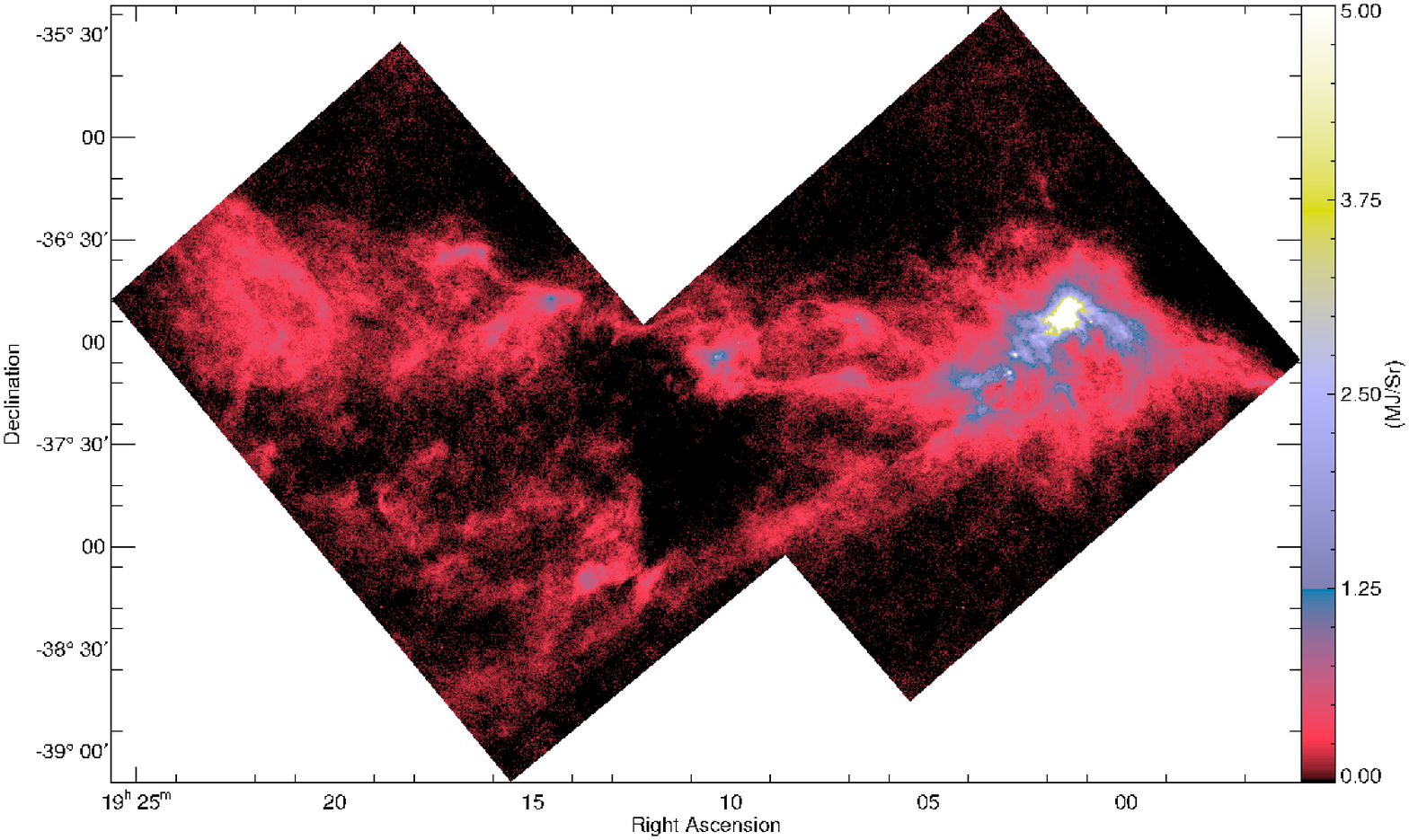}
\caption{The 160-$\mu$m flux density measured in Corona Australis with PACS.}
\end{figure*}

\newpage
\begin{figure*}
	\centering
	\includegraphics[ width=\vsize, angle=90, trim={4cm 0cm 1cm 0cm}]{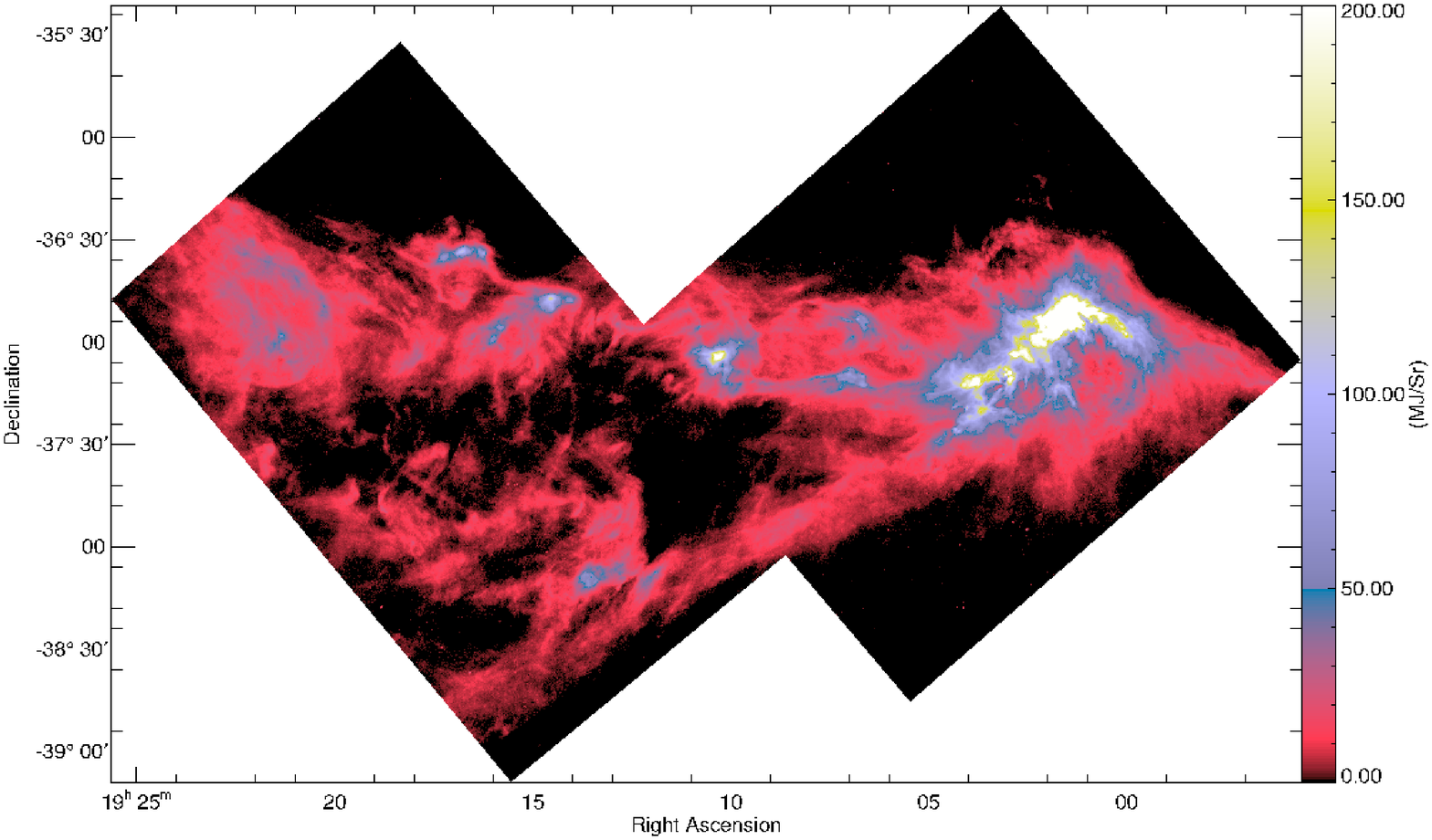}
\caption{The 250-$\mu$m flux density measured in Corona Australis with SPIRE.}
\end{figure*}

\newpage
\begin{figure*}
	\centering
	\includegraphics[ width=\vsize, angle=90, trim={4cm 0cm 1cm 0cm}]{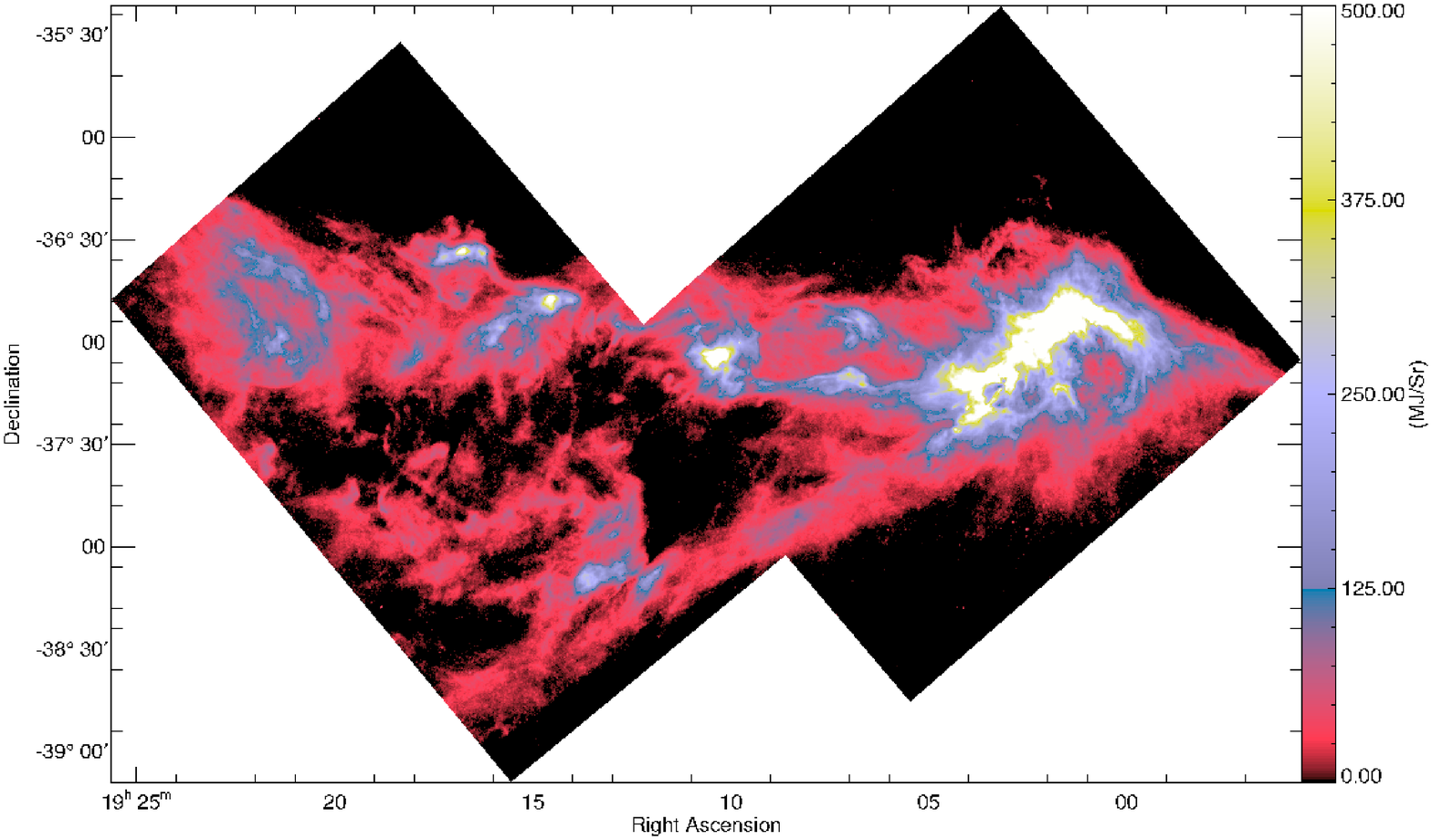}
\caption{The 350-$\mu$m flux density measured in Corona Australis with SPIRE}
\end{figure*}

\newpage
\begin{figure*}
	\centering
	\includegraphics[ width=\vsize, angle=90, trim={4cm 0cm 1cm 0cm}]{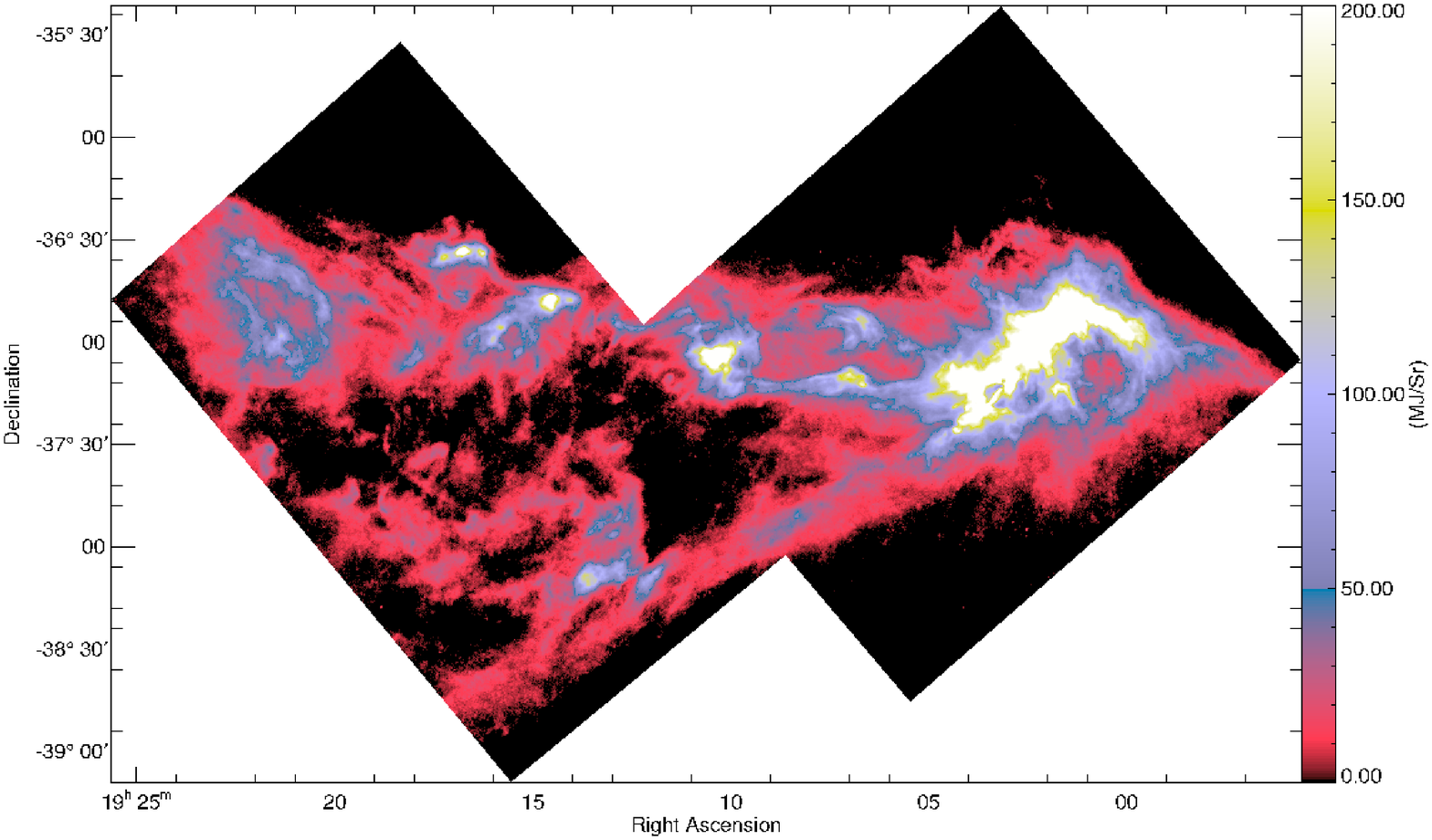}
\caption{The 500-$\mu$m flux density measured in Corona Australis with SPIRE.}
\end{figure*}
\clearpage
\newpage

\end{document}